\documentstyle[preprint,aps,tighten]{revtex}

\input epsf

\def\Journal#1#2#3#4{{#1} {\bf #2}, #3 (#4)}

\def\NPB{{\em Nucl. Phys.} B}
\def\NPA{{\em Nucl. Phys.} A}
\def\PLB{{\em Phys. Lett.}  B}
\def\PRL{\em Phys. Rev. Lett.}
\def\PRD{{\em Phys. Rev.} D}
\def\PRC{{\em Phys. Rev.} C}
\def\ZPC{{\em Z. Phys.} C}

\newcommand{\expl}{\langle \!\langle}
\newcommand{\expr}{\rangle \!\rangle}

\begin{document}
\draft

\preprint{UGI-99-28}

\title{Stochastic treatment of Disoriented Chiral Condensates
     \\within a Langevin description}

\author{Zhe Xu\footnote{E-mail: Zhe.Xu@theo.physik.uni-giessen.de}
and Carsten Greiner\footnote{E-mail: Carsten.Greiner@theo.physik.uni-giessen.de}}

\address{Institut f\"ur Theoretische Physik, Justus-Liebig-Universit\"at
Giessen,\\
D-35392 Giessen, Germany}

\date{November 1999}

\maketitle

\begin{abstract}
Applying a microscopically motivated semi-classical Langevin description
of the linear sigma model we investigate for various different scenarios
the stochastic evolution of a disoriented chiral
condensate (DCC) in a rapidly expanding system. Some particular emphasize is
put on the numerical realisation of colored noise in order to treat the
underlying dissipative and nonmarkovian stochastic equations of motion.
A comparison with an approximate markovian (i.e. instantaneous) treatment
of dissipation and noise will be made in order to identify the
possible influence of memory effects in the evolution of the
chiral order parameter.
Assuming a standard
Rayleigh cooling term to simulate a D-dimensional scaling expansion
we present the probability distribution in the low momentum pion number
stemming from the relaxing zero mode component of the chiral
field.
The best DCC signal is expected for
initial conditions centered around
$\langle \sigma \rangle \approx 0 $ as would be the case of effective
light `pions' close to the phase transition.
By choosing appropriate idealized global parameters for the expansion
our findings show that an experimentally feasible DCC, if it does exist
in nature,
has to be a rare event with some finite probability
following a nontrivial and nonpoissonian distribution on an event by event
basis. DCCs might then be identified experimentally by inspecting
higher order factorial cumulants $\theta _m$ ($m\ge 3$)
in the sampled distribution.

\end{abstract}

\section{Introduction and Motivation}
\label{sec:intro}

The prime intention for ultrarelativistic heavy ion collisions
is to study the behaviour of nuclear or hadronic matter at extreme
conditions like very high temperatures and energy densities.
One of the major goals, particular at the upcoming RHIC facilities,
is to find evidence for a new state of
deconfined partonic matter, the quark gluon plasma (QGP) \cite{Mul85}.
Besides the confinement-deconfinement transition one also expects
a transition of hot hadronic matter, where chiral symmetry is being restored.
Lattice calculations of quantum chromodynamics (QCD) give the belief,
that both transitions do occur at the same critical temperature $T_c$
at vanishing net-baryon densities \cite{DeTar}.

The formation of so-called disoriented chiral condensate (DCC) \cite{Raja}
has been considered as the maybe most prominent signature
for the restoration of
chiral symmetry occuring in the ongoing evolution of the hot matter
from the chirally restored to the
chirally broken phase. The idea is here that in the course of the evolution
of the system from the initially (and only transiently existing)
unbroken phase with (the order parameter being) $\langle \bar{q} q \rangle \approx 0$
to the true
groundstate with $\langle \bar{q} q \rangle \neq 0$ the pseudo-scalar
condensate $\langle \bar{q} {\bf \tau} \gamma_5 q \rangle $ might assume
temporarily non-vanishing values. This misaligned condensate has the same quark content
and quantum numbers as do pions and thus essentially constitutes a classical
pion field. The subsequent relaxation of this field back to the
alignment of the outside vacuum could then lead to an excess
of low momentum pions in a single direction in isospin space.

The possible occurrence of a semi-classical and coherent pion field
was first raised in a work of Anselm \cite{Ans}
but the idea of forming DCC was made widely known due to Bjorken
\cite{Bj92} and Rajagopal and Wilczek \cite{RW93}.
Since then many works appeared on various aspects of DCC
formation in heavy ion collisions. As the microscopic physics governing
the chiral phase transition is not known well enough, one typically
employs effective field theories like the linear $\sigma $-model \cite{RW93}
in order to describe this non-equilibrium phenomenon. On the other
hand the description of quantum field theory out of equilibrium is interesting
in its own right and thus has given raise
to a major attraction for theoretical studies in order to describe
the evolution of disoriented chiral condensates \cite{Raja1}, like
e.g. standard Hartree factorization or large $N_c$-expansion methods
\cite{Boy95}.
Usually these considerations assume an initial state
at high temperature in which chiral symmetry is
restored by vanishing collective fields.
Independent thermal fluctuations in each isospin direction
of the $O(4)$ $\sigma $-model are present.
This configuration sits on the top of the barrier of
the potential energy at zero temperature, so a sudden
cooling of the system supposedly brings it into an
unstable state. This picture is referred to as
the quenched situation \cite{RW93}.
The spontaneous growth and subsequent
decay of these configurations would give raise to large collective fluctuations
in the number of produced neutral pions compared to charged pions,
and thus could provide a mechanism explaining a family of peculiar
cosmic ray events, the Centauros \cite{La80}. A deeper reason for these
strong fluctuations lies in the fact that all pions
constituting the classical and coherent field are sitting in the
same momentum state and the overall wavefunction can carry no isospin
\cite{Gr93}. 

The proposed quench scenario \cite{RW93} assumes
that the effective potential governing the evolution
of the long wavelength modes immediately turns
to the classical one at zero temperature.
This is a very drastic assumption as the soft classical
modes completely decouple from the residual thermal
fluctuations at the chiral phase transition temperature.
Such an idealized scenario of immediate decoupling
might hold effectively if the expansion and the
associated cooling of the fireball occurs sufficiently fast \cite{Ra96a}.
An alternative, the annealing scenario \cite{GM94}, was suggested by
Gavin and M\"uller. They used the one-loop
effective potential instead of the classical one including thermal
fluctuations. For moderate expansion and cooling it was shown that
the system can exhibit longer instable periods and thus should
lead even to a stronger enhancement of the soft pionic fields.
On the other side, both scenarios assume that
the initial fluctuation of the order parameter at the beginning
of the DCC formation are centered around zero with a sufficiently
small width in a rather
ad hoc manner. Preparing the initial configuration with stronger initial fluctuations,
no DCC formation has been observed \cite{As95}.
If the soft field remains in thermal contact with
the fluctuations giving raise to the one-loop potential,
then one also has to allow for appropriate thermal fluctuations
in the initial conditions \cite{Bi96,Ra97}.
The proposed quenched initial conditions within the linear sigma model
seem statistically unlikely.

The likeliness of an instability leading potentially to a DCC
event during the evolution with a continuous contact 
with the heat bath of thermal pions was investigated
by Biro and one of us by means of simple Langevin equations \cite{PRL}.
There the average and statistical properties of
individual solutions
were studied with the emphasis on such periods of
the time evolution when the transverse mass $\mu_{\bot }$
of the pionic modes becomes imaginary
and therefore an exponential growth of unstable fluctuations in the
collective fields might be expected.
It was found that for different realistic initial volumes
individual events of an ensemble lead to sometimes significant growth
of fluctuations \cite{PRL,Ch99}. Subsequent investigation by us
in fact leads to the idea of stochastic formation of DCC
for particular special stochastic evolution of the order parameter
\cite{GXB98}.

This idea is what we want to detail in the present study in more depth.
Our main conception is that the order parameter before and after the onset
of the chiral phase transition still interacts (dissipatively) with
its (nearly) thermal surrounding of thermal
(or `hard') pions, which then give raise also to large
fluctuations in the evolution.
This one can interpret as a breakdown of the standard mean-field approximation.
Applying a microscopically motivated semi-classical Langevin description
of the linear sigma model we investigate for various different scenarios
the stochastic evolution of a single disoriented chiral
condensate in a rapidly expanding system assuming a
D-dimensional scaling expansion \cite{Ra96a,GM94,Bi96,PRL}.
Our stochastic description will allow for a systematic recording
of the statistically possible initial configurations of the order parameter.
Furthermore it also describes
the nontrivial influence of dissipation and fluctuations on the
nonequilibrium evolution and the coherent amplification
on the collective pionic zero mode fields during and after the onset
of the phase transition.

It remains to answer the important question of
how likely particular nonequilbrium evolutions of a statistically generated
ensemble will lead to the formation of a `large' DCC-domain, which will
depend nontrivially on the initial and subsequent fluctuations
suffered by the surrounding in the course of the evolution.
For this we calculate the effective pion number contained in the pionic
collective field emerging by the rolling down of the chiral
fields to its true vacuum values,
which by subsequent emission will be freed as low momentum
pions. With this number at hand we can make the decision whether
accordingly these pions can contribute to an experimentally
measurable enhancement of low momentum pions and thus might
provide indeed a signal for the occuring chiral phase transition.
As it turns out the probability distribution in the pion number
contains interesting new information for the characteristics
of the non-equilibrium evolution
stemming from the relaxing zero mode component of the chiral
field. In the interesting cases the to be expected yield
in low momentum pions do not follow
an usual and simple statistical distribution, but posesses large and nontrivial
(non-poissonian) fluctuations.
The best DCC signal is expected for
initial conditions centered around
$\langle \sigma \rangle \approx 0 $ as would be the case of effective
light `pions' close to the phase transition.
By choosing some idealized global parameters for a
$(D=)\, $3-dimensional, spherical expansion,
our findings show that an experimentally feasible DCC, if it does exist
in nature,
has to be a rare event with some finite probability
following a nontrivial and nonpoissonian distribution on an event by event
basis.
Comparing with an additional incoherent background the fluctuations in
the low momentum pion number might be revealed in the nonvanishing
of higher order factorial cumulants $\theta _m$ ($m\ge 3$).
Admittingly, we have to say that although we do
stress a new physical picture our study has still to be seen as a fairly
idealized scenario. Nevertheless, we believe that our results are interesting
in their own right and should serve as a simplified estimate for the
nontrivial late dynamics encountered in an ultrarelativistic
heavy ion collision.

In the next section \ref{sec:Lanlinsig} we describe the linear $\sigma $-model
within a Langevin treatment. For this we will first summarize
the theoretical ideas behind a semi-classical Langevin
description for the soft (i.e. low momentum) fields in thermal quantum field
theory. The hard modes are treated as thermal quasiparticles which constitute a
surrounding, open heat bath.
We then discuss the model introduced in \cite{PRL} in more depth for simulating
the evolution of the order parameter and the collective zero mode pionic fields.
The damping term entering the dynamical evolution will be discussed
and a systematic recording
of the statistically possible initial configurations of the order parameter
for finite volumes will be given.
The final equations of motion to be used
for the dynamical evolution
including a D-dimensional scaling expansion for modeling the possible formation
of DCC are then stated.
As a characteristic for describing the `strength' of a DCC we consider
the effective pion number content of the evolving domain.
Some particular emphasize will then be
put first in section \ref{sec:nonmarkov}
on the numerical realization of colored noise in
order to treat the
underlying dissipative and nonmarkovian stochastic equations of motion.
This, to the best of our knowledge, is the first
numerical treatment of
nonmarkovian Langevin equations in thermal quantum field theory
and might be of importance for other related topics.
A comparison with a standard markovian (i.e. instantaneous) treatment
of dissipation and noise will be made. For the later simulations it shows that
it is sufficient to consider only the markovian approximation, which is
numerically much easier to handle.
In section \ref{sec:stochDCC} we finally present
numerical results of the simulation
on the evolution of a coherent pionic field.
Four different scenarios,
annealing or quench with initial conditions governed by effective `light'
or physical mass pions, will be investigated. We calculate the pion
number for a single domain and the distribution of the pion number,
which are the observables relevant to the experimental detection of DCC,
and which also will give quantitative predicition on the
possibility of forming an experimentally accessible DCC.
The unusual distribution in the number of low momentum pions is further
analyzed by means of a cumulant expansion.
To be more realistic we also take into account an
additional incoherent (poissonian) contribution on the production
of soft pions.
Inspecting the resulting distribution for the low momentum pion number
we show that the high order factorial cumulants can still be large.
This provides a new signature to identify possible DCC formation.
Some conclusions for possible experimental searches are drawn.
We close our findings with
a summary. In appendix \ref{app:Dissip} we give a microscopic derivation of
the frequency dependence of the dissipation kernel being employed. Appendix
\ref{app:Noise} describes in more detail our strategy for simulating
gaussian nonwhite, colored noise for an arbitrary noise kernel.
Appendix \ref{app:Cumulant} gives a brief reminder on the cumulant expansion for
statistical distributions.

\section{Langevin description of linear sigma model}
\label{sec:Lanlinsig}

In this section we develop in some more detail the Langevin description
of the linear $\sigma$-model introduced in \cite{PRL,GXB98}.
The starting point is the phenomenological Lagrangian which is given by
\begin{equation}
\label{Lagrlinsig}
{\cal L} \, = \, \frac{1}{2}\,\partial_{\mu}\phi_a\partial^{\mu}\phi^a
\, - \,
\frac{\lambda}{4}
(\phi_a\phi^a  -v^2)^2 \, + \, H\phi_0 \, ,
\end{equation}
where $\phi _a = (\sigma, \pi _1, \pi _2, \pi _3)$. We employ the
standard parameter $f_{\pi } = 93 $ MeV for the pion decay constant,
$m_\pi=140$ MeV for the vacuum mass of the pion and $m_\sigma\approx 600$
MeV for the `mass' of the $\sigma $-meson.
For the three parameters in (\ref{Lagrlinsig}) one then finds
\begin{equation}
\label{O4parameter}
\lambda \, = \, \frac{m_{\sigma}^2-m_{\pi}^2}{2 f_{\pi}^2}  \approx  20 , \,
v^2=f_{\pi}^2-\frac{m_{\pi}^2}{\lambda}
=(87 \, \mbox{MeV})^2 ,
H=f_{\pi} m_{\pi}^2\
=(122 \, \mbox{MeV})^3  .
\end{equation}
The linear $\sigma$-model represents an effective chiral theory of the
low energy properties of QCD. It can be motivated in more
theoretical depth from QCD by the modern methods of the renormalization group
\cite{JW99}.
At finite temperature, to leading order in $ \lambda $,
the thermal fluctuations $ \langle \delta \phi ^2 (\vec{x},t) \rangle $
of the pions and $\sigma $-mesons do generate an effective Hartree type
dynamical mass giving raise to an effective temperature dependent potential.
In the high temperature expansion this results in \cite{BK96}
\begin{equation}
\label{O4thermmass}
m_{th}^2 \, \stackrel{(T \gg m_{\pi }, m_{\sigma })}{\longrightarrow } \,
\frac{\lambda }{4} \left( \frac{N+2}{3} \right) T^2
\stackrel{N=4}{\longrightarrow } \,
\frac{\lambda }{2} \, T^2 \, .
\end{equation}
The resulting chiral phase transition is compatible with the expectations
of lattice gauge QCD calculations \cite{G91}. There exist also convincing
theoretical arguments \cite{PW84} that the chiral phase transition
near the critical temperature $T_c$ of QCD with two massless quarks
lies in the same universality class as an $O(4)$-Heisenberg magnet
and thus (in this idealized case of massless quarks)
exhibits a true second order phase transition which can
be described in the Landau-Ginzburg theory by means of an effective
linear $\sigma $-model. In this sense one considers the
linear $\sigma $-model as an appropriate realisation of the
chiral behaviour of QCD over the whole range in temperature, though
the effective parameters near $T\approx T_c$ need not really be equivalent
to those at $T=0$.

We now address in an intuitive model how one can go beyond
the mean field level for the semi-classical chiral collective fields.
Our main physical conception will be that the order parameter
and the collective fields before and after the onset
of the chiral phase transition still interacts (dissipatively) with
its (nearly) thermal surrounding of thermal
(or `hard') particles.
To outline these ideas more conceptually we will first summarize
in the following subsection the theoretical reasonings behind
a semi-classical Langevin description of the soft, i.e. low momentum fields.

\subsection{Equations of motion for long wavelength modes in a heat bath }
\label{sec:softmode}

One of the recent topics in especially nonabelian
massless quantum field theory at finite temperature
or near thermal equilibrium concerns the evolution and behaviour of the long
wavelength modes. These modes often lie entirely in the non-perturbative regime.
Therefore solutions of the classical field equations in Minkowski space have
been widely used in recent years to describe long-distance properties
of quantum fields that require a non-perturbative analysis.
A justification of the classical treatment of the long-distance dynamics
of weakly coupled bosonic quantum
fields at high temperature is based on the observation that the average 
thermal amplitude of low-momentum modes is large and
approaches the classical equipartition limit
\begin{equation}
n(\omega_{\bf p}) = \left(e^{\hbar\omega_{\bf p}/T} -1 \right)^{-1}
\buildrel \vert {\bf p}\vert \to 0 \over \longrightarrow
{T\over m^*} \gg 1 \label{climit}
\end{equation}
in the case for a sufficiently small generated dynamical mass $m^* \ll T$.
On the other hand the thermodynamics of a classical
field is only defined if
an ultraviolet cut-off $k_c$ is imposed on the momentum {\bf p} such as a
finite lattice spacing $a$.
In a recent paper \cite{GMu97} it was shown, at least principally,
how to construct an effective semi-classical action
for describing not only the classical
behaviour of the long wavelength modes below some given
cutoff $k_c$, but taking into account also perturbatively the interaction among the soft
and hard modes. The resulting effective action $S_{{\rm eff}}[{\rm soft}]$,
which one has to interpret as a stochastic, dissipative action
\cite{GMu97,GL98a},
turns out to be complex,
leading to a stochastic equation of motion for the soft modes.
If the hard modes are already in thermal equilibrium then
the evolution of the soft modes is described by a set
of generalized Langevin equations - the equations of motion
corresponding to the above complex effective action.

\begin{figure}[h]
\centerline{\epsfysize=3cm \epsfbox{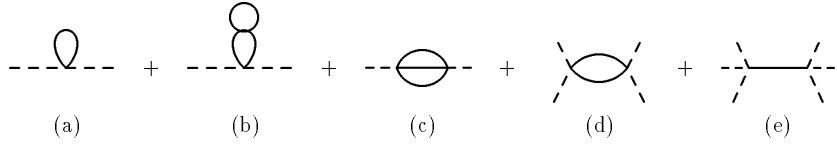}}
\caption{Feynman diagrams contributing to
the influence action $S_{IF}$ up to second order ${\cal O}(g^4)$
in $S_{{\rm int}}$.}
\label{DCCfig1}
\end{figure}

We briefly sketch the main strategy following
\cite{GMu97} by considering a scalar field with interaction
$ {\cal L}_{{\rm int}} = \frac{g^2}{4!} \, \tilde{\phi }^4$.
The splitting of the Fourier-components,
$ \tilde{\phi }(p,t) = \phi(p \le k_c,t) \, + \, \varphi(p>k_c,t)$,
leads to the following interaction part in the action
\begin{equation}
 S_{{\rm int}}[\phi,\varphi] = - \int_{t_0}^t \! d^4x \,\,
     \left( \frac{g^2}{4!}\varphi^4 + \frac{g^2}{3!} \left(
     \phi^3\varphi + \frac{3}{2} \phi^2 \varphi^2 
     +\phi\varphi^3 \right) \right).
\end{equation}
By integrating out the hard modes up
to second order in the interaction, one obtains the effective action
(or influence functional) $S_{IF}[\phi,\phi']$ for the soft modes
following the Feynman-Vernon approach \cite{Fe63}.
Fig. \ref{DCCfig1}
shows the resulting non-vanishing diagrams contributing to $S_{IF}$.
The contributions from diagram (a) and (b) are real and
generate the Hartree like dynamical mass term.
Moreover one notices that
Feynman graphs contributing at order ${\cal O}(g^4)$
(diagrams (c), (d) and (e))
to the effective action contain imaginary contributions.
Their real part leads
to dissipation (like in linear response theory) whereas
the imaginary part drives the fluctuations of the hard particles on the
soft modes.
From the effective action semi-classical, stochastic equations of motion result,
which have the general shape
\begin{equation}
\label{GLangEoM}
\Box \phi + \bar{m}^2 \phi + \frac{g^2}{3!} \phi^3
+ \sum_{N=1}^{3} \frac{1}{(2N-1)!} \phi^{N-1}
({\rm Re} \Gamma_{2N} ) \phi^N =
\sum_{N=1}^3 \phi^{N-1}  \xi_N \, .
\end{equation}
Here $\Gamma_{2N}$ denotes the effective contribution with
$2N$ soft legs, $\bar{m}^2$ the resummed Hartree-Fock
self energy (cactus graphs) and $\xi_N$ are associated
noise-variables with a correlation
$ \langle \xi_N \, \xi_N' \rangle = {\rm Im}\Gamma_{2N} $.
These generalized Langevin equations (\ref{GLangEoM}) are similar in spirit
to those obtained by
Caldeira and Leggett in their discussion of quantum Brownian motion \cite{CL83}.

For the sake of simplicity we concentrate from now only on the
contribution of the sunset diagram (c) of Fig. (\ref{DCCfig1}),
i.e. the N=1 contribution of (\ref{GLangEoM}). Performing a Fourier transformation
to (\ref{GLangEoM})
yields the semi-classical stochastic field equation
for a soft mode with momenta ${\bf k}$ \cite{GMu97,GL98a}
\begin{eqnarray}\label{sm}
\!\!\!& &\!\!\!\ddot \phi({\bf k},t)+\left ( {\bf k}^2+m^{*^2}\right )
\phi({\bf k},t)+\frac{\tilde g^2}{6}
\int^{k_c}\!\frac{d^3k_1 d^3k_2}{(2\pi)^6}\,\,
\theta(k_c-|{\bf k}-{\bf k}_1-{\bf k}_2|)\phi({\bf k}_1,t)\times\nonumber\\
\!\!\!& &\!\!\!\quad \phi({\bf k}_2,t)
\phi({\bf k}-{\bf k}_1-{\bf k}_2,t)+2\int_{-\infty}^t\!\!dt'\,\,
\Gamma({\bf k},t-t')\,\dot \phi({\bf k},t')=\xi({\bf k},t)\,.
\end{eqnarray}
$\Gamma({\bf k},t-t')$ and $\xi({\bf k},t)$ denote the (real valued) dissipation kernel
and the noise source, respectively, due to the thermal fluctuations
of the integrated out hard particles.
The dissipation kernel is related to the standard imaginary part of the
sunset diagram via \cite{GL98a}
\begin{equation}
\label{GamSret}
\Gamma({\bf k},\omega) \,
\equiv \, \frac{-{\mbox{Im}} \Sigma^{\mbox{ret}}({\bf k},\omega)}{\omega}
\, ,
\end{equation}
which follows by a partial integration of
\begin{equation}
\label{partint}
\int_{-\infty }^t \, dt' \, \Sigma_{{\rm ret}}({\bf k},t-t') \,
\phi ({\bf k},t') \, = \,
- 2\Gamma ({\bf k},\Delta t=0) \phi ({\bf k},t) +
 2\int\limits_{-\infty}^{t} dt' \, \Gamma ({\bf k},t-t') \,
 \dot{\phi} ({\bf k},t') \, .
\end{equation}
(The integration constant represents an additional momentum dependent shift
in the dynamically generated mass and will be neglected further on.)
The explicit calculation of the dissipation kernel is given in the appendix
\ref{app:Dissip}.

Within the present treatment the noise turns out to be gaussian, but colored,
characterized by the (ensemble averaged) correlation function \cite{GMu97,GL98a}
\begin{equation}\label{dft1}
\expl \xi({\bf k},t) \xi({\bf k}',t')  \expr \, = \,
(2\pi )^3 \delta ^3 ({\bf k} + {\bf k}') I({\bf k},t-t')
\end{equation}
or
\begin{equation}\label{dft2}
\expl \xi({\bf k},t) \xi(-{\bf k},t')  \expr \, \equiv \,
V I({\bf k},t-t') \, ,
\end{equation}
where the noise correlation strength is related via a generalized
fluctuation dissipation relation to the dissipation kernel as
\begin{equation}
I (\omega )
\, = \,  \omega \,
\frac{\exp (\hbar \omega /T) +1}{\exp (\hbar \omega /T) -1} \, \Gamma (\omega )
\,
\stackrel{\omega \ll T}{\longrightarrow} \,2\, T  \, \Gamma (\omega ) \, .
\label{Hu5}
\end{equation}
In the high-temperature limit $\omega \ll T$ the noise acting on the dynamics of the
soft modes then fulfills the (entirely) classical relation
\begin{equation}\label{dft}
\expl \xi({\bf k},t) \xi(-{\bf k},t') \expr \, = \, 2TV \Gamma({\bf k},t-t')\,.
\end{equation}
The fluctuation-dissipation-theorem ensures that
the soft modes approach thermal equilibrium precisely
at the temperature $T$ of the hard modes.

When the characteristic time scale in the evolution of
hard modes in the heat bath is short compared to the one of the
soft fields and its coupling to the soft fields is sufficiently weak,
the appropriate (`instantaneous') markovian limit then has a form
\cite{GMu97}
\begin{equation}
\label{Markovlim}
 2\int\limits_{-\infty}^{t} dt' \, \Gamma ({\bf k},t-t') \,
 \dot{\phi} ({\bf k},t') \, \approx \,
 \eta
\dot{\phi } \, ,
\end{equation}
where $\eta=\Gamma(\omega_k=\sqrt{m^2+k^2}) $ in the linear, harmonic
approximation describes
the familiar on-shell plasmon damping rate
(see also appendix \ref{app:Dissip}).
In the semi-classical, high temperature limit
and within the markovian approximation the noise becomes white, i.e.
\begin{equation}
\expl \xi({\bf k},t) \xi(-{\bf k},t') \expr \, =
\,  2T V \eta  \delta(t-t') \, .
\label{noisemarkov}
\end{equation}

\subsection{Effective description of zero mode in the linear $\sigma $-model }
\label{sec:zeromode}

In an ultrarelativistic heavy ion collision the idealized onset
of a `quench', as assumed in \cite{RW93}, is not really be given.
Instead, one expects that the most dominant particles to be freed
after the onset of the transition are the light pions, which represent
a thermalized, further evolving system. Their occupation in phase space
is described via a Bose distribution and cannot be correctly
taken care of in a purely classical field description.
This environmental pion gas may then actually expand rapidly enough
(in longitudinal and transversal directions) to allow for a nonequilibrium rolling down
of the chiral order parameter and giving potentially raise to the formation of
a DCC. In any case this gas of `hard' pions does represent a heat bath
with which the order parameter and the long-wavelength coherent pionic
fields do interact. In this sense these collective modes represent
an open system, which acts dissipatively and fluctuatively with the 
environment.
Referring to the general ideas outlined in the previous subsection one thus
expects that the (assumed semi-classical) dynamics of those modes
can be described by means of appropriate Langevin equations.
This intuition gave the phenomenological basis for the equations
of motion used in \cite{PRL}.

As we will argue in subsection \ref{sec:modelDCC} we expect
that for realistic initial (small) volumes $V(\tau_0)$ the zero mode
(${\bf k}=0$) pionic fields will cover the dominant coherent
pion modes to be possibly amplified in the course of a sufficient
rapid evolution of the system. These three modes in fact do represent the
pionic portion to the zero mode
\begin{equation}
\label{Ordpara}
\Phi ^a (t)
\, := \,
\frac{1}{V}\int d^3x \phi ^a(\vec{x},t) \, \equiv \,
(\sigma , \pi ^1, \pi^2 , \pi ^3)(t) \, .
\end{equation}
In the following we want to restrict ourselves to the effective description
of this zero mode field $\Phi^a$, which formally corresponds to the limit
$k_c \rightarrow 0$ in the previous discussion.

In analogy to (\ref{sm}) we now propose the following
effective Langevin equation of motions for the zero mode
\begin{eqnarray}
\ddot{\Phi}_0 + \Gamma [\dot{\Phi}_0 ]
+ \mu _{\bot }^2 \Phi_0 & = & f_{\pi}m_{\pi}^2 + \xi_0, \nonumber \\[5mm]
\ddot{\Phi}_i + \Gamma[\dot{\Phi}_i ]
+ \mu _{\bot }^2 \Phi_i & = & \xi_i \, .
\label{EOMO4eff}
\end{eqnarray}
The temperature dependent one-loop transversal (`pion') and the longitudinal
(`$\sigma $'-meson) mass for the respective fluctuations
(see also (\ref{O4thermmass})) are given by \cite{GM94,Bi96,Ra97}
\begin{eqnarray}
 \mu _{\bot }^2 & = & \lambda \left( \Phi_0^2 + \sum_i \Phi_i^2
- f_{\pi}^2 \right) + m_{\pi}^2 + m_{th}^2  \nonumber
\\
\label{O4mpi}
& = & \lambda \left( \Phi_0^2 + \sum_i \Phi_i^2 + \frac{1}{2} T^2
- f_{\pi}^2 \right) + m_{\pi}^2  \, ,
\\
\label{O4msig}
\mu_{\| }^2 & = & \mu_{\bot }^2 +
2\lambda \left( \Phi_0^2 + \sum_i \Phi_i^2  \right) \, .
\end{eqnarray}
The dissipation functional $\Gamma [\dot{\Phi }]$
as well as the (semi-classical) noise will be treated either in the
markovian approximation or within the full nonmarkovian expression
\begin{eqnarray}
\label{dissipfunc}
\Gamma [\dot{\Phi }] & = &
\left\{ \begin{array}{cr}
\eta \, \dot{\Phi } & \mbox{(markovian appr.)} \\[2mm]
 2\int\limits_{-\infty}^{t} dt' \, \Gamma (t-t') \,
 \dot{\Phi} (t') & \mbox{(nonmarkovian) }
 \end{array} \right. \, ,  \\[4mm]
\label{noisefunc}
\expl \xi _a  (t) \expr = 0  & , &
\expl \xi _a  (t_1) \xi _b (t_2) \expr \, = \,
\left\{ \begin{array}{cr}
\frac{2 T}{V} \eta \delta _{ab} \delta (t_1-t_2)
& \mbox{(markovian appr.)} \\[2mm]
\frac{2 T}{V} \Gamma (t_1-t_2) \delta _{ab}
& \mbox{(nonmarkovian)}
\end{array}  \right. \, .
\end{eqnarray}
Here $T$ denotes the temperature and $V$ the size of the volume of the considered
system. It will be the major point of
section \ref{sec:nonmarkov} to simulate nonmarkovian Langevin
equations and to compare them with the appropriate markovian treatment.

These coupled Langevin equations (\ref{EOMO4eff}) resemble
in its structure a stochastic Ginzburg-Landau description of phase transition
\cite{HH77}, especially for an overdamped situation \cite{BGR98},
where the $\ddot{\Phi }$-term can then be neglected.
On the other hand,
with $\lambda \approx 20 $ we are obviously not in a weak coupling regime,
so that the formal apparatus layed down in the previous section
\ref{sec:softmode} can only serve as a basic motivation.
Semi-classical Langevin equations may not hold for a strongly
interacting theory as for highly non trivial dispersion relations 
the frequencies of the
long wavelength modes are not necessarily much smaller than the temperature.
Still, when the soft modes become tremendously populated
one can argue that the long wavelength modes
being coherently amplified behave classically \cite{RW93}.
Aside from a theoretical justification one can  regard the Langevin
equation as a practical tool to study the effect of thermalization
on a subsystem, to sample a large set of possible trajectories
in the evolution, and to address also the question of all thermodynamically
possible initial configurations in a systematic manner.

\begin{figure}[h]
\centerline{\epsfysize=10cm \epsfbox{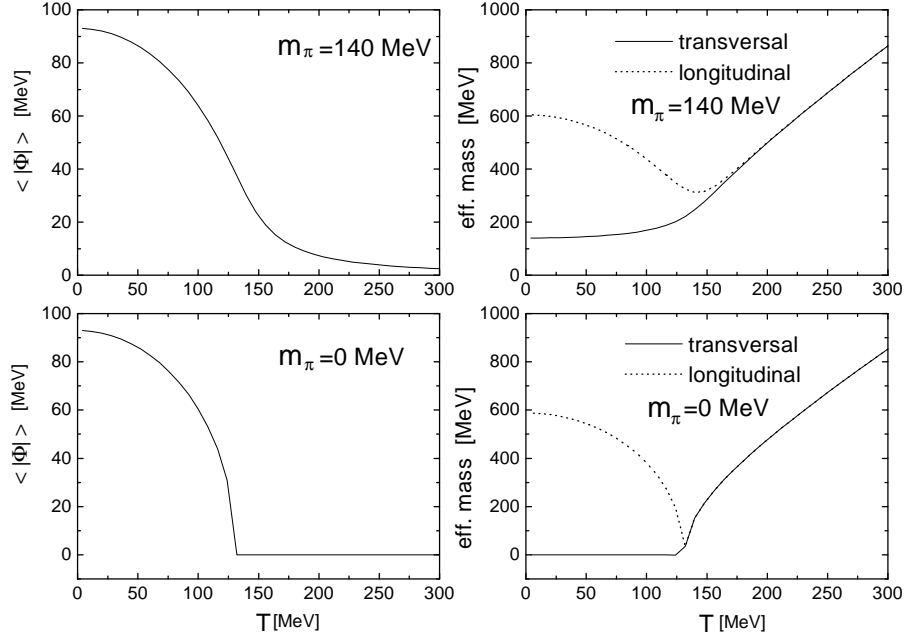}}
\caption{The temperature dependence of the magnitude of the order parameter 
$|\Phi|=\sqrt{\sigma^2+\vec\pi^2}$,
the transversal (pion-like) and longitudinal ($\sigma $-like) mass at thermal
equilibrium for the physical case of a non-vanishing pion mass and the case
without explicit chiral symmetry breaking ($H=0$).
The averages are obtained over an ensemble of $10^3$ realizations.}
\label{DCCfig2}
\end{figure}

A physically motivated choice for the damping coefficient
and the dissipation kernel $\Gamma $ we will state immediately below.
For the moment we stay to the markovian case and take $\eta $ as an
appropriate free parameter.
The `Brownian' motion of the soft field configuration
leads to equipartition of the energy at constant temperature.
In Fig. \ref{DCCfig2} we show the effective transversal masses $\mu_{\bot }$
of the pion modes and $\mu_{\| }$
of the $\sigma $ mode as a function of the temperature obtained by
solving eqs. (\ref{EOMO4eff}) at fixed
temperature $T$ and sufficiently large volume $V$. The masses shown
are thus taken as an ensemble average of the different realizations
within the Langevin scheme. For larger volumes the fluctuations
in the obtained masses are of the order $1/V$ and thus small.
For the situation that the vacuum pion
mass is assumed to be zero (no explicit symmetry breaking)
one can realize from Fig. \ref{DCCfig2} the situation for a true second order
phase transition occuring at the transition temperature
$T=T_c\equiv \sqrt{2 f_{\pi}^2 - 2m_{\pi }^2/\lambda } \approx  125$ MeV.
On the other hand for the physical situation of a nonvanishing pion mass
of $m_{\pi}=140 $ MeV the `phase transition' resembles the form
of a smooth crossover. In this case, at $T \approx T_c$, the
$\sigma $-field still posseses a nonvanishing value of
$\langle \sigma (T\approx T_c)\rangle \approx f_{\pi }/2 \approx \sigma _{vac}/2$.
(In the large volume limit one has
$\langle \sigma \rangle= \langle |\Phi| \rangle$ at fixed temperature, 
where $|\Phi|=\sqrt{\sigma^2+\vec\pi^2}$ denotes the magnitude of the order
parameter.)
Comparing with results of lattice QCD calculations
the transition temperature of
$T_c\approx 125 $ MeV is considerably smaller than the typical ones
of $T_c\approx 150-200$ MeV. This one might correct by using instead of
(\ref{O4thermmass}) the value obtained by the large N-expansion \cite{Boy95}
as $m_{th}^2 \equiv \frac{\lambda }{3} T^2$ resulting in $T_c \approx 154 $ MeV.
On a qualitative level the present description of the chiral phase transition
is compatible with the expectation of lattice calculations. However,
for the later one finds that the phase transition occurs in a much
sharper window around the critical temperature $T_c$: Slightly above $T_c$
the order parameter $\langle \bar{q}q \rangle \sim \langle \sigma \rangle$
already nearly vanishes; furthermore, sufficiently below $T_c$, the
order parameter has merely changed from its vacuum value.
This abrupt behaviour around the critical temperature is not realized within
the present treatment of the linear $\sigma $-model, which obviously shows
a much smoother dependence with temperature.
A more refined analysis within the linear $\sigma $-model might account
for this behaviour \cite{BS96}.

\begin{figure}[h]
\centerline{\epsfysize=8cm \epsfbox{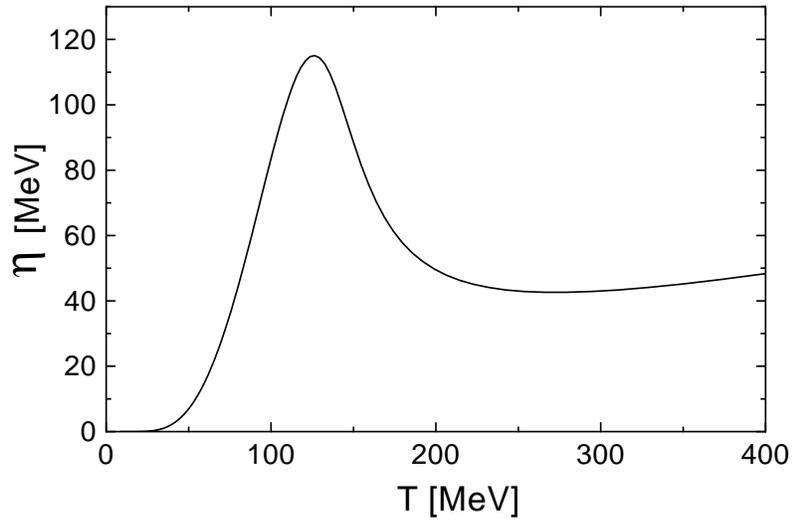}}
\caption{The dependence of the friction coefficient $\eta$ on the temperature.}
\label{DCCfig3}
\end{figure}

We now turn our attention to specify the dissipation coefficient $\eta $
or damping kernel $\Gamma $ of (\ref{dissipfunc}) entering the Langevin equations
(\ref{EOMO4eff}). From a physical point of view they should incorporate
the net effect of the dissipative scattering of the thermal (`hard') pions
with the collective fields. Its value is thus also of principal interest
for DCC formation as a (too) large damping of the collective pionic fields
would subsequently reduce significantly
the amplitude of any coherently amplified pionic field \cite{Ko98}
and thus might destroy any possible DCC. Being consistent within
the linear $\sigma $-model we consider here the `sunset'-contribution
(see Fig. \ref{DCCfig1}(c)) as the dominant term for the dissipation,
as it incorporates the net effect due to scattering of a soft mode
on a hard particle into two
hard particles and vice versa. The on-shell plasmon damping rate can then easily
be evaluated in analogy to standard $\phi^4$-theory to be
\begin{equation}
\label{O4diskoef}
\eta \, =
\frac{9}{16 \pi ^3} \lambda ^2  \frac{T^2}{m_p} \, f_{Sp} (1-e^{-\frac{m_p}{T}})
\, \, \, ,
\end{equation}
where $f_{Sp}(x) = -\int_{1}^{x} dt \frac{\ln t}{t-1} $ (see appendix
\ref{app:Dissip}). As emphasized in \cite{GMu97}, the appropriate
markovian approximation in a weakly coupled theory just corresponds
to this on-shell approximation.
At first sight, in the present situation of a strongly coupled theory,
one might think that this `choice'
can only provide a rather crude estimate as
the zero mode do not evolve on-shell during the (possibly unstable)
evolution. Hence the dissipation and noise correlation should better
be described by nonmarkovian terms including memory effects.
For this we have to evaluate the complete (off-shell) frequency dependence of the
dissipation kernel. This calculation we have shifted to appendix
\ref{app:Dissip}. As a further assumption we now take for
the plasmon mass $m_p$ the `pionic' mass $\mu_{\bot }(T)$ for the
transversal fluctuations depicted
in the right upper plot of Fig. \ref{DCCfig2}. This choice should be valid
near or above $T_c$ as the transversal and longitudinal masses becomes
nearly degenerate. The thus resulting dissipation coefficient $\eta $ of
(\ref{O4diskoef}) is shown in Fig. \ref{DCCfig3} as a function of the
temperature $T$  (see also \cite{IT99}).
With this prescription one notes that $\eta $ posesses a maximum value
of $\approx 100$ MeV near the critical temperature, which will result
in relaxation (or equilibration) times of roughly $2$ fm/c
(compare also with Fig. \ref{DCCfig7}).
For sufficiently smaller temperatures $\eta $ decreases fast to a neglible small
value as the density of the thermal pions as potential scattering centers
also falls rapidly with decreasing temperature. This behaviour is in line
with findings in \cite{Ko98}, where the on-shell damping
coefficient has been calculated by means of standard
chiral pion scattering amplitudes in the vacuum.

Some critical remarks are in order:
(1) It is questionable that above the critical temperature
all contributing degrees of freedom are being considered. Above $T_c$ one
expects that due to the deconfinement transition occuring at the same
critical temperature quarks and gluons are freed and thus might have
a considerable influence on the damping coefficient of the collective, mesonic
excitations.
(2) The damping coefficient $\eta $ introduced in (\ref{O4diskoef})
should be appropriate for temperatures close to $T_c$, where spontaneous
symmetry breaking has just emerged. On the other hand,
for a deeply broken phase ($T\ll T_c$), the $\pi + \pi \rightarrow \pi + \pi $
scattering amplitude will become significantly reduced by the additional
t-channel exchange of a $\sigma $-meson, leading to the well known
chiral derivative coupling for lower transferred momenta. This additional
contribution for the deeply broken phase we have not taken into account
and we thus overestimate the damping associated with the thermal scattering
especially for low temperatures (see eg for comparison the damping coefficient
given in \cite{Ko98}).
(3) Moreover, for temperatures much below $T_c$
the O(4) transverse and longitudinal mass for the fluctuations are not equal
anymore. From Fig. \ref{DCCfig2} one recognizes that below $T \approx 100$ MeV
the longitudinal mass $\mu_{\|}$ exceeds
two times the transversal mass $\mu_{\bot }$, so that
the decay of the longitudinal mode into two transversal particles becomes possible.
In vacuum this just corresponds to the decay
$\sigma  \rightarrow \pi \pi$ \cite{DR98,CEJK}
with a width on the order of a few hundred MeV.
This would give raise to an additional temperature dependent dissipation
in longitudinal direction for the evolving order parameter and might
have also interesting consequences for the DCC formation investigated
in section \ref{sec:stochDCC}.
Qualitatively one expects that the associated damping will then
effectively slow down considerably the rolling down in longitudinal
(i.e. `radial') direction of the order parameter
along the effective potential.
We leave an implementation of
this kind of longitudinal damping for future work.
(4) A final problem, which we briefly mention, concerns
the chiral limit $m_{\pi }=0$. Below $T_c$ the pions remain as massless
Goldstone bosons (see also Fig. \ref{DCCfig2}) and the $\sigma $-meson becomes degenerate
with the pion at and above the critical temperature.
Taking the expression (\ref{O4diskoef}) one notices that
the dissipation coefficient $\eta $ diverges like $1/m_p$.
On the other hand one expects for a true second order phase transition
a critical slowing down of the excitations near the critical temperature
and thus a vanishing of the dissipation coefficient \cite{Pi98}.
This then implies that a perturbative evaluation is not valid but
requires a nonpertubative analysis via e.g. renormalization group methods
\cite{Pi98}.

Our discussion should demonstrate that a precise determination
of the description of the dissipation functional $\Gamma [\dot{\Phi }]$
near the critical temperature is far from being settled.
We consider our choice as a physical motivation, and which is also
numerically tractable.

\subsection{Fluctuations of initial conditions at critical temperature}
\label{sec:initfluc}

As a first and straightforward application we address the important question
for the possible distribution of the order parameter (\ref{Ordpara})
at the critical temperature for a {\em finite} system with fixed size $V$.
With the noise fluctuating according to (\ref{noisefunc}) we expect
(similarly like in Brownian motion) that the chiral fields
do fluctuate thermally around its mean as well.
Assuming that slightly above the transition temperature the system is near
thermal equilibrium, generating an ensemble
distribution then offers a systematic sampling of all possible
`initial' configurations for the later dynamical evolution of
the order fields, which then lead to a stochastic formation of DCC.

\begin{figure}[h]
\centerline{\epsfysize=8cm \epsfbox{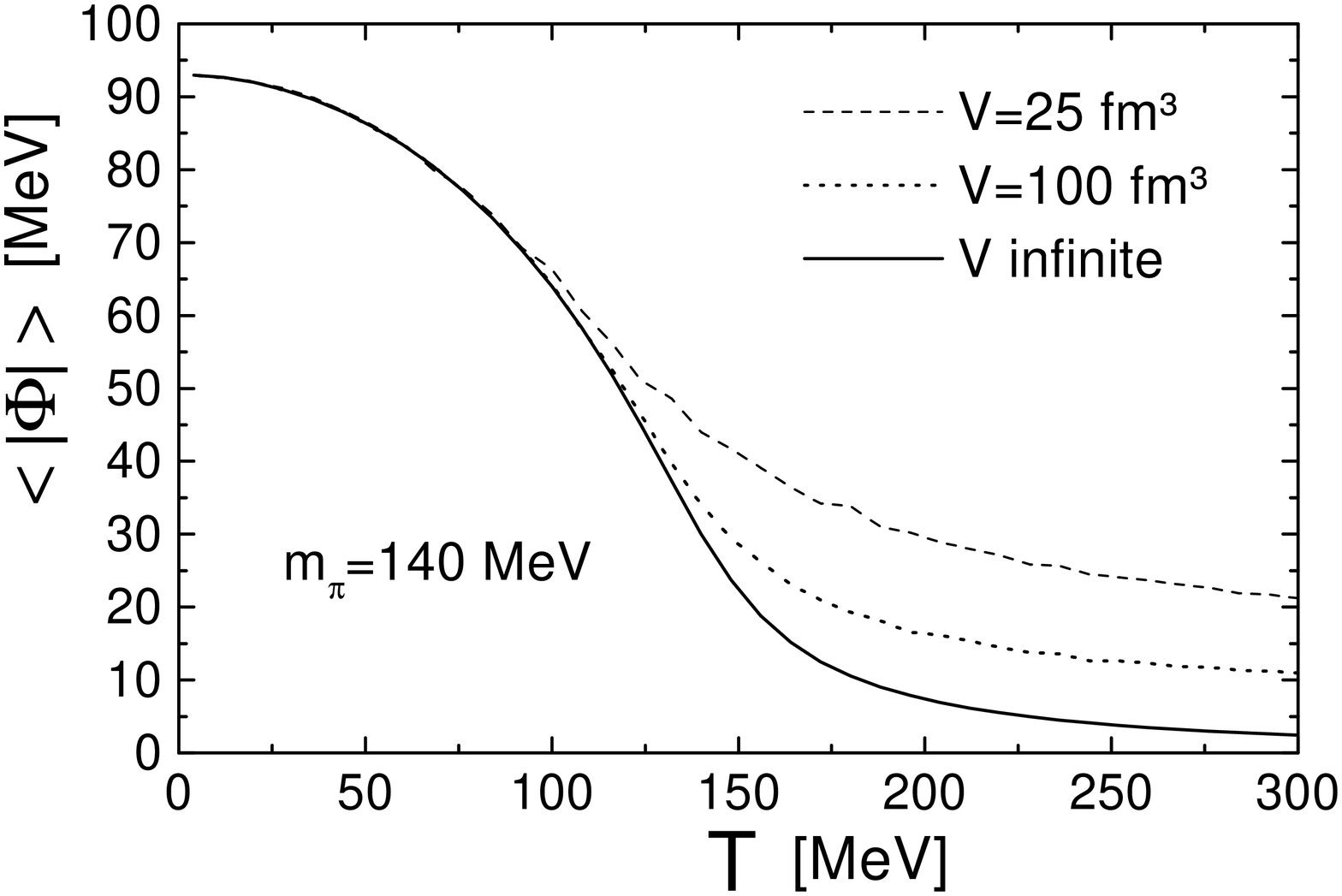}}
\centerline{\epsfysize=8cm \epsfbox{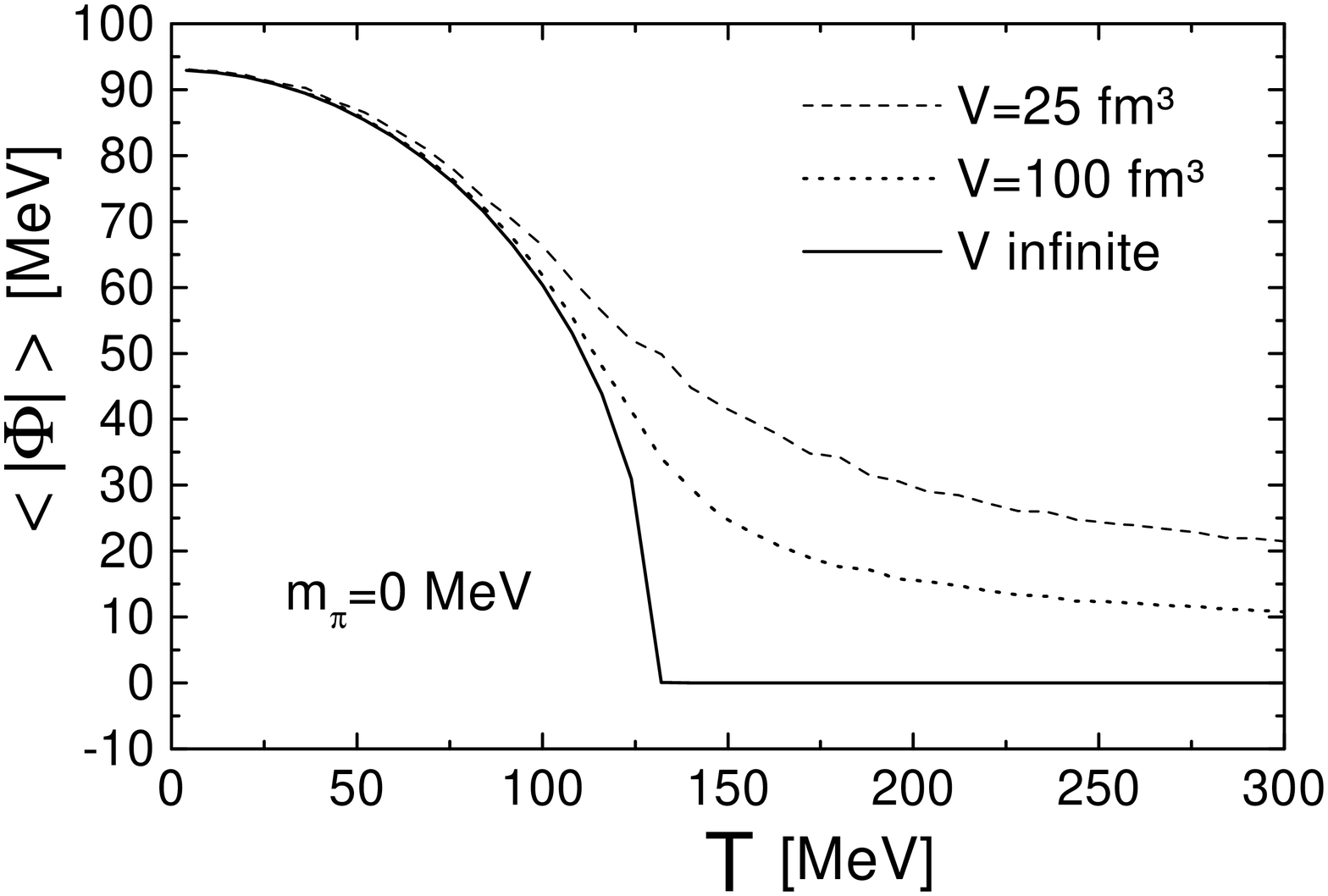}}
\caption{The temperature dependence of the magnitude of the order parameter 
$\Phi$ at
thermal equilibrium for different volumes.
The averages are obtained over an
ensemble of $10^3$ realizations.}
\label{DCCfig4}
\end{figure}

In Fig. \ref{DCCfig4} we show first the sensitivity of the (ensemble) averaged
value of the order parameter
$\expl |\Phi| \expr= \expl \sqrt{\sigma^2+\vec\pi^2} \expr $ on various sizes
$V$ as function of the temperature.
As expected, finite sizes lead to a positive shift of the order parameter
and to a (further) rounding of the phase transition.
In Fig. \ref{DCCfig5} the characteristic distributions
of the chiral fields and their `velocities' at the critical
temperature are depicted. The average width scales like $1/\sqrt{V}$.
Such a behaviour has been reported already within an independent approach
in \cite{MSL98}. One might also employ the quantal version of the noise
fluctuations according to (\ref{Hu5}), which in the markovian
on-shell treatment one would approximate as
\begin{equation}
\label{QuantFDR}
I \, \rightarrow \, \frac{m_p}{V}\,  \eta \, {\rm coth} ({m_p \over 2T} )
\delta (t_1-t_2) \, .
\end{equation}
Such a prescription results in even larger fluctuations.
It is also interesting to look at the situation in the chiral limit
$m_{\pi } =0$. The characteristic distribution $P(\sigma )$ is given in
Fig. \ref{DCCfig6}. In this case the fluctuations are even larger
and scale effectively with $1/V^{1/4}$. (One can find analytically \cite{PRL}
that for this case $\expl \sigma^2 \expr  = 1/2 \sqrt{T_c/(\lambda V)}$,
so that the width in the distribution $P(\sigma )$
thus has to scale with $V^{-1/4}$.)

In the next subsection we will now turn to the description of the chiral
fields for an expanding environment leading then to stochastic
individual trajectories with considerable fluctuations and thus also
for particular events out of an ensemble possibly to experimentally accessible
DCC candidates. In a sense the `faith' of all individual
trajectories (entering to some amount the unstable region
$\mu_{\bot }^2<0$ \cite{PRL}) is not really predictable and has to be sampled
in some quantitative way as within our proposed Langevin picture.
We have to admit that one can certainly improve in various ways on many
aspects in describing phase transitions out of equilibrium.
Much retains to be learned about how these condensates
evolve in out-of-equilibrium. Probably the most ambitious
description on the quantal evolution of the chiral fields
in out-of-equilibrium has been developed by Niegawa \cite{Ni98} employing
the powerful closed-time-path (CTP) real-time Greens function technique.
It has to be seen whether
this formal development can be used for practical simulations concerning
DCC formation. Using the CTP technique, this approach (as well as earlier
developments in the same direction
\cite{Boy95}) is, by construction, an ensemble averaged description \cite{GL98a},
which can thus describe within sophisticated methods the dynamical evolution
of (ensemble averaged) expectation values. Unusual fluctuations, like e.g.
in the pion number, as shown later here, can only be accounted for
by higher order correlation functions. These are
typically not considered. Our approach, we believe, states thus a fresh
new way in order to account in a simple transparent manner for such unusual
strong fluctuations and being far from a simple gaussian mean field treatment.

\subsection{Modelling the evolution of potential DCC}
\label{sec:modelDCC}

In the following we will state the final equations of motion for simulating
the stochastic formation of possible DCC. In the markovian approximation
these corresponds to the ones proposed originally in \cite{PRL}.
As a later characteristic quantity we will consider the
pion number of the zero mode contained in the evolving domain,
which is assumed to roughly correspond to the effective number of soft pions
freed from the subsequent decay of the pionic fluctuations, i.e. the final
decay of the DCC.

It is instructive to first outline how possible DCCs
would be formed in a heavy ion collision.
This intuitive and idealized physical scenario
will give some insight for the choices of the value of
the free parameters to be specified and will also give a perspective to
understand the physical matters to be discussed in the following sections.
Our picture of a possible DCC formation in high energy
heavy ion collisions is as follows:
\\[2mm]
{\large $\bullet $}
In the first stage of the collision (at proper times
$\tau = 0.3 \ldots 0.5$ fm/c in the respective subvolume of the system)
a parton gas is formed with a temperature $T\gg T_c$
well above the chiral restoration point.
Chiral symmetry is completely restored in this hot region.
\\[2mm]
{\large $\bullet $}
In the following ($\tau = 2 \ldots 3$ fm/c), because of the
subsequent collective expansion (longitudinally or later even transversally)
the temperature drops to around the critical one ($ T \approx T_c$) and
some small chirally restored or already slightly disoriented
domains of collective
pionic fields start to form together with a thermalized
background of (quasi-)pions and possibly other thermal excitations
within the respective subsystem. The individual subsystems are assumed
to evolve independently as they are spatially separated and might be
separated in rapidity. The possible distribution of the chiral (mesonic)
order parameter then depends on the size of the volume $V(\tau )$
of the individual domain, as shown in the previous subsection.
\\[2mm]
{\large $\bullet $}
At a further time ($\tau_0 = 3 \ldots 7$ fm/c) the temperature
of a (rapidly) expanding domain crosses the critical temperature $T_c$,
having a certain volume $V(\tau _0)$.
At the same time the partonic gas
would undergo the deconfinement/confinement phase transition into the mesonic
freedoms. The temperature of the surrounding
`heat bath' further decreases as the volume increases due to the
collective expansion.
At this stage chiral symmetry becomes spontaneously broken.
The stable point of the order
parameter characterizing the broken phase moves from 
$(\sigma \approx 0, \vec\pi \approx 0)$ in the symmetric phase towards 
$(\sigma \approx f_{\pi}, \vec\pi \approx 0)$ in vacuum. This change happens
fast if the system expands sufficiently rapidly. A possible
(but not necessary)
instability might arise depending on the actual (`initial') values of
the order fields \cite{PRL}.
In certain cases, depending crucially on the `appropriate'
initial configuration, the order parameter can `roll down'
in a `disoriented' direction with a fixed orientation in isospin space,
giving raise to a large coherent collective pion mode.
A potential DCC is formed.
Possible DCC domains differ from each other in the orientation in isospin space,
in the size and in the pionic content. A large DCC domain denotes here
a large pionic content.
Intuitively the order parameter in such a large DCC domain will go
through a trajectory deviating strongly from the $\sigma $-direction
during the roll-down period.
In any case a sufficiently fast expansion and cooling
is mandatory for the possible formation of larger DCCs. (Because of the explicit
symmetry breaking term $H \sigma $, which, in analogy to a ferromagnet,
acts as an external and rather strong constant magnetic field,
together with the dissipative interaction with the heat bath,
the order parameter will
otherwise align more or less quasi adiabatically
at its thermally dictated equilibrium value along the $\sigma $-direction,
if the experienced cooling is not fast enough.)
\\[2mm]
{\large $\bullet $} With the ongoing (radial) expansion
($\tau \ge 10$ fm/c) and due to the explicit symmetry breaking
the order parameter will oscillate with decreasing amplitude
around the stable point $<\sigma>=f_{\pi}$ along the chiral circle 
$(\sigma^2+\vec\pi^2=f_{\pi}^2)$. The expansion will come to a halt
at some freezeout time, the fireball breaks off.
The coherent semi-classical pion state within the possible DCC domain
decays by the emission of long wavelength pions,
with isospin distribution characteristic to DCC, and which in number
correspond approximately to the effective pion number stored originally
in the coherent state.
If this number of the coherently produced
low momentum pions is not too small compared with
incoherent low momentum pions from other, random sources,
constituting the `background', a careful
event-by-event analysis can provide identification of the DCC formation.

In the following we want to investigate the evolution of the
zero mode chiral fields in contact with the heat bath
constituted by all the other modes
solely in one single domain being created out of the initially hot
region by means of equations of motion analogous to
(\ref{EOMO4eff}).
As outlined above, of course, many of such domains might well
be created. We assume, for simplicity, that these individual domains
are independent and do not further interact.

The (rapid) expansion can be incorporated effectively
by means of the boost-invariant Bjorken scaling expansion \cite{Bj83}
assuming that the order fields
$\Phi_a \equiv \Phi_a (\tau )$ depend on time only implicitly
via the proper time variable $\tau = \sqrt{t^2-x_{eff}^2}$, where
$x_{eff}:=z$ for $(D=)$ 1-dimensional longitudinal expansion and
$x_{eff}:=r$ for $(D=)$ 3-dimensional radial expansion
\cite{Ra96a,GM94,Bi96,PRL,MSL98,Bj83,La96}.
In the equations of motion the d'Alembertian is then replaced
by $\partial^2/\partial \tau^2 + (D/\tau) \partial / \partial \tau$,
giving raise to an effective Raleigh damping coefficient $D/\tau $.
This one might also interpret as an effective Hubble constant \cite{Du99}
due to the volume dilution
\begin{equation}
\label{Volexpan}
\frac{\dot{V}}{V} - \frac{D}{\tau}  = 0 \,
\longrightarrow
V(\tau )=V(\tau_0) \left ( \frac{\tau}{\tau_0} \right )^D
\end{equation}
for the expanding volume $V(\tau )$ of the domain.
In the quasi-free regime of a freely moving bosonic field the
amplitude then decreases in (proper) time with $\sim \tau ^{-D/2}$.

From (\ref{EOMO4eff}) we then receive the equations of motion for the
zero mode fields in an expanding environment as
\begin{eqnarray}
\ddot{\Phi}_0 + \frac{D}{\tau}\,\dot{\Phi}_0+\Gamma [\dot{\Phi}_0 ]
+ \mu _{\bot }^2 \Phi_0 & = & f_{\pi}m_{\pi}^2 + \xi_0, \nonumber \\[5mm]
\ddot{\Phi}_i + \frac{D}{\tau}\,\dot{\Phi}_i+\Gamma[\dot{\Phi}_i ]
+ \mu _{\bot }^2 \Phi_i & = & \xi_i \, .
\label{EOMDCC}
\end{eqnarray}
The dissipation functional $\Gamma [\dot{\Phi }]$ as well as the transversal
mass $\mu_{\bot }$ do both depend on the temperature $T(\tau )$.
The stochastic noise fields obey (\ref{noisefunc}).
One therefore also needs to know how the local temperature
evolves with time. In principle one has to ask for the equation of state
of the system and solve for the hydrodynamic equations within the
(assumed) D-dimensional scaling expansion. For the ideal case of a massless
gas (which is not a too bad approximation
for pions) an isentropic expansion results in
\begin{equation}
\label{Texpan}
\frac{\dot{T}}{T} + \frac{D}{3\tau}  = 0 \,
\longrightarrow
T(\tau )=T(\tau_0) \left ( \frac{\tau_0}{\tau } \right )^{D/3} \, .
\end{equation}
We take this as an idealized guide for the temperature profile
$T(\tau )$ with proper time. (One should note, however, that
for temperatures above $T_c$ partonic degrees of freedom
contribute significantly to the equation of state and thus might modify
the here assumed profile substantially, if the initial temperature is chosen
above the critical temperature.)

We note  that the initial proper time $\tau_0$ and the dimension
$D$ of the expansion are
here the important parameters determining the dynamics of the expansion: 
Large $D$ and small $\tau_0$ lead to a more rapid expansion and cooling.
`Initial' is meant here as the proper time $\tau_0$
where the partonic gas confines into the
mesonic degrees of freedom and before the roll-down. We thus choose the critical
temperature $T_c$ as the initial temperature. (We will also later comment
briefly for cases where we have chosen higher initial temperatures.)
For the (unknown) initial volume $V(\tau_0)$ we will take
$V_0 = 10 - 200 \, fm^3$ as a reasonable range,
which implies a spherical initial domain of radius $r=1.4 - 3.6$ fm.
(Later we will see that varying the initial volume will not lead to a major
change in the final results within our model.)

In order to make a statistical analysis we need to sample the initial 
configurations $\Phi_a(\tau_0)$ and $\dot\Phi_a(\tau_0)$ at the initial
temperature in a systematic manner. As demonstrated in the last
subsection we let the chiral fields propagate at thermal equilibrium
for sufficiently long hypothetical times
at the initial temperature in order to generate a
consistent ensemble of possible initial
configurations for $\Phi_a$ and $\dot\Phi_a$. The main assumption here
is thus then the hypothesis of (nearly) perfect thermal
equilibrium for the initial chiral order fields before the possible
roll-down period.

In \cite{PRL} the average and statistical properties of
individual solutions of the above Langevin equations (\ref{EOMDCC})
within the markovian approximation (cf. (\ref{dissipfunc}) and (\ref{noisefunc}))
have been studied with the emphasis on such periods of
the time evolution when the transverse mass $\mu_{\bot }$ becomes imaginary
and therefore an exponential growth of unstable fluctuations in the
collective fields might be expected.
It was found that for different realistic initial volumes
individual events lead to sometimes significant growth
of fluctuations. For the quantification of the resulting strength
of the coherent pionic zero mode fields and as an
experimentally more direct and relevant quantity
we consider in the following the effective pion number content of these
chiral pion fields.
In the semi-classical approximation this number is given
by
\begin{equation}
\label{O4pionnumb}
n_{\pi} = \frac{1}{2} m_{\pi}
\left( \vec{\pi }^2(\tau) + \frac{1}{m_{\pi}^2} \dot{\vec{\pi }}^2(\tau)
\right) \, V(\tau )\,.
\end{equation}
This expression can be most simply obtained
by considering the energy density of the zero mode
$\epsilon _{\pi ; k=0} = 1/2\,(m_{\pi}^2\,\vec{\pi }^2 + \dot{\vec{\pi }}^2)$.
As $\vec{\pi}^2(\tau)$ will be proportional to
$1/V(\tau)$ at the late stage of the evolution after
the roll-down period, $n_{\pi}(\tau)$ then becomes constant at
late proper times when the effective pion mass $\mu_{\bot }$
relaxes towards its physical vacuum value.
This constant number will be extracted from the simulations
as the total pion number freed
from the DCC decay.
For this effective pion number
one crucial point is then how large the evolving
volume $V(\tau )$ of the DCC domain has increased when the pion oscillations
have emerged.

We will now first employ our model to
understand the effect of the dissipation and the
possible role of memory effects on the evolution.
We then further investigate
within different scenarios the
statistical distribution in the resulting pion number
(\ref{O4pionnumb}) and will
propose a new signature of stochastic DCC formation
based on the cumulant expansion.

At this point one might indeed worry why we only consider
the $k=0$ zero mode and not also some other long wavelength pionic excitations,
which should also experience some unusual amplification
according to the general wisdom of DCC formation.
From a principle point of view our model could be worked out or generalized
to take into account also some more long wavelength modes. The cutoff momentum
should then be taken as
$0< k_c \stackrel{< }{\sim } \sqrt{\lambda }f_\pi \approx 400 $ MeV
to account for the pionic modes who could possibly become unstable and thus
amplified. From the power spectrum shown in the work of Rajagopal and Wilczek
\cite{RW93} one notices that even within
the drastic quenched situation of instantaneous cooling only the lowest
discretized momentum mode becomes dominantly amplified, whereas
the next higher lying pionic modes only show
some moderate behaviour. In the more physical situation
the inclusion of a thermally generated mass term
$\lambda /2 T^2(\tau )$ in the effective potential will
cut down even further the low momentum range for possible unstable modes,
i.e. $k_c \ll \sqrt{\lambda }f_{\pi }$.
Furthermore also the volume
$V(\tau_0)$ of an initial domain as chosen by us
(at $T\approx T_c$) is much smaller than in \cite{RW93},
with a radius between 1.4 - 3.6 fm. Hence, in such a quantized picture
of a finite volume only a few
Fourier modes except the zero mode could really become unstable.
We therefore expect that only the pionic zero mode can
predominantly be amplified.

\section{Dissipation: Markovian vs nonmarkovian \\
description}
\label{sec:nonmarkov}

In this following section we address
on a quantitative level the possible differences between the full
nonmarkovian treatment and the markovian (`instantaneous') approximation
for the dissipative (\ref{dissipfunc}) and fluctuating dynamics (\ref{noisefunc})
within the Langevin model.

The exact nonmarkovian functional $\Gamma [\dot{\Phi }]$ of (\ref{dissipfunc})
at a given temperature $T$ and plasmon mass $m_p$ has been worked out
in appendix \ref{app:Dissip}. (As also stated in the appendix we
only consider in the present investigation the contribution
of thermal scattering to the dissipation functional,
i.e. the part denoted as $\gamma_1$ in the appendix.)
As ellaborated in \cite{GMu97} and stated in the equation (\ref{Markovlim})
the appropriate markovian limit for a sufficiently weakly dissipatively
interacting system results in the on-shell dissipation or viscosity coefficient
$\eta \equiv \Gamma (\omega = m_p)$, i.e. (\ref{O4diskoef}).
For the nonmarkovian dissipational functional
we therefore consistently choose for the plasmon mass $m_p$ the temperature
dependent transversal mass $\mu_{\bot }\equiv m_{\pi}(T)$ of
the right upper picture of Fig. \ref{DCCfig2}.
Besides of evaluating a history dependent memory
functional $\Gamma [\dot{\Phi }]$ to treat the full nonmarkovian dissipative dynamics,
as a further complication one also has to face the problem of how to
simulate colored (i.e. non-white) gaussian noise for the fluctuating
forces in order to be consistent with the underlying fluctuation-dissipation
relation (\ref{dft}) or (\ref{noisefunc}). Our strategy for achieving
a numerical realization of colored gaussian noise is briefly outlined
in appendix \ref{app:Noise}. With this we can then numerically
solve the full nonmarkovian equations of motion. The Langevin
equations (\ref{EOMO4eff}) or (\ref{EOMDCC})
are then solved for both cases by means of a standard third order
multistep scheme, the Adams-Bashforth method \cite{NRec}.

In a strong coupling theory like the linear $\sigma $-model
and also for instable situations encountered in describing possible
DCCs the magnitude of the soft modes
$\sqrt{|\phi |^2}$ might vary sufficiently fast so that
no dominant oscillatory frequency of the fields does occur and thus
the Markov approximation should not hold.
This gave the motivation for this particular study.
As it turns out, and as we will argue in the following, however, for situations
(and thus appropriately chosen parameters for $D$ and $\tau_0$), where
{\em large} and experimentally significant DCC can occur, the
distinction between the two cases becomes
more or less irrelevant. One can then incorporate the
numerically much simpler markovian treatment.
On the other hand, to the best of our knowledge, our study represents
the first numerical treatment of
nonmarkovian Langevin equations in thermal quantum field theory
and might certainly be of importance for other related topics, e.g.
in the description of phase transitions in cosmological settings
by means of Langevin equations \cite{BG95}.

A general expectation for the possible difference is
that the rate of thermalization, i.e. how fast the considered relevant
modes do approach their thermal equilibrium properties
within the heat bath, might be substantially affected.
This is best and most straightforwardly demonstrated for very simple classical
examples like Brownian motion of a diffusive particle or oscillator.
For a more systematic investigation in this respect we refer to a future
publication \cite{XGL99}, where also the difference between `weak' and
`strong' dissipative Langevin behaviour for diffusive processes
will be discussed.

\begin{figure}[h]
\centerline{\epsfysize=6cm \epsfbox{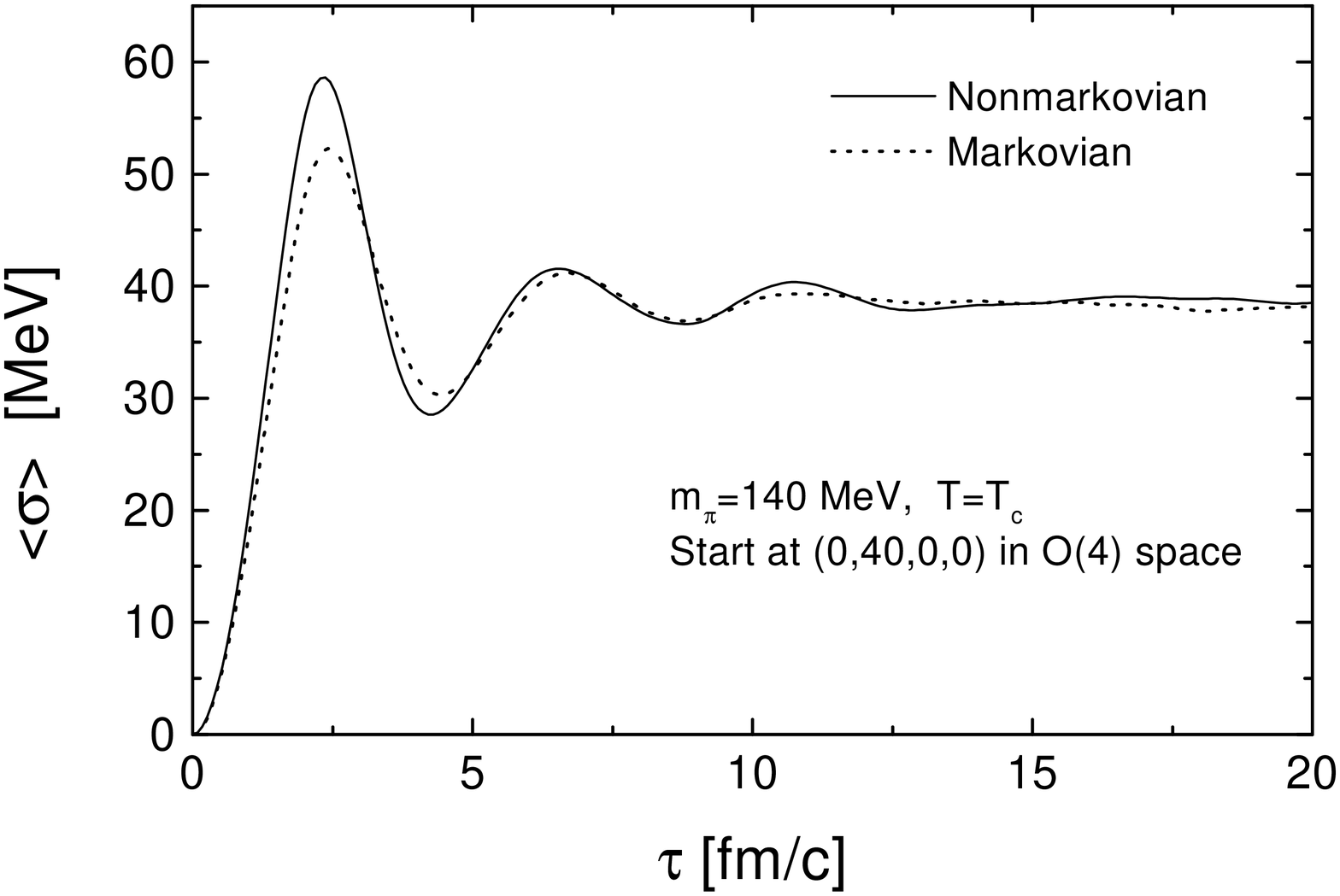}}
\centerline{\epsfysize=6cm \epsfbox{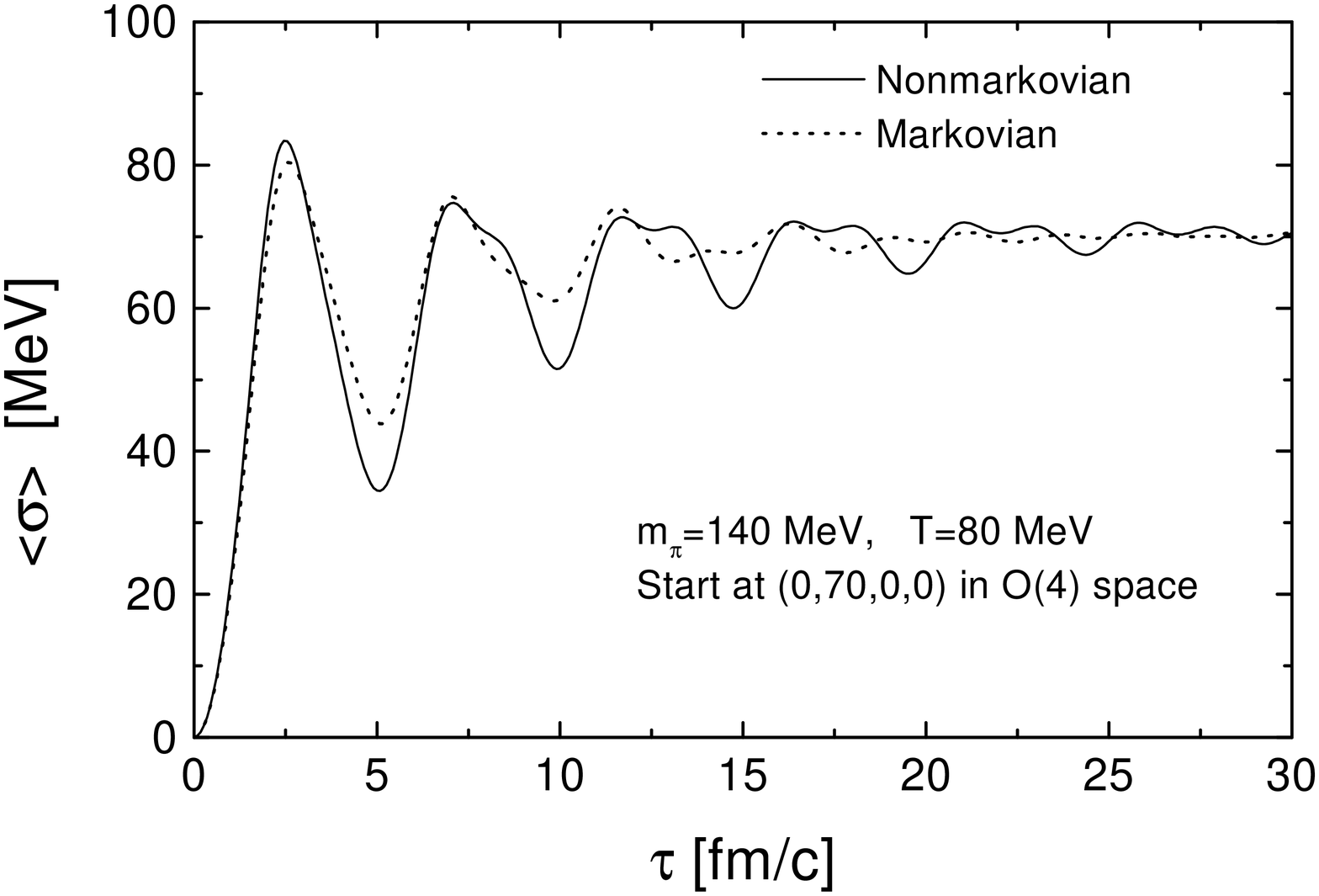}}
\centerline{\epsfysize=6cm \epsfbox{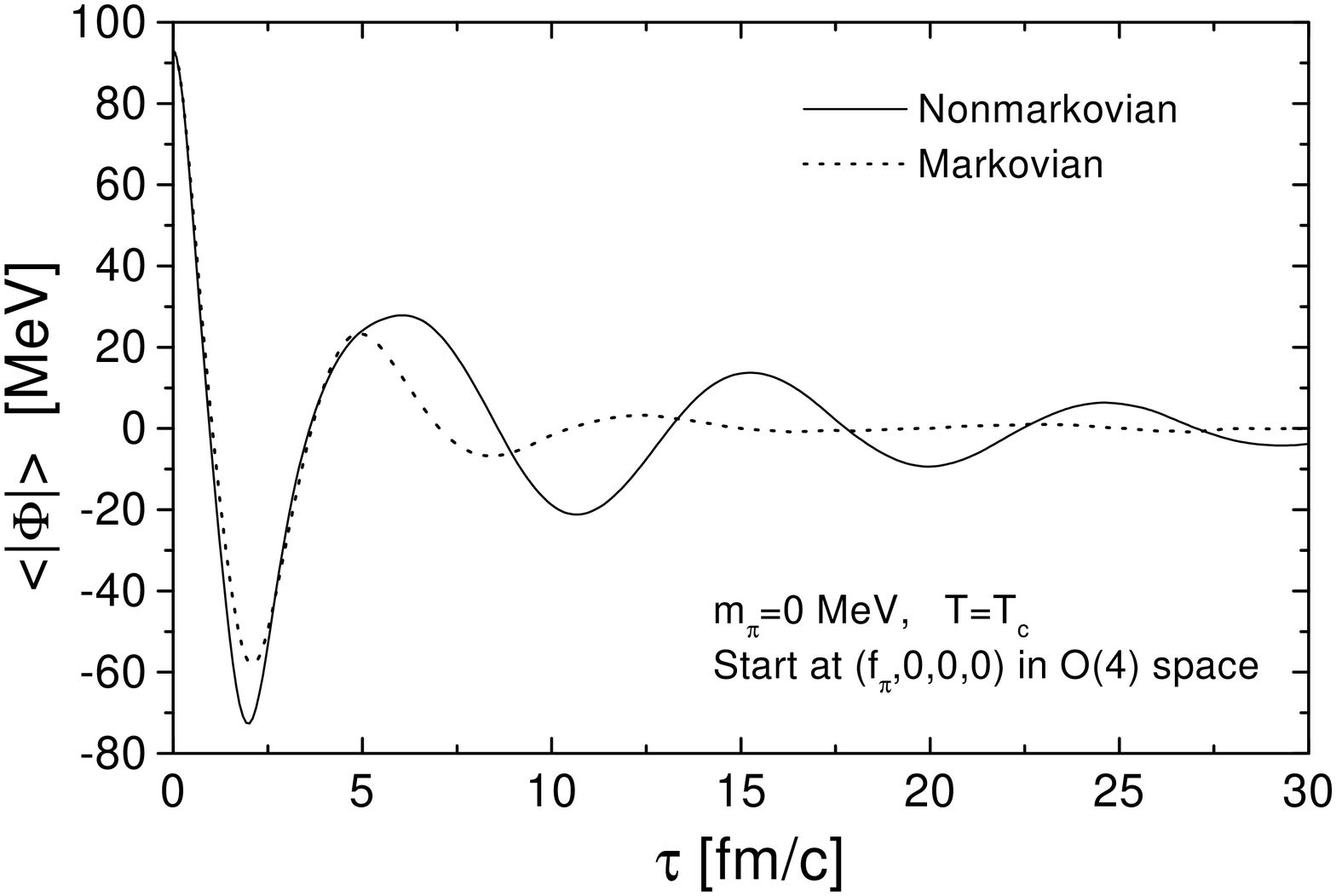}}
\caption{Relaxation of the ensemble averaged $\sigma $ field within
a heat bath at finite temperature for the
nonmarkovian and markovian case. The upper and middle part correspond
to the situation of a
physical pion mass, whereas the bottom one corresponds to the case
without explicit chiral symmetry breaking ($m_\pi=0$). In this case
we depict the relaxation of the ensemble averaged magnitude of the order
parameter.
The averages are taken over $10^3$ realizations.}
\label{DCCfig7}
\end{figure}

\begin{figure}[h]
\centerline{\epsfysize=7cm \epsfbox{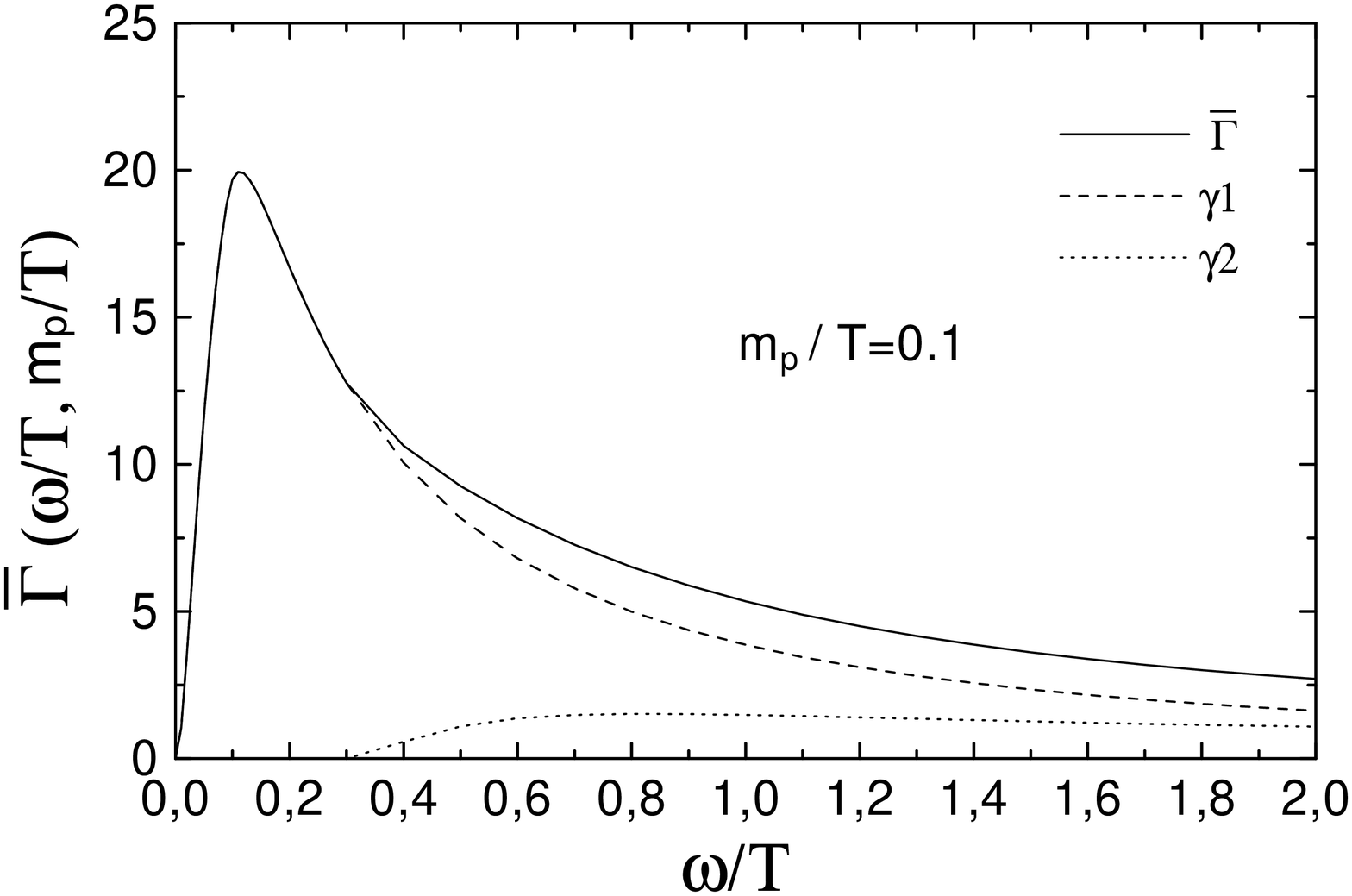}}
\centerline{\epsfysize=7cm \epsfbox{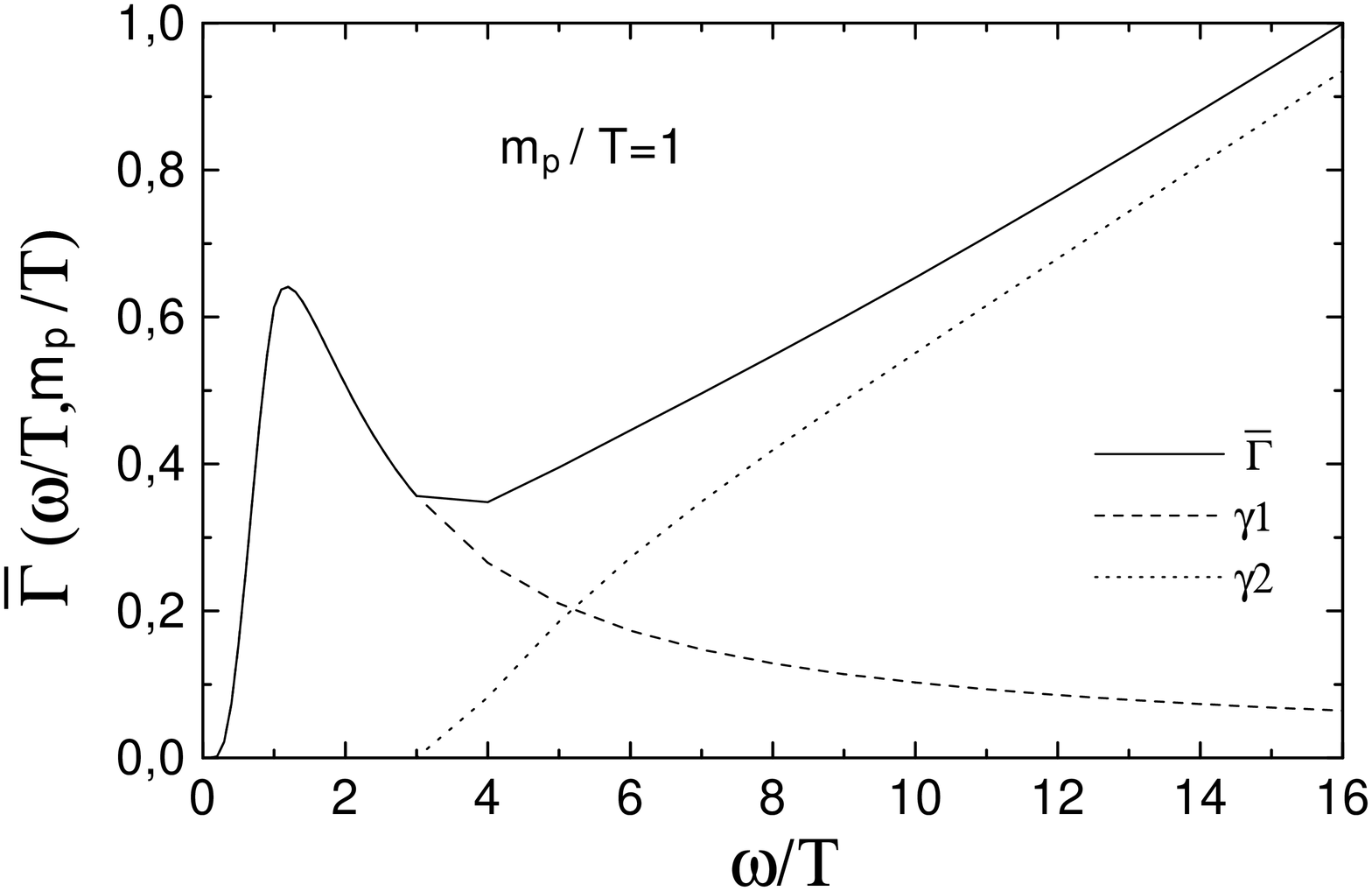}}
\caption{The reduced dissipation kernel for $m_p/T=0.1$ and $m_p/T=1$. 
$\gamma_1$ and $\gamma_2$ are the contributions to ${\bar \Gamma}$ from the
contributing scattering and decay process (see text).}
\label{DCCfig19}
\end{figure}

Referring to our present model it is certainly interesting to
study how fast the order fields can move (or `diffuse')
towards their equilibrium properties discussed in the previous
sections \ref{sec:zeromode} and \ref{sec:initfluc}.
(A somewhat similar study for simple markovian dissipation has been
previously carried out in \cite{IT99}.)
In Fig. \ref{DCCfig7} we show for various cases the relaxation of the
ensemble averaged $\sigma$ field $\expl \sigma \expr (t)$,
being initially distorted by hand, towards its equilibrium value
$\langle \sigma \rangle _{eq}(T) $
(compare Figs. \ref{DCCfig2} and \ref{DCCfig4})
in a surrounding heat bath at fixed temperature.
As constant volume we haven taken $V_0=100 \, fm^3$.
In the two upper figures we haven chosen as initial values
$\Phi_a(t=0)=(0,\pi_1\approx \langle  |\Phi| \rangle _{eq}(T) ,0,0)$ and
$\dot\Phi_a(t=0)=(0,0,0,0)$, i.e. an initial distortion
of the chiral zero mode fields in one particular pion direction
along the effective finite temperature dependent chiral circle
$\sigma^2+\vec{\pi}^2=\langle |\Phi|^2 \rangle _{eq}(T)$.
In the upper figure the situation
is depicted at the critical temperature $T_c$,
whereas for the middle figure
we have taken $T=80 \, MeV$. For this investigation we consider $10^3$
independent simulations for taking the ensemble average.
For both cases the averaged $\sigma$ field follows a damped
oscillation along the effective chiral circle.
For the markovian simulation one sees that the relaxation towards
equilibrium goes in accordance with the value of the dissipation
coefficient (\ref{O4diskoef}) depicted in Fig. \ref{DCCfig3}.
The nonmarkovian evolution now shows a slightly less damped relaxation
towards the equilibrium value, the difference being more pronounced
for the lower temperature.
Qualitatively one can understand this behaviour by comparing
the frequency spectrum of the dissipation kernel $\Gamma(\omega)$ (its
reduced form is shown in Fig. \ref{DCCfig19}) with the on-shell damping
coefficient used in the markovian approximation 
$\Gamma(\omega)=\Gamma(m_p)=\eta $.
This spectrum has its maximum in frequency more or less exactly at the on-shell
frequency, so that simulation carried out within the markovian
approximation will result in an effectively larger damping and thus faster
relaxation, since the effectively contributing
frequency modes $\Phi(\omega)$ of the motion $\Phi(t)$ in the full 
treatment are less damped (for $\omega\neq m_p$) than those in the
markovian approximation.

Another interesting example
is shown in the lowest part of Fig. \ref{DCCfig7}.
Here we consider the relaxation of the order parameter, initially being
distorted to its vacuum value, towards its equilibrium value
$\langle \sigma \rangle (T_c) =\langle \vec \pi \rangle (T_c)=0$ at the chiral
phase transition without
explicit symmetry breaking. Here both the effective masses
$\mu_{\bot }$ and $\mu_{\| }$ of
the chiral fields vanish, so that the effective potential
does not posess any quadratic term. Looking again on
Fig. \ref{DCCfig19} one would expect from the behaviour
$\bar \Gamma (\omega \rightarrow 0 )
\rightarrow 0 $ of the dissipation kernel for low frequencies that
within the nonmarkovian treatment the relaxation towards equilibrium
will be much prolonged. This trend can certainly be seen from
inspecting the figure. However,
the nonlinear effective $\phi^4$ potential drives the initial
relaxation comparable to the simple markovian treatment. A
significant and steadily increasing
reduction of the relaxation rate sets in only at later stages of the evolution,
when the effective potential really becomes flat. The complete relaxation
within the nonmarkovian scheme shows thus a highly nonlinear behaviour.

We now go over to discuss the possible differences for the dynamics
of the order parameter including the simple D-dimensional
expansion and cooling scenario discussed in section \ref{sec:modelDCC}
in view of possible DCC formation. As the characteristic quantity we concentrate
on the effective final pion number $n_\pi$ of (\ref{O4pionnumb}).
Potential DCC pionic modes are driven by the initial as well as the intermediate
fluctuations experienced in the evolution.

\begin{figure}[h]
\centerline{\epsfysize=8cm \epsfbox{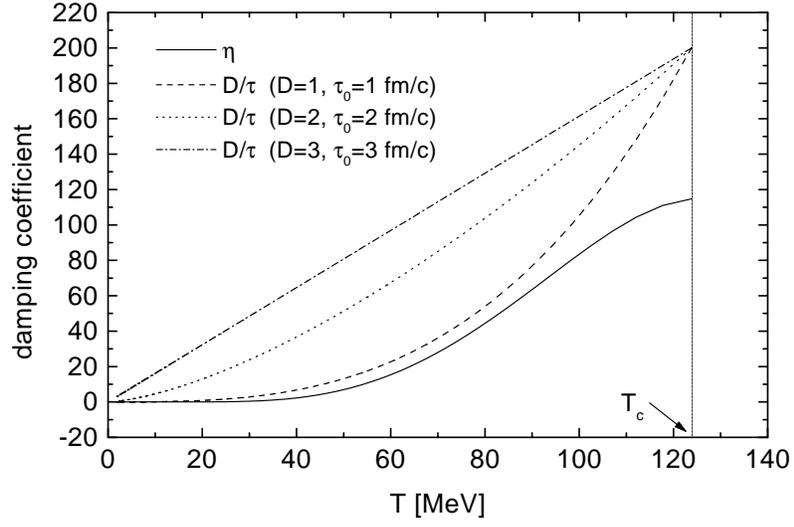}}
\caption{Comparison of the friction coefficient $\eta (T) $
with the Raleigh damping term $D/\tau$.}
\label{DCCfig8}
\end{figure}

Generally it is clear that dissipation will subsequently
diminish potential large pionic fluctuations and thus also decreases
the strength, i.e. the pion number, of the potential DCC candidate.
Only a sufficiently fast expansion and cooling, where the
expansion and cooling rate is comparable or larger then the experienced
damping rate, can counterbalance the effect of dissipation on the heat bath.
To start to be more quantitative let us consider
first the markovian description. One has to compare the Raleigh damping term
$D/\tau$ (the effective `Hubble' parameter) with the dissipation or viscosity
coefficient $\eta$. Both associated terms in the equations of motion
(\ref{EOMDCC}) will diminish the amplitude of any pionic fluctuations being
buildt up during the roll-down period. On the other hand
the effect of the Raleigh damping on the pion number content $n_\pi $
is exactly counterbalanced by the volume dilution (\ref{Volexpan}).
$n_\pi $ being buildt up during the roll-down period can thus
physically only be decreased by the `true' dissipation experienced
from the heat bath. Whether this dissipation can act substantially
depends on whether
the damping coefficient is comparable in magnitude to
the Hubble parameter
\[
\frac{D}{\tau}=\frac{D}{\tau_0}\,\left ( \frac{T}{T_c} \right )^{3/D}\,.
\]
In Fig. \ref{DCCfig8} we compare $\eta(T)$ with the
Raleigh coefficient $D/\tau(T)$ for 3 set of parameters of dimensionality
$D$ of the expansion and initial proper time $\tau_0$.
This serves as a rough illustration how fast the expansion has actually
to be for any potential DCC candidates to appear.
For some reasonable choices
of $D$ and $\tau_0$ one can see that the Raleigh damping $D/\tau$ will be
sufficiently larger than $\eta$, at least for later temperatures below about $70$ MeV.
For a sufficient fast
expansion $D/\tau$ will be much larger than $\eta$, so that the dissipation
due to the interaction with the heat bath can not have any tremendous
effect on the potential DCC candidates
except for a slight hindrance on the evolution.
The important thing during the roll-down is that the fluctuation due
to the noise will be large and can eventually enable a large disorientation
of the order parameter.
For moderate or slower expansion, however, when both damping coefficients
becomes comparable in magnitude after the roll-down period even for later times,
the dissipation will lead
to an additional strong reduction for the pionic fluctuations and thus
for the pion number, making DCC formation physically impossible.

In order to support these qualitative arguments we calculate
the average pion number (i.e. the sum of the pion number of each
individual event divided by the total number of events) and
the pion number of the `most prominent' event
within $10^3$ independent events
by solving the markovian Langevin equation (\ref{EOMDCC}) and
compare those with the result obtained by solving the same equation
but without the damping term $\eta $ and the fluctuating noise.
(The thermally distributed initial configurations are the
same for both cases.) The most prominent event is meant here and in the 
following sections as the one where the final pion number
is the largest within the generated, finite ensemble.
The `most prominent' event is at first, of course, of no direct statistical
significance. The error of its occurrence for a finite ensemble
will indeed be very large. We explicitely show it for the reason
to simply see what maximum magnitude in the pion number is possible
within a finite total number of generated events within one particular chosen
ensemble.

The calculations are performed for
different parameters $D$ and $\tau_0$. Table \ref{dcctable1} shows the results.
For a discussion and possible motivation for the various parameters
and their actual physical relevance we refer at this point to the next
section \ref{sec:stochDCC}. Here we want to stress that the
results of table \ref{dcctable1} confirm our arguments: For the
relative slower expansion the dissipation due to the interaction with the heat
bath destroys any possible large pionic oscillations and
therefore leads only to a small total pion yield. In contrast, for a fast
expansion the dissipation has only a minor influence on the DCC formation.
For these cases the damping coefficient $\eta$ is indeed rather small
compared to the Raleigh coefficient $D/\tau$.

Now we can answer our primary question: Is there any difference
between the full nonmarkovian treatment of the dissipative dynamics
compared to the approximate markovian treatment on the possible formation
of DCC. From our findings at the beginning of this section we expect that
the effective damping experienced by the memory effects
within the nonmarkovian case is moderately, but not significantly diminished.
(`Memory' indicates that the earlier
stages of the evolution influence
the present motion of the order parameter.)
The answer is `frustrating' and simple.
For a moderate expansion the pion yield from the DCC decay will
increase compared to the markovian treatment, but only slightly.
In any case for such a situation the possible pion yield obtained within
the simulations are too small to have any experimentally relevant
consequence! On the other hand, for a sufficiently fast expansion
(, which might be speculative or not to be realized in a relativistic
heavy ion collision,)
and for which more prominent DCC candidates will show up
(compare the next section), the memory effects of the treatment
of the dissipation and noise are not of particular significance
for the final pion number distribution.

As one particular example we show the outcome of a simulation
for both cases in Fig. \ref{DCCfig9}.
We take $D=1$ and $\tau_0=0.5$ fm/c to simulate
a somewhat moderate expansion.( The dimensionality parameter
$D=1$ for longitudinal expansion simulates a rather slow expansion.
On the other hand the here chosen value of $\tau_0$ is
very small so that the initial cooling and expansion after the onset
of the phase transition is still rather
fast. This value is definitely too small
to be realized in nature. Typically one expects a few $fm/c$ for
the onset of the phase transition. In the next section we will see that only
a $D=3$-dimensional expansion can lead to any prominent DCC candidates
for reasonable choices of $\tau_0$. Therefore one should not consider
this present example as a physically relevant scenario.
Its purpose is merely to be an example which does indicate
some differences between the nonmarkovian and markovian treatment.)
In the upper part the sampled distribution of the final
pion number is shown within $10^4$ independent events.
In the lower part the individual trajectories
$(\sigma,\pi_i)(\tau )$, i=1,2,3, for the most prominent event
out of each ensemble are depicted. Clearly these trajectories
represent the ones expected intuitively for a true DCC event.
This intuitive picture is further strengthened
when examing Fig. \ref{DCCfig10}, where the evolution in time of
the transversal mass $\mu_{\bot }(\tau )$ and the effective
pion number $n_{\pi }(\tau )$
are depicted for the most prominent candidate within the markovian simulation.
One clearly recognizes that for this candidate the evolution
starts with an unstable situation where the pion fields
and thus also the pion number will be amplified significantly in
the very first stage.
The subsequent minor decrease in the pion number is
then attributed due to the further experienced damping.

The point to make here is that the pion yield of the most prominent
event as well as the average pion number in the ensemble
are somewhat larger within the nonmarkovian simulation.
However, in sake of the variety of choices for the parameters $D$ and $\tau_0$
and the associated wildly differing outcomes
(compare e.g. tables \ref{dcctable1} and \ref{dcctable2}),
a modification as presented here is only of minor importance.

We certainly have now to answer what this torturing enterpise
for achieving a simulation of nonmarkovian Langevin equations
was good for!
In general dissipation as well as the associated noisy fluctuations
are non-local phenomena in time leading to a memory functional over the
past history of the system for describing the dissipation
as well as to a finite correlation in time of the noise.
The more phenomenological markovian and white noise approximation
are generally used in one way or the other as their numerical realization are
considerably more simple. Typically such an approximation is justified
in a loose sense when there exist a clear separation of timescales
between the slow degrees of freedom under consideration and the ones integrated out.
At first sight this is not really given for our situation, though
our investigation shows that one might indeed work with the much simpler
markovian approximation. It is also easy to imagine
that such a separation is not given either for a variety of
interesting problems
in other areas of physics where one wants to describe the effective
dynamics of a system in terms of only a few `relevant' degrees of freedom.
We therefore believe that our investigation and in particular the numerical
realization of nonmarkovian Langevin equations with colored noise
is of general and principle relevance for similar problems
of classical or quantum dissipative systems in other areas.
In addition, our extensive discusssion here underlines the importance
of understanding the certainly complex dissipative nature of the chiral phase
transition in more detail. We believe that our `choice' for describing
the dissipation for the pionic fluctuations of the zero mode
with the surrounding heat bath is motivated by an intuitive physical picture.
If, on the other hand, one can show
that the experienced dissipation for the pionic (transversal)
fluctuations is in fact much stronger,
then there is no chance at all for any DCCs to be formed
in heavy ion collisions!

\section{Stochastic formation of DCC}
\label{sec:stochDCC}

Although the inclusion of dissipation as discussed in the last
section \ref{sec:nonmarkov}
gradually destroys on general grounds
any possible large oscillations of the coherent pionic fields
during and after the roll-down period, there
still should be a chance for a particular large pion yield
originating from some `appropriate' initial fluctuations of the order parameter and
as well as from the subsequent fluctuations experienced.
If the initial fluctuations are large, and if the subsequent
expansion during the roll-down of the order parameter
is sufficiently fast, there should indeed be a considerable probability for a
long-time large disorientation of the chiral fields away from the
$\sigma $-direction during and shortly after the roll-down period.
This would then lead to a particular large final pion yield.
As emphasized already in \cite{PRL}, individual statistical events
will lead to sometimes significant growth of pionic fluctuations.
In this sense the formation of a particular `large' DCC, i.e.
with a sizeable amount of low momentum pions being emitted,
can follow some unusual distribution to occur because of the special
stochastical and nonlinear dynamical nature with a possible, temporarily onsetting instability.
To anwser this question of how often particular events
might occur with some unusual large pion
yield, we investigate in the following
the distribution of the pion number for
different DCC scenarios which differ from each other in the cooling or/and the
sampling of the initial fluctuations.
We will see that the distribution in the final
pion number takes a nontrivial and nonpoissonian form,
at least for the more speculative scenarios or parameters employed where
one might expect larger DCCs to occur.
By means of the
cumulant expansion of the resulting distributions we will then show
that the higher order factorial cumulants are even still moderately large
when allowing for an additional and incoherent realistic background
of low momentum pions. Therefore these unusual fluctuations might
indeed be observed experimentally and thus provide a very interesting
new signature for a nonequilibrium chiral phase
transition and the associated formation of DCC.

As a crude estimate for the maximum soft pion number
to occur from the decay of a DCC one can think of a `true' DCC
where the chiral order fields `circle' around with the maximum
amplitude as $ \langle \vec{\pi }^2 + \frac{1}{m_{\pi}^2} \dot{\vec{\pi }}^2 \rangle \,
\leq f_\pi ^2 $ along the chiral circle $ \sigma ^2 + \vec{\pi }^2  \,
\approx f_\pi ^2 $ at some intermediate stage
$\bar{\tau }$ after the roll down in the evolution
(compare with the lower part of Fig. \ref{DCCfig9}). (Due to the ongoing
expansion in our model the amplitude will then subsequently decrease
due to the experienced Raleigh damping, so that at late times
the chiral fields will then only fluctuate around the
vacuum value $\langle \sigma \rangle = f_{\pi }$.)
This will result in a coherent pion number density
of $n_{\pi }/V \approx f_\pi^2 m_\pi /2 \approx  0.08 \, fm^{-3}$.
For the total pion number the crucial question is then how large
has the evolving
volume $V(\bar{\tau })$ of the DCC domain increased when the pion oscillations
have emerged.

At this point we should give also another rough estimate of how many
low momentum pions should emerge out of decaying DCC in order for
a chance of experimental detection. In a relativistic
heavy ion collision at RHIC one typically expects around 1000
pions being produced per unit in rapidity.
On this `background' one has to look for a peculiar and unusual enhancement
in the pion spectrum at low transverse momentum to identify
possible DCC formation. It is thus clear that the number
of emitted pions out of a domain should be somehow comparable
to the number of background pions for a particular small window
of low transverse momentum. The expectation is that one should
have a surplus of at least 50 pions stemming from a DCC per unit rapidity
in a window of $p_t<200 \, MeV$
in order to allow for a promising detection \cite{BGH98} (see also the
schematic Fig. \ref{DCCfig16}). This number should thus
serve in the following as a rough guide. We will come back to
the experimental detection possibilities at the end of this section.
For our above estimate for the maximum number of pions out of a
domain this would mean that the intermediate volume
$V(\bar{\tau })$ has to have increased up to a value of about $10^3$ $fm^3$
when the order parameter has reached the chiral circle.
This again implies to consider (or demand) a rapid expansion,
i.e. to consider (D=)3-dimensional expansion and sufficiently
small initial time $\tau_0$, as already demonstrated in the last section
(see table \ref{dcctable1})
and which first was emphasized in other studies \cite{Ra96a,La96}.

\subsection{Different scenarios}
\label{sec:scenarios}

In the following we will present numerical results for
the formation of DCC, i.e. the coherent amplification of the pionic
chiral order fields resulting in a final
pion number (\ref{O4pionnumb}) being effectively emitted by the domain,
for various parameter sets and also for
four somewhat different scenarios (see also \cite{GXB98}).

The first scenario we want to discuss is the `normal' one already described
in section \ref{sec:modelDCC}:
\\[2mm]
{\large $\bullet $}
{\it Langevin} or `annealing' scenario \cite{GM94,PRL}:
Like the dynamical calculations in the
last section \ref{sec:nonmarkov} the initial configurations
of the chiral fields (and their velocities) are
sampled statistically for an assumed thermal equilibrium
at the initial temperature $T(\tau_0)=T_c$ and an initial
volume $V(\tau_0)$, thus covering a (nearly) complete set of possible
initial thermal conditions.
For $\tau > \tau_0$ the subsequent evolution
of the chiral order fields
for each individual realization of the sample
is then described by the
(markovian) equations of motion (\ref{EOMDCC}) within a
D-dimensional scaling expansion according to (\ref{Texpan}) and (\ref{Volexpan}).

\begin{figure}[h]
\centerline{\epsfysize=12cm \epsfbox{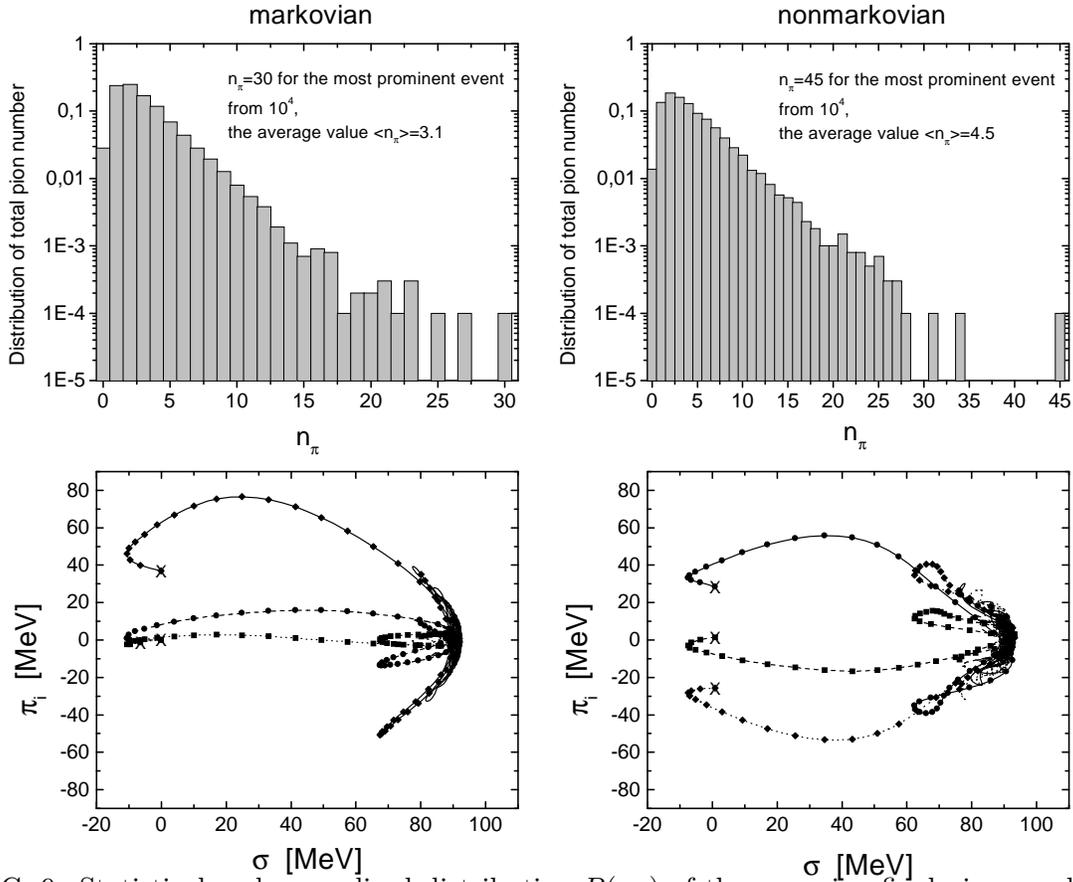}}
\caption{Statistical and normalized distribution $P(n_\pi )$
of the emerging final pion number
and the time evolution of the
three pionic trajectories ($(\sigma(\tau),\pi_i(\tau))$, i=1,2,3,)
for the most
prominent event within an ensemble of $10^4$ realizations in both a
markovian and
a nonmarkovian simulation. The trajectories start at the initial proper time
$\tau_0=0.5$ fm/c. The starting points are marked with `X'. The marks along 
the trajectories are positioned at time 
intervals of $\Delta \tau =0.21$ fm/c.}
\label{DCCfig9}
\end{figure}

\begin{figure}[h]
\centerline{\epsfysize=8cm \epsfbox{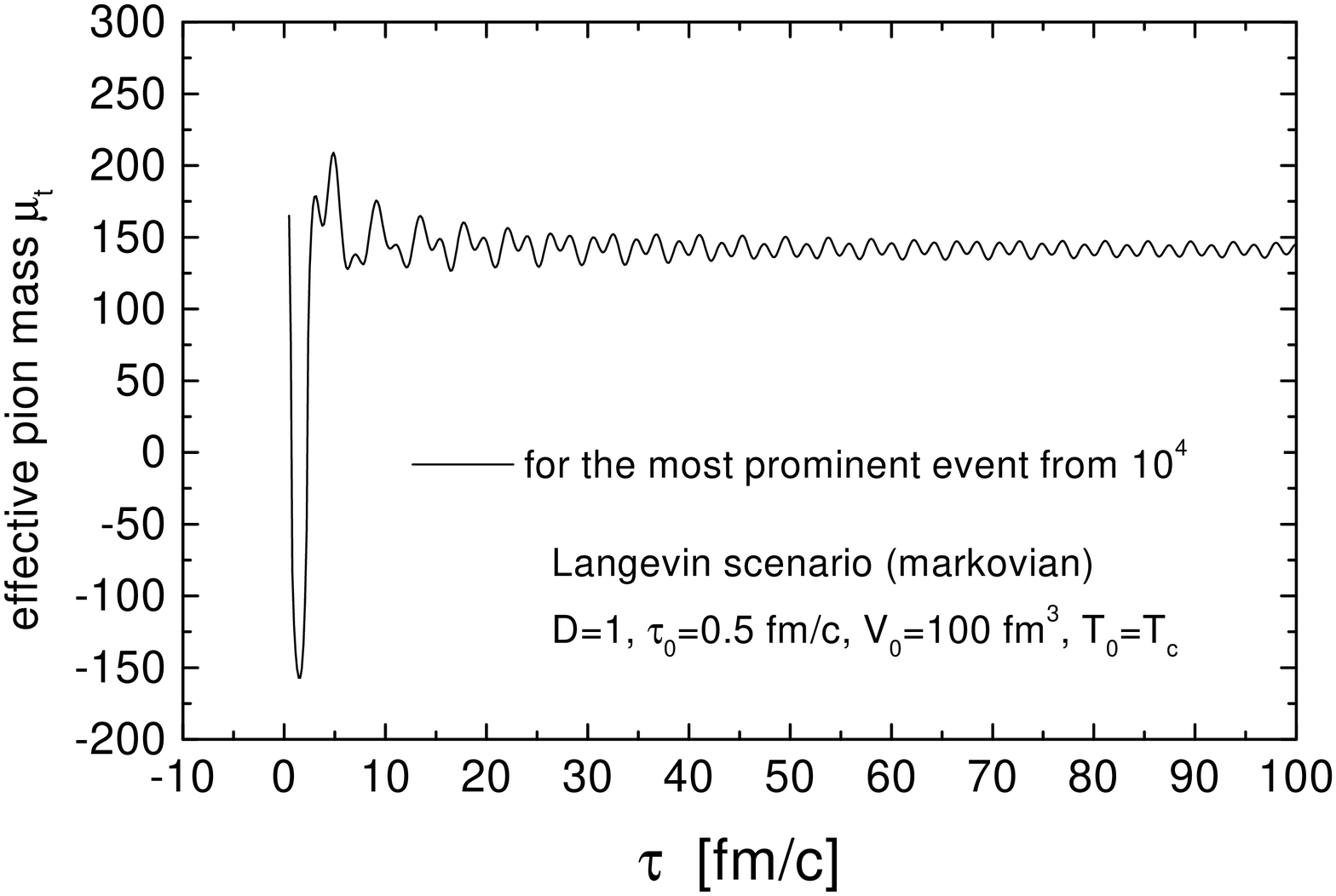}}
\centerline{\epsfysize=8cm \epsfbox{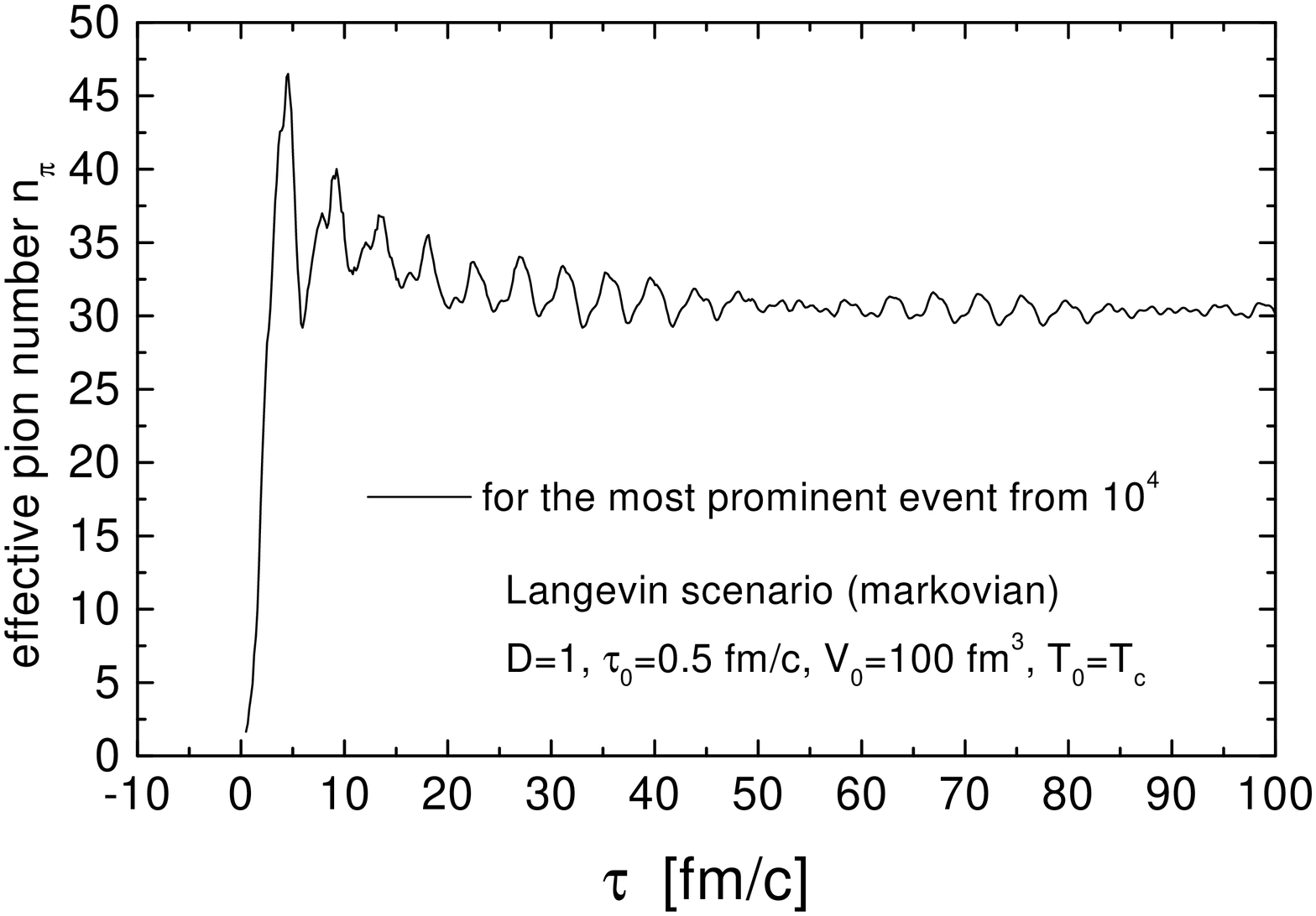}}
\caption{The time evolution of the effective pion mass
$\mu_{\bot }(\tau )$ and the effective pion
number $n_\pi (\tau )$ for the most prominent event
obtained in the markovian simulation of
Fig. \ref{DCCfig9}}.
\label{DCCfig10}
\end{figure}

\begin{figure}[h]
\centerline{\epsfysize=10cm \epsfbox{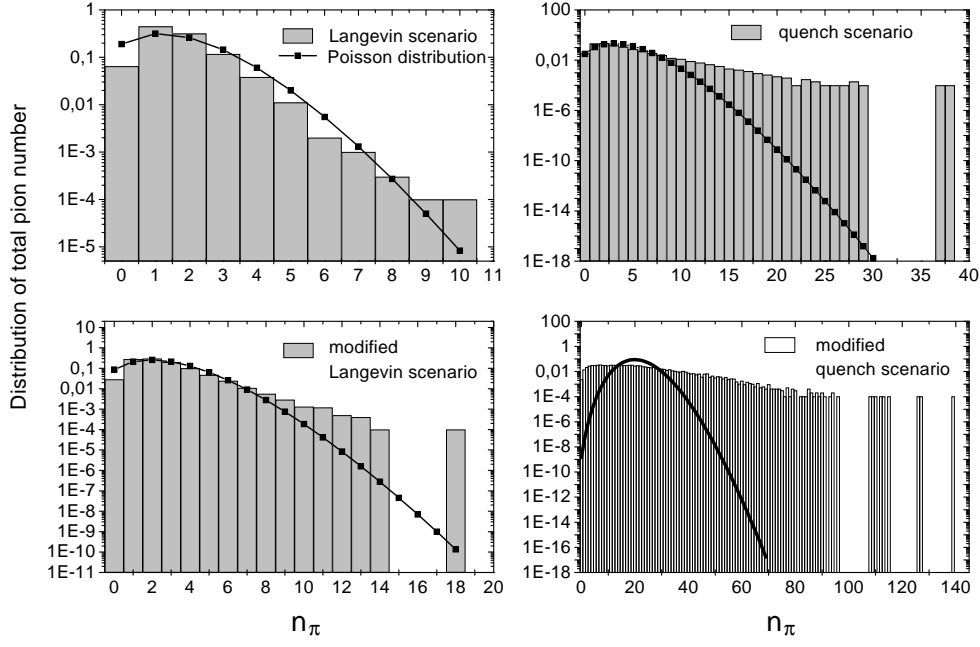}}
\caption{Statistical distribution of the final yield in the pion number
for four different scenarios (see text) within a (D=)3-dimensional
scaling expansion. Each simulation
has been performed with $10^4$ independent events.
The initial volume $V(\tau_0)=100 \, fm^3$ and the initial proper
time is taken as $\tau_0=7 \, fm/c$.
The distributions are compared with the corresponding
Poisson distributions. The averaged pion number in the
Langevin, modified Langevin, quench and modified quench scenario are
1.66, 2.45,
3.46 and 20.36 respectively.}
\label{DCCfig11}
\end{figure}

\begin{figure}[h]
\centerline{\epsfysize=8cm \epsfbox{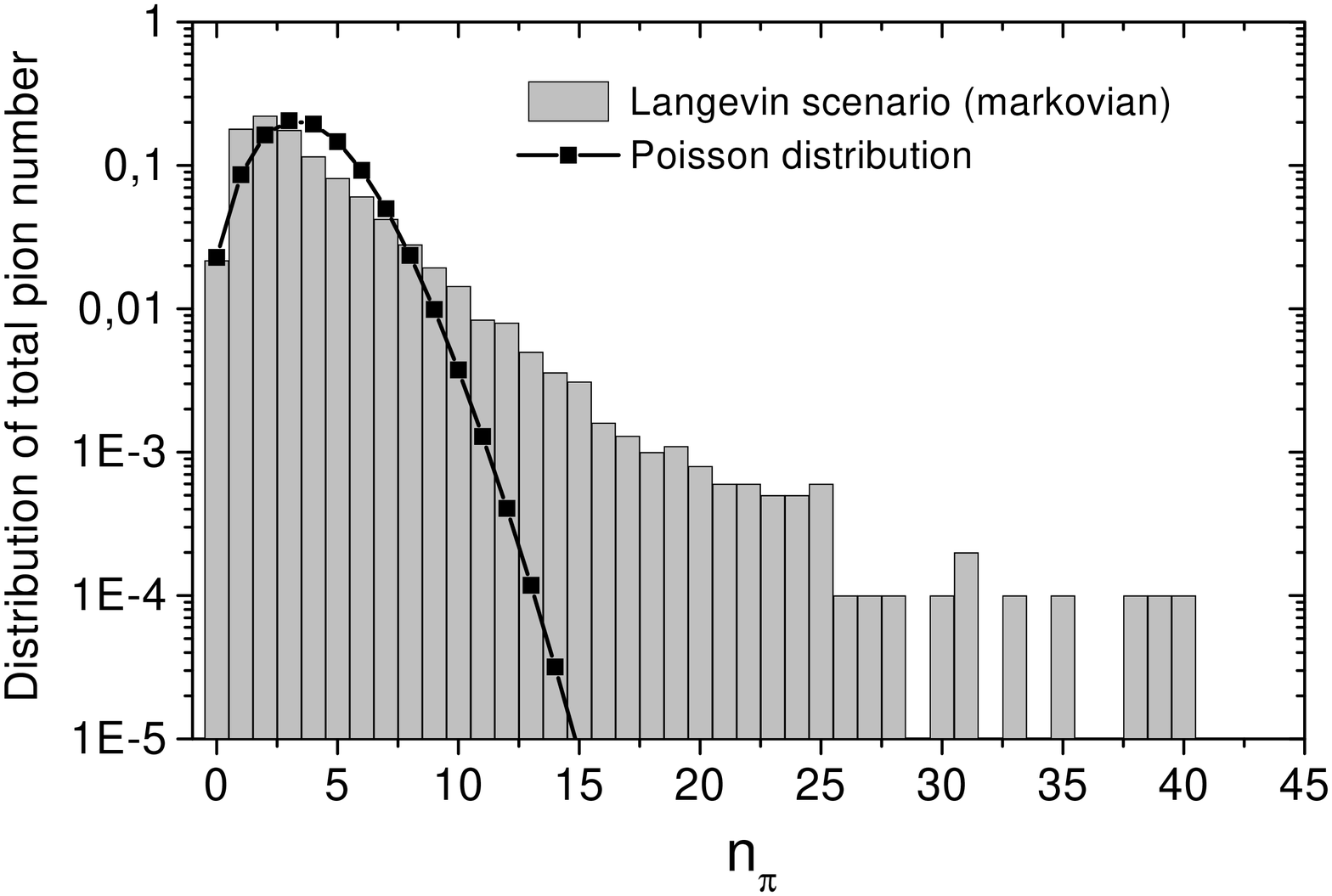}}
\centerline{\epsfysize=8cm \epsfbox{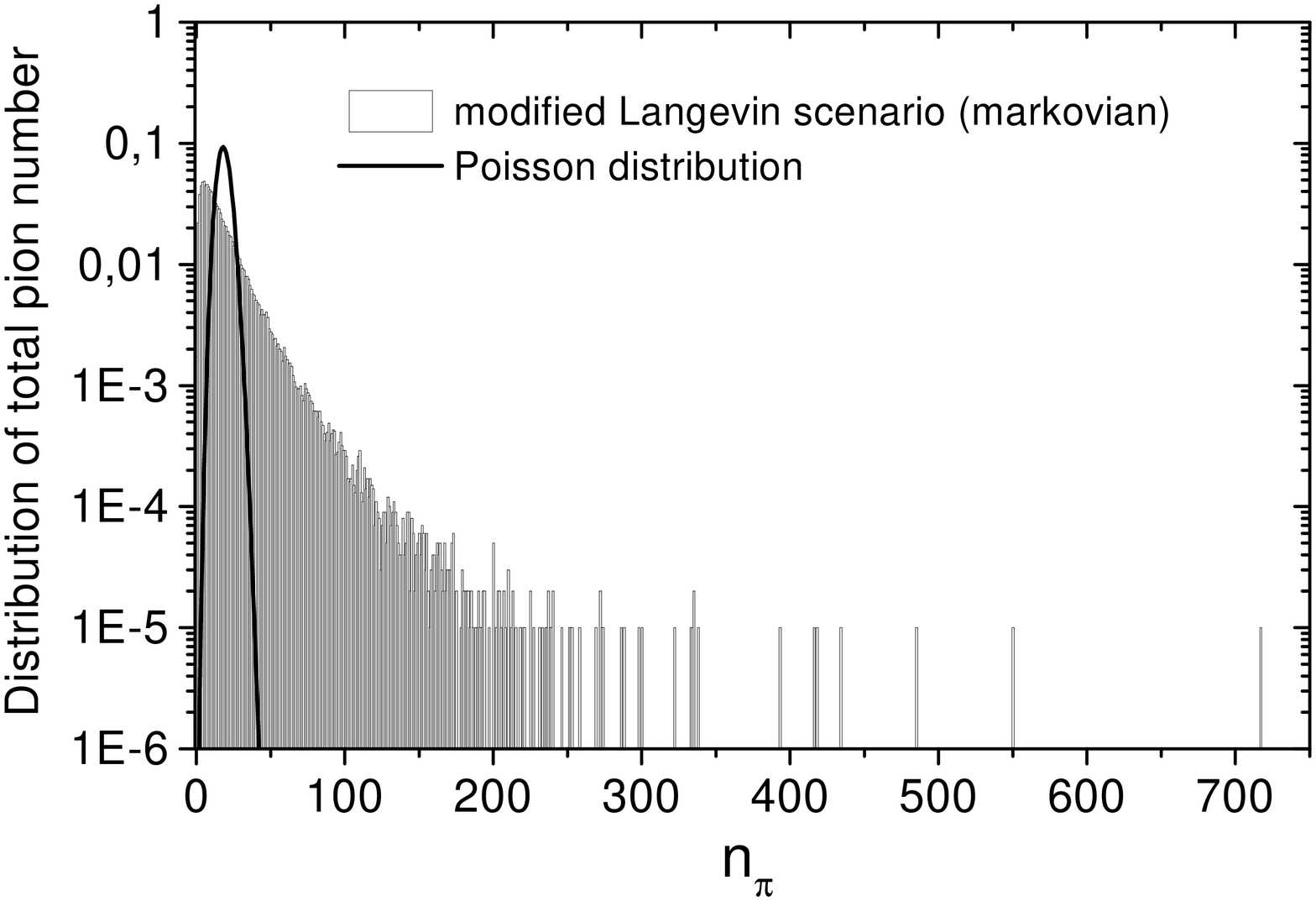}}
\caption{Statistical distribution of the final yield in low momentum
pion number within the Langevin (upper figure) and modified
Langevin (lower figure) scenario compared with the corresponding poissonian 
distribution.
A fast expansion is simulated by choosing $D=3$ and $\tau_0=3$ fm/c.
The upper distribution is calculated within $10^4$ events,
whereas for the lower a sample of $10^5$ independent events has been chosen.}
\label{DCCfig12}
\end{figure}

Particular examples are listed in the tables \ref{dcctable1} and \ref{dcctable2}
and in the Figures \ref{DCCfig9}-\ref{DCCfig12} for various
parameters D, $\tau_0$ and $V(\tau_0)$. As outlined at the beginning of section
\ref{sec:modelDCC} one expects that the chiral phase might set in
at proper times $\tau_0 \approx 3 - 7 \, fm/c$.
Inspecting table \ref{dcctable1} one recognizes that employing a $D=1$ or $D=2$
dimensional scaling ansatz either the average outcome $\expl n_\pi \expr$
or also the outcome for the most prominent candidate of the sample are
unacceptable small for experimental detection. Only for the $D=3$ case
(see the tables and the Figures listed above) individual and unusual
events might occur for a small initial proper time $\tau_0 \leq 3 fm/c $
and which might be detectable. This is the situation for a very rapid expansion
and cooling as noted the first time by Randrup \cite{Ra96a}.
We note, however, that the average $\expl n_\pi \expr $ is still only moderate
even for this rapid scenarios, i.e. $\expl n_\pi \expr \approx 3-4$,
and thus also unacceptable small (see e.g. the upper part of Fig. \ref{DCCfig12}).
As already stressed in \cite{PRL},
for an experimental identification this would imply to look for
(rather) rare and unusual strong
fluctuations on an event by event analysis in certain rapidity and low
$p_t - $ windows.

On the other hand one can clearly recognize from the outcome that
the original annealing picture proposed by Gavin and M\"uller
\cite{GM94} and assuming there a rather moderate expansion and cooling
($\tau_0 \approx 7-10 \, fm/c)$ does not work as the final
pion number is by far too small (confer table \ref{dcctable2}).
Experimentally significant DCCs cannot happen for this picture
according to our calculations.

\begin{figure}[h]
\centerline{\epsfysize=10cm \epsfbox{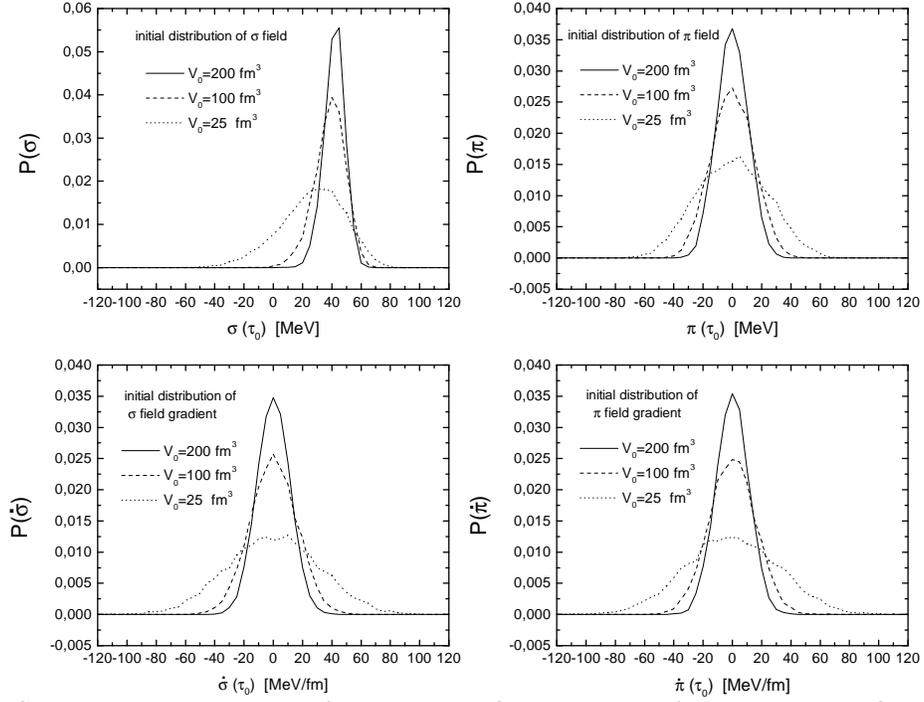}}
\caption{ Statistical distribution
of the chiral $\sigma$
field and one of the three pion fields for different finite volumes.
The temperature is taken as the critical temperature
$T_c$. The distributions are obtained from an ensemble of $10^4$
independent realizations.}
\label{DCCfig5}
\end{figure}

Inspecting table \ref{dcctable2} more closely, one recognizes
the at first sight maybe somewhat paradox behaviour, that
the average as well as the pion yield for the most prominent
candidate of each numerically generated ensemble do not show
any strong sensitivity on the chosen initial volume $V(\tau_0$).
This result one can at least qualitatively understand as follows:
The initial fluctuations of the chiral fields depend on
the initial volume as discussed in
section \ref{sec:initfluc} (see also Fig. \ref{DCCfig5}).
For a smaller initial volume the initial fluctuations
become stronger. Hence there is a larger probability
for the order parameter to
start to evolve against the positive $\sigma $-direction into the
`backward hemisphere'.
(For this see e.g. the two examples shown in Fig. \ref{DCCfig9}.
All `more prominent' candidates do show a time evolution for the chiral
fields akin to the ones depicted there.)
For such a case the order parameter
has to turn back during the roll-down so that period of large pionic
fluctuations would be
prolonged. This then gives raise to a larger disorientation of the
order parameter. On the other hand, however, the volume of a DCC domain at the 
freeze-out time is accordingly smaller for the initial volume being smaller.
Referring to (\ref{O4pionnumb}) both trends seem to nearly exactly
counterbalance each other and hence leading to this peculiar behaviour.

\begin{figure}[h]
\centerline{\epsfysize=10cm \epsfbox{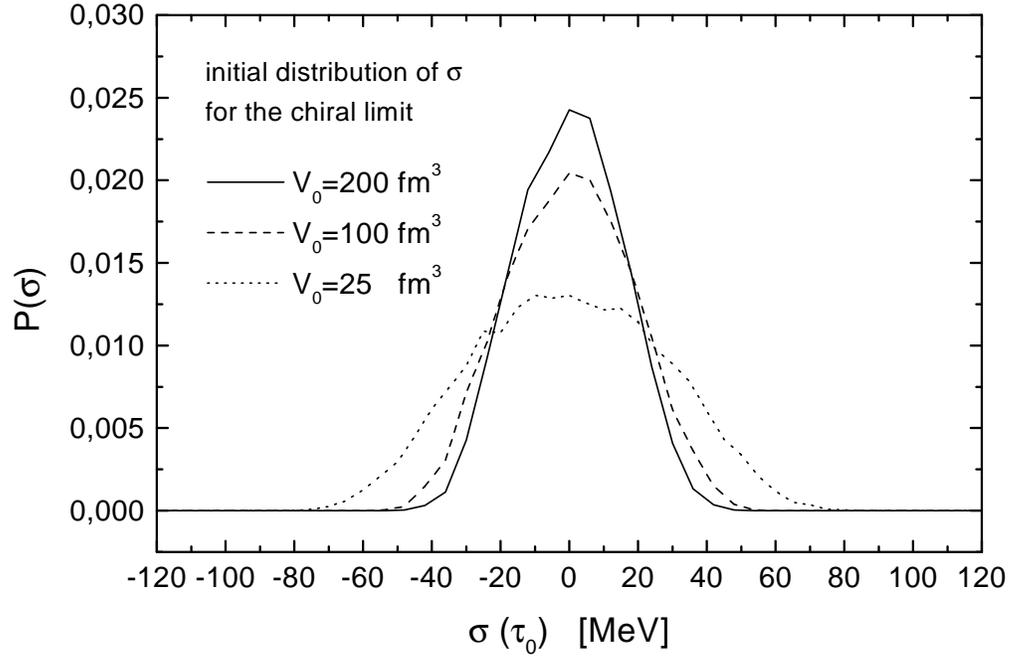}}
\caption{ Statistical distribution
of the chiral $\sigma$ field for different finite
volumes for the situation without explicit chiral symmetry breaking.
The temperature is taken as the critical temperature $T_c$. The
distributions are obtained from $10^4$ realizations.}
\label{DCCfig6}
\end{figure}

We now turn to present a few results obtained for three other, more speculative
scenarios:
\\
{\large $\bullet $}
{\it quench} scenario:
The initialisation at $T=T_c$ follows completely analogous to
the {\it Langevin} scenario. However, when switching on to the evolution
for $\tau > \tau_0$ including the volume dilution (\ref{Volexpan}), we
demand that during the expansion the term  $\lambda T^2/2$ of the
effective potential in the equations of motion (\ref{EOMDCC})
is being omitted as well as also the dissipative term and the noise
in order to mimic an abrupt occurence of the zero temperature vacuum potential.
Consequently the dissipation and the fluctuation vanish during the roll-down 
and the oscillation of the order parameter.
Within this scenario we try to simulate somewhat the picture proposed
in \cite{RW93,Raja1}, where it is assumed that the effective
potential below $T_c$ changes quasi abruptly to the vacuum potential
for $T=0$. However, the initial conditions, on the other hand,
are sampled at thermal equilibrium
at critical temperature $T_c$. We believe that this picture
represents a strong idealization and probably is
not likely to happen in an ultrarelativistic heavy ion collision. Due to the
abrupt cooling there is likely more instablity for the order parameter
allowing for stronger final fluctuations.
\\[2mm]
{\large $\bullet $}
{\it modified Langevin} scenario:
This scenario differs from the
Langevin scenario only in the sampling of the initial configurations. For the
sampling we neglect the explicit chiral symmetry breaking (i.e. $H=0$;
see Fig. \ref{DCCfig6}).
On the other hand, for the evolution at $\tau > \tau_0$ we employ the
same equations of motion (\ref{EOMDCC}) including the
explicit symmetry breaking.
In the modified Langevin scenario the initial fluctuations are stronger
than for the other two scenarios since the most probable initial value of
the order parameter is centered around
$(\expl \sigma \expr =0, \expl \vec\pi \expr =0)$ and the
effective potential for this case is more flat.
Hence the possibility for the order
parameter to start its evolution towards the backward hemisphere is
more likely to occur.
One might argue that the use of the initial conditions prepared within this
picture is inconsistent within the
linear sigma model with a physical pion mass. When discussing
Fig. \ref{DCCfig2} we noted that the phase transition
resembles a smooth crossover.
From QCD lattice calculations, however, one knows that the chiral transitions
happens much sharper within a very narrow window close to $T=T_c$.
This means that it might very well be that the order parameter
will (strongly) fluctuate around zero near the critical temperature
as mimiced by the present realization of the initial conditions.
With this in mind one might consider this present scenario
even more realistic than the Langevin scenario following
the simple minded linear $\sigma $-model.
\\[2mm]
{\large $\bullet $}
{\it modified quench} scenario:
The initial configurations of
the order parameter are sampled as in the modified Langevin scenario. The 
dynamical evolution corresponds to the quench scenario.

In Fig. \ref{DCCfig11} we depict the distribution
$P(n_{\pi})$ of produced pions logarithmically within
$10^4$ events within the four different DCC scenarios.
As parameters we choose $D=3$ and $\tau_0 =7\, fm/c$ \cite{GM94} and
$V(\tau_0 )=100 \, fm^3$, i.e. still only a rather moderate expansion.
As expected, the pion yields in the modified scenarios are larger than in the
normal scenarios for the most prominent event as well as for the average.
A comparison of annealing and quench scenarios, both
with finite and vanishing pion mass
(for generating the initial conditions)
reveals that
the most productive DCC events would lead
for this set of parameters to a few (6-8 in
annealing scenario with finite pion mass), to a moderate number
(20 - 40 in annealing scenario with zero-centered
initial conditions or quench with
massive pions) or to about 140 long wavelength pions (in quench scenario
with initial conditions generated by massless pions), respectively.
The final results, of course, majorly depend
on how fast the effective cooling and expansion
proceeds, i.e. on the value
of the initial time $\tau _0$
and thus the overall initial Hubble constant $D/\tau_0$
(see also Fig. \ref{DCCfig12}).
In general one finds that
for sufficiently fast expansion individual unusual strong
fluctuations of the order of 50 - 200 pions might occur in all the four
scenarios, although the average number $\expl n_\pi \expr $
of the emerging
long wavelength pions only posesses a rather moderate (and likely undetectable)
value of 5 -20.

For a direct comparison we
depict a poissonian distribution
\begin{equation}
\label{poisdistr}
P(n)\, = \, \frac{\bar{n}^n}{n!}\,e^{-\bar{n}} \, ,
\end{equation}
where the mean value $\bar{n}$ is equal the averaged pion number
$\expl n_\pi \expr $ obtained
numerically for each sample.
With the chosen parameters of Fig. \ref{DCCfig11}
the distribution of the pion number for the Langevin scenario
is indeed still similar to a simple poissonian distribution.
(As mentioned above, for such a slow expansion the coherent
pions produced from a DCC decay would be washed out by the
background of incoherent pions and thus could not provide any signature.)
However, for the other three  cases the
final distribution does {\em not} follow an usual poissonian distribution.
This represents a very important outcome of our previous \cite{GXB98}
and the present, more
detailed investigation!
Fluctuations with a large number of produced pions are still likely
with some small but finite probability! In principle, an ensemble averaged
description of potential DCC formation carried out
within the mean field approximation,
as presented in the various literature, can not account
for such fluctuations and thus has to fail at some point.
We remark further that also the so called isospin ratio signal
is close to that expected
for a DCC event.

To demonstrate this interesting behaviour of strongly
nonpoissonian fluctuations even more pronounced, we show
in Fig. \ref{DCCfig12} the pion number distribution obtained
within the Langevin and modified Langevin scenario for a rather
fast expansion ($D=3$ and $\tau_0=3\,  fm/c$).
This parameters are in line with the ones used in other studies
\cite{Boy95,Ra96a,MSL98,La96}.
Both distributions differ strongly from their corresponding poissonian
distributions. The averaged pion number $\expl n_\pi \expr $
are 3.9 and 18.5, respectively, and are both comparable to the values
obtained within the quench and modified quench scenario of Fig. \ref{DCCfig11}.
The appearance of particular events with very
large pion number (more than 200) is hereby attributed to the
initial fluctuations and the ones experienced during the
roll-down periode.

This special and unusual statistical distribution
(obtained within the modified Langevin scenario) will be further analyzed
in the next section \ref{sec:critfluc}. We want to note at this point, that
Bjorken and collaborators \cite{ABL97} had speculated that the final
distribution in the pion number will arguably go beyond a gaussian (or poissonian)
distribution even when the DCC fluctuations are generated by a gaussian distribution
for the initial condition parameters. This `expectation' is what we have demonstrated now.
The initial conditions (compare Figs. \ref{DCCfig5} and \ref{DCCfig6})
follow, more or less, a gaussian distribution, whereas the final
occuring distribution in e.g. the pion number strongly deviates
from a gaussian behaviour for the
(assumed) nonequilibrium situations,
where the occuring DCC phenomena could be considered as
experimentally be detectable.
We also like to mention
that Krzywicki and Serreau
had recently found in a somewhat similar setting, following the
model of \cite{La96}, that the so called enhancement factor
for the final fluctuations also will follow some unusual and
nonpoissonian distribution \cite{KS99} (see also \cite{PRL}).

\begin{figure}[h]
\centerline{\epsfysize=8cm \epsfbox{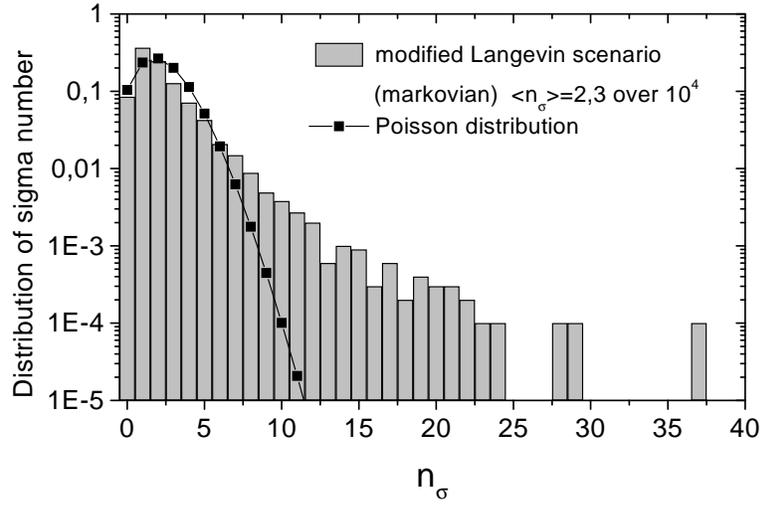}}
\caption{Statistical distribution of the final number of $\sigma$-mesonic
excitations within the modified
Langevin scenario of Fig. \ref{DCCfig12}
obtained
within $10^4$ events.}
\label{DCCfig13}
\end{figure}

There also had been the conjecture in the literature that the
long wavelength amplification of the pionic fluctuations is not
really driven by the `true' DCC phenomenon, but actually could
be attributed to a parametric resonance behaviour
driven by the late and final oscillations of the coherent $\sigma $-field
\cite{MM95,Boy95}. This alternative idea we can at least
qualitatively address. In Fig. \ref{DCCfig13}
we show the statistical distribution of the final number for
the `$\sigma $'-quanta by means of an analogous expression as (\ref{O4pionnumb}),
i.e.
\begin{equation}
\label{O4signumb}
n_{\sigma } = \frac{1}{2} m_{\sigma }
\left( ( \sigma (\tau ) - \langle \sigma \rangle _{vac} )^2  +
\frac{1}{m_{\sigma }^2} \dot{\sigma }^2 (\tau)
\right) \, V(\tau )\,,
\end{equation}
relaxing to a constant value
in the late oscillations of the $\sigma $-field in longitudinal
direction around its vacuum value.
The scenario and parameters chosen are the ones for the very pronounced situation
of the lower part of Fig. \ref{DCCfig12}.
On the average about $\expl n_\sigma \expr \approx 3 $ $\sigma $-particles are
produced, but within the sample also some events with more than 30
$\sigma $-particles can occur. Due to the potential vacuum decay
$\sigma  \rightarrow \pi \pi$
(with a width on the order of a few hundred MeV)
the existence of these $\sigma $-quanta
would result on the average in 6 additional pions or, for the more pronounced
events, up to more than 60 additional pions. This is a quite reasonable number,
however, it is still considerably smaller than the direct pions stemming of the true
DCC as seen from the lower part of Fig. \ref{DCCfig12}. So, if these late
$\sigma $-oscillations really do exist, because of energy conservation,
the amount of pions being produced out of them, either
thinking in a pertubative way as a result of an individual decay of a $\sigma $-quantum
or within the nonpertubative mechanism of parametric resonance, the numbers
of produced pions is found to be significantly less than the direct ones
of `true' DCC in all simulations carried out.
We therefore are tempted to conclude
that parametric behaviour is not as efficient compared to the `true'
DCC phenomenon.
On the other hand, we want to stress here, that these late
$\sigma $-oscillations obtained in our simulations are actually a caveat
of our model. As already pointed out at the end of section \ref{sec:zeromode},
exactly because of the possible decay mode
$\sigma  \rightarrow \pi \pi$, one has in principle to
account for an additional temperature dependent dissipative term
in longitudinal direction for the evolving order parameter.
An inclusion would in fact then accordingly continously decrease these
`radial' oscillations because of the decay into pions \cite{DR98,CEJK}.

\begin{figure}[h]
\centerline{\epsfysize=8cm \epsfbox{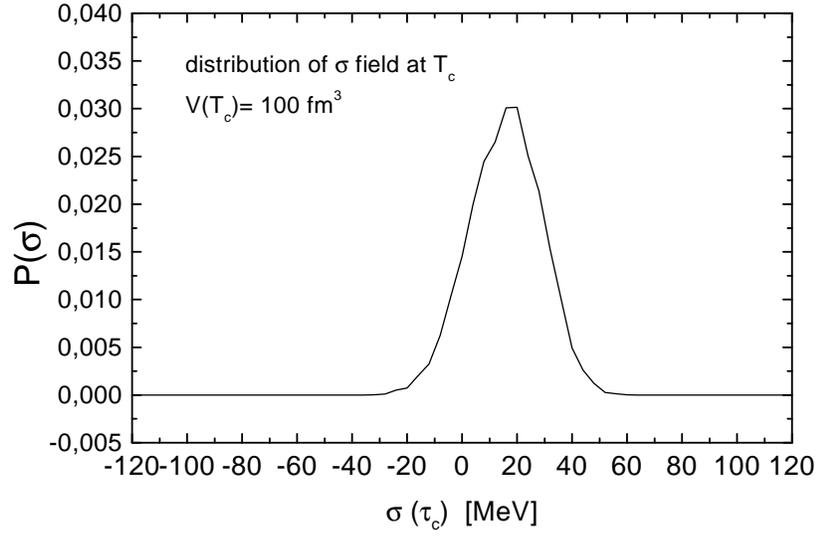}}
\centerline{\epsfysize=8cm \epsfbox{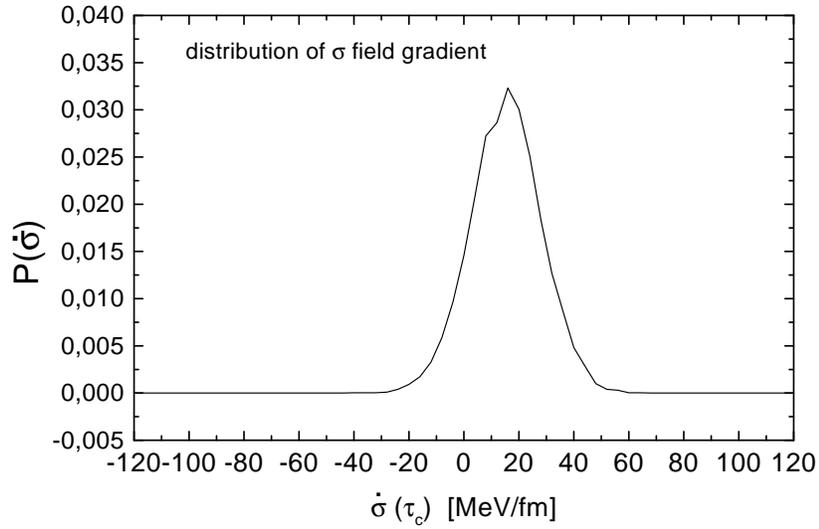}}
\caption{Statistical distribution of the $\sigma$ field and its
temporal gradient at the time $\tau_c= 3 \, fm/c$ when the
critical temperature $T_c$ is reached.
The time evolution starts at a higher temperature
$T_i=300$ MeV with a 3-dimensional expansion in the Langevin scenario.
The volume at $\tau_c$ is 100 $fm^3$.}
\label{DCCfig14}
\end{figure}

As a last investigation we consider the possibility that one might
prepare the initial conditions for the fluctuating
chiral fields at some higher initial temperature $T_i(\tau_i) \gg T_c$ within the
Langevin scenario. The order parameters are then centered more or less
around zero \cite{Ra96a}. Switching on to a rapid 3-dimensional scaling
expansion one intuitively would expect that the chiral fields still would fluctuate
around zero when the system cools down at and below
the critical temperature $T_c$ and thus providing somehow similar initial
conditions like in the modified Langevin scenario. Accordingly one would
expect a more dramatic yield in the pion numbers comparable to the
one obtained within the modified Langevin scenario.
It turns out that this is not the case.
We find that the final yield follows more closely the result of the
Langevin scenario with the initial conditions sampled at $T=T_c$, if
the parameters chosen for the initial time $\tau_i$ and initial volume $V(\tau_i)$
are adjusted in such a way that they exactly
coincide with $\tau_0$ and $V(\tau_0)$
for the standard Langevin case
when $T(\tau = \tau_0)$ becomes $T_c$. In Fig. \ref{DCCfig14} we show the
statistical distribution of the $\sigma $-field and its temporal gradient
at $\tau=\tau_c =3 \, fm/c$ for the case when the system was prepared
at an initial temperature $T_i=300 \, MeV$. The final pion number distribution
(not shown) looks more or less identical to the upper part of Fig.
\ref{DCCfig12}, i.e. to the corresponding Langevin scenario being
prepared at $T=T_c$. From Fig. \ref{DCCfig14} one notices that indeed the
$\sigma $-field is still more centered around zero
than within the standard case (compare with Fig. \ref{DCCfig5}).
However, the distribution of the gradient $\dot{\sigma }$
has shifted towards a nonvanishing positive value
because of the forward drift experienced by the explicit
symmetry breaking term. This shift in the later distribution at $T\approx T_c$
then effectively influences the outcome especially in the final
pion yield distribution in a counteracting way compared to
the naive expectation.

Summarizing this subsection let us highlight once more the main outcome
of our investigations:
If in a heavy ion collision it will come to the formation of
an experimentally detectable DCC domain - say with at least 50 low momentum
pions - it very likely has to be a rare event with
the average yield $\expl n_{\pi } \expr $,
stemming from the zero mode fluctuations from a single domain,
still being considerably smaller than $50$.
A dedicated event-by-event analysis is then unalterable.
If, on the other hand, nature is more `obliging', it might also be that
$\expl n_{\pi } \expr \stackrel{<}{\sim} 50$ (as in some very speculative
quench scenarios), there exist again with some finite probabilities
on the percent level some events which contain a multiple in the number
of pions compared to the average.
One can consider those particular events as really unusual `pion bursts'.
Also for such a situation a dedicated event-by-event analysis is
definitely desirable.
For both cases, a possible detection of unusual fluctuations
would provide nontrivial evidence for the formation of DCCs
and the existence of the chiral phase transition.
Of course, all this is speculation, as also the whole issue of possible
DCC formation is. Which of the scenarios or assumed parameters
are realized in nature one does not know. A slow or moderate
expansion of the system within the Langevin scenario, which one may
consider as the most physical one, will indeed not result to
any verifiable signal.
In the next section we will discuss
in more detail on the statistical nature of the unusual distributions
found and on their experimental detection possibilities.

\subsection{Critical dynamical fluctuations}
\label{sec:critfluc}

In the last subsection \ref{sec:scenarios} we have
demonstrated that the probability distribution in
the number of the coherently emitted soft pions from the DCC decay is
nontrivial and nonpoissonian for a
sufficiently fast expansion in the various scenarios presented.
The statistical facets of these unusual probability distributions
are what we want to explore in more detail in this last subsection.

Although in fact the distributions in the pion number from a DCC might
be realized as such, one very likely can not
prove these directly from the experimental
measurement of the unusual pion number abundances, as there are
much more pions emitted independently at the late stage of
an ultrarelativistic heavy ion collision \cite{BGH98}.
Since one expects that the emission of the soft pions would be
affected most significantly compared to the moderate or
high (transverse) momentum pions, one has to consider to allow for a
low momentum $p_t$-cut in the data to enhance
significantly (and sufficiently) the signal to background ratio.
With this at hand
we will then show in the later part of this subsection
that indeed the unusual fluctuations might still be
clearly visible and thus provide a very interesting
and new event by event signature for DCC formation to be analyzed via a
cumulant expansion in the (to be)
measured low momentum pion number distribution in a given rapidity interval.

In order to account for the true higher order correlations
of the statistical distribution $P(n_\pi)$ we consider as a characteristic
tool an expansion in factorial cumulants $\theta _m$.
For a rather brief introduction and some further properties and analysis
we refer to appendix \ref{app:Cumulant}.
The factorial cumulant $\theta_m$ of order $m=1,2,3, \ldots$
represent the non-trivial statistical $m$-point correlations
of the distribution.

In \cite{ES92} it was stated
that so called bin-averaged factorial reduced cumulants
for higher than two (i.e. $m\ge 3$)
are consistent with zero when analyzing
the particle multiplicity of (lighter) nucleus-nucleus collisions
(and contrary to hadronic collisions). From this fact Elze and Sarcevic
then motivated to describe the occurring multiparticle density flucutations
in such reactions by means of a (gaussian)
three dimensional statistical free field theory \cite{ES92}, and suggested
the conservative view that no first or second order phase transition
should be implied as long as there is {\it no} compelling evidence
in the data. Our situation, of course, is different as we (have to) assume
a rapid chiral phase transition to occur in order to mimic the
formation of DCCs.

\begin{figure}[h]
\centerline{\epsfysize=8cm \epsfbox{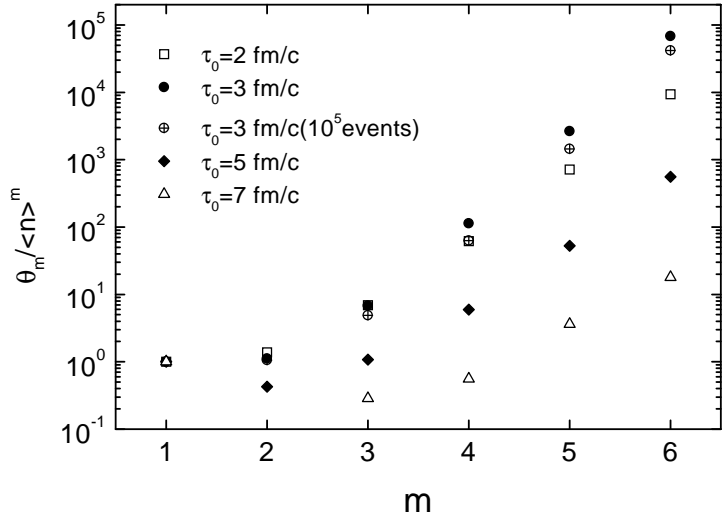}}
\caption{The reduced factorial cumulants for $m=1$ to 6
for the pion number distribution
obtained within the (markovian) modified Langevin scenario (D=3)
for different initial proper
times. The initial volume is taken as $V(\tau_0 )=100 \, fm^3$.
The average pion number $\expl n_\pi \expr$ within $10^4$ events are 54, 18.5, 4.8 and 2.4,
respectively, corresponding to $\tau_0=$ 2, 3, 5 and 7 fm/c, respectively.
In addition the cumulants obtained for a distribution
for $\tau_0=3$ fm/c within a larger sample of $10^5$ events are also shown
to estimate the numerical error.}
\label{DCCfig15}
\end{figure}

With the probability distribution of the soft pion number obtained
numerically within our model we can calculate the factorial moments and
subsequently the factorial cumulants.
In appendix \ref{app:Cumulant} we have stated explicitly the first
six factorial cumulants expressed via the usual factorial
moments of the distribution.
In Fig. \ref{DCCfig15} we show the first six {\it reduced} factorial cumulants
$\theta_m/<n_{\pi}>^m$ in a logarithmic scale for different expansions
simulated by $D=3$ and varying $\tau_0$
within the modified Langevin scenario (see also the lower part of
Fig. \ref{DCCfig12}). Here
$<n_{\pi}>\equiv \expl n_\pi \expr \equiv  \theta_1$
denotes the average pion number of the corresponding distribution.
Each distribution, except one,
was sampled by $10^4$ independent events
employing the markovian equations of motion. For the one remaining
distribution (with $\tau_0=3$ fm/c) $10^5$ events were generated
in order to estimate the possible error.
One immediately realizes the striking behaviour that the higher order
and reduced factorial cumulants $\theta_m/<n_{\pi}>^m$
with $m\ge 3$ are clearly nonvanishing
and in fact show an exponentially increasing tendency, as for each
distribution the higher order reduced cumulants lie more or less on
a straight line in this logarithmic representation!
Comparing the results for the two distributions for $\tau_0=3$ fm/c,
where one was generated with a sample of $10^4$ events and the second
one with a sample of $10^5$ events to account for higher statistics,
we can estimate the error for the higher lying factorial cumulants for a
sample of $10^4$ events to be still in the order of factor of two.
This can easily be traced back to the obvious fact that the higher
factorial cumulants depend most sensitively on the
tail of the numerically generated distribution
with large multiplicity $n$.
On the other hand the general trend of exponentially increasing reduced
factorial cumulants is not affected by the higher statistics
\footnote{
On our suggestion, an expansion in higher order cumulants in the
distribution of the so called enhancement factor $A_0$,
as given in \cite{KS99}, shows exactly the same tendency of exponentially
increasing reduced cumulants \cite{JS99}.
As the model there is different from ours,
this repeated finding points towards some `universal' behaviour.}.
This behaviour suggests a special sort of dynamical scaling as
the (at least) higher order factorial cumulants approximately take the form
\begin{equation}
\label{Cumchar}
\theta_m \, \approx \, a e^{\alpha \, m} <n>^{m} \, ,
\end{equation}
where $\alpha$ and $a $ denote constant numbers,
depending on the
parameters chosen for sampling of the distribution,
and $<n>\equiv \expl n_\pi \expr$ just represents the average number of
pions of the distribution $P(n_\pi )$.
With this asymptotic form (\ref{Cumchar}) for the factorial cumulants
$\theta_m$ (now assumed also for $m=1$ and $m=2$) one can
actually invert the expansion and find the corresponding distribution
$\bar{P}(n)$ giving raise to such characteristic factorial cumulants.
This is briefly worked out in appendix \ref{app:Cumulant}.
The resulting distribution, reflecting for the
higher order factorial cumulants,
is given by a {\em shifted} Poisson distribution
\begin{equation}
\label{Distrchar}
\bar{P}(n) \, = \, \frac{a^{n'}}{(n')!}\, e^{-a} \,
\end{equation}
where $n= (<n> e^{\alpha })\cdot n'$ and $n'=0,1,2,\ldots $.
As generally $a$ is positive and a small number - typically from the above
slopes one has $a\ll 1$ - and $ \expl n_\pi \expr e^{\alpha }$ is
some {\em multiple} of the average number $\expl n_\pi \expr $,
the deduced distribution (\ref{Distrchar}) provides a nice intuitive
and intriguing picture for the unusual events:
Occasionally, a semi-classical `pion burst' with pion number
$n_\pi = (\expl n_\pi \expr e^{\alpha })\cdot n'$ is being emitted for
some special events. These represent rare events as the distribution in $n'$
follows a standard Poisson distribution sharply peaked at $n'=0$.
Such rare and unusual events
are then in fact quite similar to the Centauro candidates \cite{La80}!
We do not want to push this interpretation too far, as smaller
deviations from the straight exponential fit and, of course, the two lowest
factorial cumulants are not considered. Yet we believe that
this interpretation provides the right
intuitive way of describing the unusual strong fluctuations in the tail
of the distribution.

At this stage one might indeed ask for the physical origin of
such a peculiar and scaling-like behaviour of the fluctuations.
Here we can provide at present no definite answer, as we can only
rely on our numerical findings. For a given ensemble of initial configurations,
the stochastic approach presented in this work results in an ensemble of
widely differing solutions. Since a gaussian initial distribution in the fields
under the time evolution of a quadratic Hamiltonian always stays gaussian,
we believe that the unusual final fluctuations in the present case
originate due to the particular nonlinear evolution.
In principle, the
occurrence of some sort of peculiar scaling behaviour in higher
order factorial moments, is
known (or speculated) for quite a time to show up in the multiplicity
fluctuations stemming from a quark-hadron phase transition
(in hadron-hadron or heavy ion collisions) described within a simple
phenomenological Ginzbug-Landau framework \cite{Hwa98}.
In this respect our findings underline the necessity to learn more
about the possible onset of a phase transition by a careful study of
final multiplicity fluctuations.

On the other hand one also clearly recognizes that the second order factorial
cumulant
increases drastically compared to any `usual gaussian' second order cumulant
on the order of the first order cumulant and thus defines a much broader
distribution. This increase in $\theta_2$ is due to the
fact that many trajectories of the sample enter temporarily
the instable region with $\mu_{\bot }^2<0$. In fact for the more
dramatic cases we have $\theta_2 \gg \theta_1$. In appendix \ref{app:Cumulant}
we briefly show that for a situation, where $\theta_2 > \theta_1$, there
exists no simple statistical distribution which can be expressed solely in 
terms of the first two factorial cumulants.
This thus signals again that the
statistical nature of the distribution is highly non-trivial!

\begin{figure}[h]
\centerline{\epsfysize=10cm \epsfbox{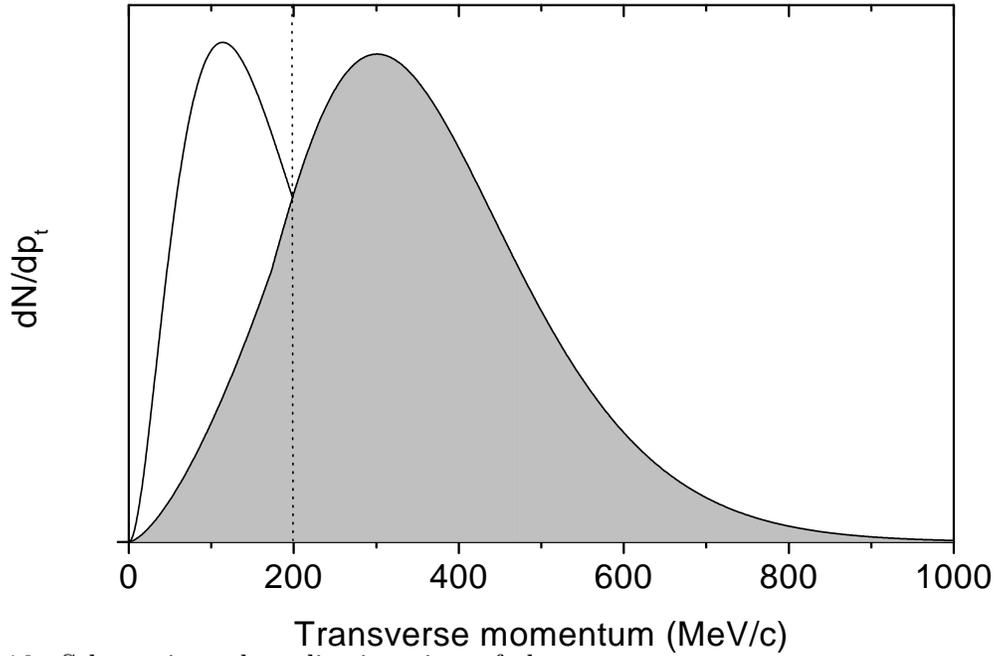}}
\caption{Schematic and qualitative view of the transverse momentum spectrum
of charged pions for one single event within some definite rapidity interval
including a single hypothetical and sufficiently prominent DCC candidate.
(This spectrum has been schematically redrawn from a simulated
event of background pions to be expected at RHIC energies \protect\cite{BGH98}.)}
\label{DCCfig16}
\end{figure}

It remains to shed light on the possibility whether such unusual fluctuations
can indeed also be reflected in the cumulant expansion based on the data
measured in real ultrarelativistic heavy ion collision experiments. In
the real world one expects a huge `background' of pions not coming from the
decay of a `large' DCC domain, as already outlined at the beginning of this
section  \ref{sec:stochDCC}. To illustrate such a background we show in 
Fig. \ref{DCCfig16} a schematic and qualitative expectation of a single
event of the transverse momentum spectrum of
charged pions within some definite rapidity interval including a single
hypothetical and sufficiently prominent DCC candidate. Such a spectrum has been
schematically redrawn from a single simulated event of background pions to be
expected at RHIC energies \cite{BGH98}. A `large' DCC domain would eventually
enhance the number of soft pions in the pion spectrum at sufficiently low
momenta (see Fig. \ref{DCCfig16}). The authors of \cite{BGH98}
provide a detailed analysis that if allowing for a low momentum $p_t$-cut of
$p_t<200\, MeV$ in some small and definite interval of rapidity (of order one)
the expectation is that one should have a surplus of at least 50 pions stemming
from a DCC per unit rapidity in such a window for a possible `direct'
observation. Even
in such a small window, however, if supposedly large DCC domain occurs,
there will be still a background
of `normal' pions of the order of 50 in average. Therefore the inherent
fluctuations of the background pion number in low $p_t$ makes it rather
difficult to find out a clear trace of the DCC formation in the soft pion
enhancement within one event.
One then has to go to an appropriate statitical analysis for discovering 
possible unusual fluctuations. More importantly,
as we have stated in the last subsection, `larger' DCC domains
are more likely to be some rare events. We thus want to pursue in the following
by means of the cumulant expansion whether there is a possibility to look
experimentally for unusual fluctuations when allowing for some
additional incoherent and simple fluctuating poissonian source producing also
low momentum pions and thus providing the background.

By now there have been two experimental investigations
to look for DCC events either in heavy ion reactions
at the CERN-SPS \cite{WA98}
or in $p-\bar{p}$-collisions at the Tevatron at
Fermilab \cite{Br96}. Both programs had so far a negative outcome in their
searches. This might still well be due to the fact that up to now no
analysis employing a sufficiently small low momentum cut
has been carried out. In addition, also a wavelet-type analysis, as originally
been proposed by Huang and coworkers \cite{HSTW96}, might further help to
look for the occurrence of unusual events or - in respect to our
present work - of unusual fluctuations in sufficiently small
rapidity and momentum windows.
There exist also other clever suggestions how to filter for the DCC events, see
e.g. \cite{CC99}.

At this step we want to provide
a rough estimate in what range the typical rapidity interval is to be
expected for the pions to be emitted out of a single DCC domain.
The spherical (D=3-)scaling expansion ansatz in proper time
was chosen to mimic for the rapid expansion. In strict terms
such a scenario is fueled by everlasting sources and thus should break down
at some later decoupling time as the whole collision of two heavy ions
does last only a finite time. Before "freeze-out" the domains are separated
from the outside or exteriour vacuum by the surrounding and expanding matter.
This deficiency of everlasting sources
can be circumvented by a mapping of the idealized 3-dimiensional boost-invariant
evolution to quasifree, truncated sources evolving in normal time
at some decoupling
time as shown by Bjorken and coworkers \cite{ABL97}. Such a truncation of
the evolution modifies somewhat the final momentum spectrum of the emitted pions
\cite{ABL97}. In any case, within the idealized scenario evolving
solely in proper time at least at the beginning of the evolution, a simple
estimate for the rapidity interval is given by
$$
\Delta \eta \, \approx \, \frac{1}{2} \ln \left( \frac{1+v_c}{1-v_c} \right)
\, \, ,
$$
where $v_c= \frac{r(\tau_0)}{\sqrt{\tau_0^2 + r^2(\tau_0)}}$ and
$r(\tau_0)=(\frac{3}{4\pi} V(\tau_0))^{1/3}$. For the parameters
employed ($\tau_0=2-7\,  fm/c $ and $V(\tau_0)=10-200 \, fm^3$)
this estimate implies a rapidity interval of $\Delta \eta \approx 0.2 -1$
for the low momentum pions to be emitted, in agreement with general expectation.

\begin{figure}[h]
\centerline{\epsfysize=8cm \epsfbox{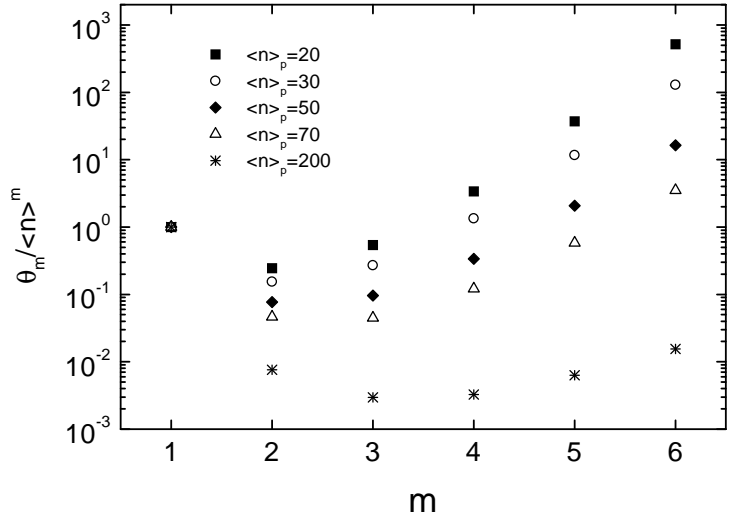}}
\caption{The reduced factorial cumulants for $m=1$ to 6 for the pion
number distribution of low momentum
pions stemming from two independent sources: A distribution
stemming from a single emerging DCC taken
from the lower part of Fig. \ref{DCCfig12} (modified Langevin scenario
with $D=3$ and $\tau_0=3$ fm/c) and a poissonian distributed background pion
source with different mean values $\langle n \rangle_P=20-200$.}
\label{DCCfig17}
\end{figure}

Suppose now that we have the following situation:
The soft pions are coming from either
a DCC domain or, independently, from an incoherent (`chaotic') source 
(background). Furthermore we assume the
emission of the incoherent soft pions follows a standard poissonian
distribution with
the mean value $<n_{\pi}>_P$, i.e.
\begin{equation}
\label{incsource}
P_P^{inc}(n) \, = \,
\frac{(<n_{\pi }>_P)^n}{n!}\,e^{-
<n_{\pi }>_P} \, .
\end{equation}
As the resulting
(factorial) cumulants in the independently combined
pion number distribution are additive (see appendix \ref{app:Cumulant}),
the reduced cumulants can thus simply be written as
\begin{equation}
\label{cumDCCinc}
\frac{\theta_m}{<n_{\pi}>^m}\, =\,
\frac{\theta_m^c+\theta_m^{inc}}{( <n_{\pi}>^c+
<n_{\pi}>^{inc} )^m}\,,
\end{equation}
where `c' denotes the coherent emission by a DCC state
and `inc' the incoherent
emission by the background source.
For $m \ge 2$
the cumulants related to the incoherent pion
source do vanish by assumption of a Poisson distribution.
(We note that if there are more than one single domain
contributing within a considered rapidity and momentum window,
and if these are truely uncorrelated,
the respective cumulants of each independent source would then again simply
add up for the combined pion distribution.)
Fig. \ref{DCCfig17} depicts
the resulting {\em reduced} factorial cumulants (\ref{cumDCCinc})
obtained for a single domain simulated with a
fast expansion ($D=3,\ \tau_0=3$ fm/c, modified Langevin scenario,
compare with lower part of Fig. \ref{DCCfig12} and Fig. \ref{DCCfig15})
and superimposed by the inclusion of
a background source with different mean values
$<n_\pi >_P$ ranging from 20 to 200 incoherent additional pions.
The last numbers can either be seen simply as basic uncertainty
and/or also as a result of lowering the $p_t$-cut.
The additional poissonian source basically lowers all the reduced
factorial cumulants
with $m>1$: As in its form (\ref{incsource}),
$P_P^{inc}(n) $ has no factorial cumulants $\theta_m $ with $m>1$,
the combined reduced factorial cumulants become smaller accordingly.
However, the
higher order ones for $m\ge 3$ are still appreciably large if
the background mean pion number is less than about 100,
especially with increasing number $m$.
If we consider as an example
the situation that $<n_{\pi}>_P$ is about
70, it shows that for a slow expansion $\tau_0=7$ fm/c
the reduced factorial cumulants of higher order $m \ge 3$
are still very small ($\le 10^{-3}$). This basically reflects the
suppression of
the few coherent emitted pions compared to the large background.
In contrast, however, for a fast expansion
$\tau_0=3$ fm/c, as depicted in Fig. \ref{DCCfig17},
where `large' DCC states are more likely to occur,
the {\em reduced} factorial cumulants of higher order are in the range
$1\sim 10$ and thus should be clearly visible and detectable.

We thus find that for sufficient fast expansion the reduced higher
order cumulants are still in the order $1\sim 10$, although the number of
incoherently emitted pions might in average be $3\sim 4$ times
larger as the averaged number
of DCC pions. (In our particular last example we have
$\expl n_\pi \expr_{DCC}=18.5$ compared to $<n_\pi >_P=70$.)
We therefore conclude that an experimental analysis
by means of the higher order
factorial cumulants for the low momentum pion number distribution
provides a well-suited indication
for the possible existence
(and to some lesser extent also for the identification)
of any DCC formation on an event-by-event analysis.
Event-by-event type analysis for getting additional new insight in
the underlying
physics of heavy ion collisions
(e.g. the process of thermalization)
has become quite popular over the last two years (see e.g. \cite{BK99}
and references therein). In this respect our work can be considered
as a special and ultimate scenario of what to expect in case of a
rapidly ongoing chiral phase transition associated with possible
DCC state formation.

\section{Summary}
\label{sec:summary}

In the present work we have elaborated in
detail within
an idealized, but microscopically motivated semi-classical
Langevin description on the statistical facets of the formation of
possible disoriented chiral condensates during and after the onset of
the chiral phase transition expected to occur
in ultrarelativistic heavy ion collisions.
Within
the Langevin treatment of the standard linear $\sigma $-model,
one can simulate,
on an event by event analysis,
the possible evolution of various DCC scenarios in a rather transparent form.
Our main focus and objective has been to understand the physical role
of dissipation and noisy fluctuations on the DCC phenomenon.
The advantage of the presented approach is that in contrast
to common mean-field treatments, which can only bring about
a deterministic description for the (ensemble) averaged evolution,
it allows for any possible branching of the dynamical trajectories
being especially important in the instability region.
Our Langevin picture is based on microscopic input, although one can
interpret the presented approach more intuitively also in the spirit
of the phenomenological Landau-Ginzburg description of phase transitions.
Our ideas could also be taken over for situations advocating
a first order transition within the linear $\sigma $-model \cite{SD99},
in order to study for such parametrizations
of the effective temperature dependent potential
the influence of dissipation and fluctuation on the
evolution of the order parameter inside the nucleating and growing bubbles.

The model, originally being first proposed in \cite{PRL},
is based on the very assumption that the high-momentum
particles (`hard' fluctuations) of the chiral fields
constitute a heat bath which behaves locally
thermalized in the expanding system. The interaction of the nonequilibrated
`soft' chiral fields with this surrounding heat bath
then gives raise to their stochastic and semi-classical evolution
of Langevin type.
Our main conception is that the order parameter
as well as the pionic fluctuations before and after the onset
of the chiral phase transition still interacts (dissipatively) with
its surrounding of thermal
(or `hard') pions, which then results in large and tremendeously differing
fluctuations during the evolution.
Furthermore we have concentrated solely
on the effective dynamics of the collective zero
mode (order parameter and pionic fluctuations).
We have argued, that, if at all, the zero mode pionic fluctuations
become most unstable during the
roll-down period and thus are the ones being predominantly amplified
for realistic initially small sized and separated expanding domains.
The overall picture of possible DCC evolution
resembles the one proposed by Bjorken and coworkers \cite{Bj92,ABL97}.

As a first application we considered the finite size fluctuations of
the order parameter and the chiral pionic fields for a given volume
and temperature, resulting in a further smoothening of the crossover
behaviour around the critical temperature
within the employed linear $\sigma $-model.
The Langevin description provides a powerful and simple tool
for generating in a systematic and efficient manner a canonic
sample of statistically possible
configurations at a given temperature $T$.
As a reasonable (or minimal) assumption
we consider the order parameter and
the chiral fields to be likely thermally distributed when the
phase transition during the later expansion starts to occur.
This sampling of possible initial configurations contrasts to
the ad hoc guesses for the initial conditions made in many of the previous
works on DCC physics and thus enables us
to investigate characteristic statistical properties of DCC formation.

We have then concentrated on the dynamical evolution of one
single domain during and after the onset of spontaneous chiral
symmetry breaking at the later stages of the initially very hot
system expected to occur in ultrarelativistic heavy ion collisions.
Because of the
collective expansion at these later stages,
the temperature will subsequently drop below the critical one,
and smaller, originally chirally restored domains (assumed to be independent
being separated spatially and in rapidity)
start to form together with a thermalized
background of (quasi-)pions and possibly other hadronic excitations
within the respective expanding subsystem.
A $D$-dimensional scaling expansion was employed
to account for the collective expansion resulting
in an additional Rayleigh or Hubble like damping term within the stochastic
equations of motion.

We stressed the important issue of the physical effect of dissipation
on the pionic fluctuations for any possible DCC evolution.
For the quantification of the resulting strength
of the coherent pionic zero-mode field and as an
experimentally more direct and relevant quantity
we considered the effective pion number content $n_\pi$
(via eq. (\ref{O4pionnumb})) of the emerging final oscillations
in the chiral pionic fields.
The dissipation kernel has been calculated by means of a standard
finite temperature field theory technique and is directly associated
to the inverse thermal scattering rate of the soft mode
on the thermal particles.
Our analysis clearly shows that the (rapid) expansion, i.e.
the Hubble damping term, has to be at least
as efficient in order to compensate for the true dissipation.
For a larger dissipation coefficient $\eta$ the final yield in the pion
number would be correspondingly smaller, as
the dissipation damps accordingly faster any
large DCC like pionic fluctuations which
have possibly emerged after the roll-down.
Although our estimate for the dissipation close to the critical
point is inspired by physical
arguments, further understanding of the certainly complex dissipative nature
of the chiral phase transition is crucial:
If one can show
that the experienced dissipation for the pionic (transversal)
modes close to the transition point
is in fact much stronger than the one we have employed,
then there is definitely no chance at all for any DCC signals to be seen
in heavy ion collisions. On the other hand, as emphasized at the end of
section \ref{sec:zeromode}, in the deeply broken phase
much below the critical temperature the associated dissipation for the
pionic fluctuations will certainly be reduced by the additional chiral
$\sigma $-meson exchange for the $\pi - \pi $-scattering amplitude
as compared to our employed estimate.
For the deeply broken phase then also
the decay $\sigma \rightarrow \pi \pi$ becomes
possible giving potentially raise to a much stronger dissipation
for the fluctuations in longitudinal direction. All this might
very well at least quantitatively effect some of our results
and conclusions presented concerning the possible survival of
DCC states and would require an even more involved
and detailed calculation.

In addition, we also have described (in appendix \ref{app:Noise})
how to numerically realize colored noise in
order to treat the
underlying dissipative and nonmarkovian stochastic equations of motion.
In general dissipation as well as the associated noisy fluctuations
are non-local phenomena in time.
This, to the best of our knowledge, is the first
numerical treatment of
nonmarkovian Langevin equations in thermal quantum field theory
and might certainly be of relevance for other related topics.

In the last section we have then given
a comprehensive numerical study for
the possible formation of DCC, i.e. the coherent amplification of the pionic
chiral fields,
for various parameter sets and also for
four somewhat different scenarios.
It shows, as pointed out the first time by Randrup \cite{Ra96a},
that a rather rapid expansion is mandatory to have any significant
chance for obtaining `large' DCCs which then might lead to some
experimental consequences. On the other hand, our analysis
has provided the at first sight more pessimistic view,
that even then, a DCC event has very likely to be an unusual and rare
event. The {\em average} characteristic $\expl n_\pi \expr$,
i.e. the average number of low momentum pions
being emitted of the final pionic modes,
shows only a moderate behaviour, which then should result, on the average,
in a mild
increase of the transversal low momentum spectrum in the pions.
As we have argued, such a mild increase is probably
tremendously difficult to observe directly
and unambiguously from the average momentum spectrum of pions.

However, the statistical distribution $P(n_\pi)$ of emitted pions
shows a striking nonpoissonian and nontrivial behaviour.
There exist within some still finite probability
some rare and unusual events which contain a multiple in the number
of pions compared to the average.
As pointed out in the subsequent analysis of the
statistical nature of such distributions,
one should indeed interpret those particular events as unusual
and semi-classical `pion bursts' similar
to the mystique Centauro candidates \cite{La80}.
This result suggests a very important conclusion:
If DCCs
are being produced, an experimental finding will be a rare event
following a strikingly, nontrivial and nonpoissonian distribution.
A dedicated event-by-event analysis for the experimental
programs (e.g. the STAR TPC at RHIC) is then unalterable.

We clearly have to say once more, that, of course, all the above conclusions
represent speculation,
as also the whole issue of possible
DCC formation is. Which of the scenarios or assumed parameters
considered in our work are realized in nature one does not know.
Any slow or moderate expansion of the system
will indeed not result to any verifiable signal!
In addition, because of the various approximations and idealized scenarios
considered, our work should not be seen as directly comparable
to any experimental data. In this respect, our final last
theoretical conjecture,
which we now want to summarize, has to be seen as a fascinating and
experimentally possible guideline for a future analysis
of the pion spectra to be taken at RHIC or already taken at CERN-SPS, if
DCC like phenomena occur in ultrarelativistic heavy ion reactions.

For any meaningful experimental
identification our results imply to look for rare and unusual strong
fluctuations on an event by event analysis in certain rapidity and
sufficiently low
$\langle p_t \rangle$ windows.
The further analysis of the unusual distribution
in the pion number associated to a rapid chiral phase transition
we have invoked by means of the factorial cumulants $\theta _m$,
which represent a powerful tool, well-known in the analysis
of final multiparticle fluctuations in high energy hadronic reactions.
We have found the striking behaviour that the higher order
and reduced factorial cumulants $\theta_m/<n_{\pi}>^m$
with $m\ge 3$ show an abnormal, exponentially increasing tendency.
This we consider as the most important outcome of our extensive
investigation.
In addition, we also found that the second order factorial cumulant
$\theta_2$ increases dramatically compared to any `usual' gaussian
distribution, thus characterizing a much broader distribution.
This broadening reflects the fact
that many trajectories of the sample have entered temporarily
the instable region.
In addition, we have allowed that on top of the pions emerging from
the decay of collective pionic modes a further incoherent
and poissonian background source of low momentum pions might
in fact overshadow or even completely wash out these striking characteristics.
As it turned out, however, the reduced higher
order factorial cumulants are still of the order $1\sim 10$, if the number of
incoherently emitted pions is already in average $3\sim 4$ times
larger than the average number
of DCC pions.

We therefore strongly advocate that an analysis
by means of the higher order cumulants
serves as a new and powerful signature to identify
any unusualities associated with potential DCC formation.
Of course we are aware that our last analysis assumes, that
within each window in momentum and rapidity, where the experimental
analysis is considered, a DCC like phenomena with conditional probability
equals to one has occurred. This might not be the true case.
However, we believe that our suggestion for future experimental analysises
is in fact rather `simple' to carry out and
represents most likely the only way to find (any) evidence for unusualities
in the low momentum pion spectra. If such an analysis turns out to be negative,
there is probably no other chance to look for the DCC phenomenon.

\acknowledgments
This work has been supported by BMBF and GSI Darmstadt.
This work has also been supported by a joint project of the
Deutsche Forschungsgemeinschaft and
the Hungarian Academy of Sciences (MTA) (project No. 101/1998)
and by the Hungarian National Research Fund
(OTKA) (project No. T019700).
The authors thank U.~Mosel for constant interest throughout the work.
C.G. thanks T.S.~Biro, M.~Greiner, S.~Leupold
and D.~Rischke for enlightening discussions
and also the Institute for Nuclear Theory
at the University of Washington, the organizers of the workshop
INT-99-3 for their hospitality and the Department
of Energy for partial support during the
completion of this work. C.G. would also like to thank J.~Serreau
for analyzing the distributions
presented in \cite{KS99} within our proposed cumulant expansion
when attending the INT workshop.


\appendix

\section{The dissipation kernel $\Gamma ({\bf k}=0,\omega)$}
\label{app:Dissip}

In this appendix we evaluate the frequency dependence of the dissipation kernel
$\Gamma ({\bf k}=0,\omega)$ used in section \ref{sec:nonmarkov} for studying
the nonmarkovian dissipative evolution of the chiral fields.

For this we had employed the {\it sunset} diagram from standard $\Phi^4$-theory
generalized to the present $O(4)$-case, which will result in a different
numerical coefficient and will be specified at the end of the appendix.

In the $\Phi^4$-theory the respective dissipation kernel
for soft modes (with $|{\bf k}|\le k_c$) interacting with the hard modes
(with $|{\bf k}| > k_c$) separated by a momentum cutoff $k_c$
is given as
\begin{equation}\label{a1}
\Gamma({\bf k},\omega)=\frac{i{\cal M}({\bf k},\omega)}{2\omega}
\, \left( \,  \equiv \,
\frac{-{\mbox{Im}} \Sigma^{\mbox{ret}}({\bf k},\omega)}{\omega} \right)
\,,
\qquad |{\bf k}|\le k_c\,,
\end{equation}
where the memory kernel ${\cal M}$ introduced in \cite{GMu97} reads
\begin{eqnarray}
\label{memory}
i{\cal M}({\bf k},\omega)=\frac{\pi}{24}g^4&&\int_{k_c}
\frac{d^3q_1 d^3q_2}{(2\pi)^6}\,\,\frac{1}{\omega_1 \omega_2 \omega_3}\,
\theta(|{\bf k}-{\bf q}_1-{\bf q}_2|-k_c)\,\times \\
\!\!\!&\{& \, \!\!\![ \,  (1+n_1)(1+n_2)(1+n_3)-n_1n_2n_3 ]
\delta(\omega-\omega_1-\omega_2-\omega_3)\nonumber\\
\!\!\!&+& \, \!\!\![ \, (1+n_1)n_2(1+n_3)-n_1(1+n_2)n_3 ]
\delta(\omega-\omega_1+\omega_2-\omega_3)\nonumber\\
\!\!\!&+& \, \!\!\![ \, n_1(1+n_2)(1+n_3)-(1+n_1)n_2n_3 ]
\delta(\omega+\omega_1-\omega_2-\omega_3)\nonumber\\
\!\!\!&+& \, \!\!\![ \,  n_1n_2(1+n_3)-(1+n_1)(1+n_2)n_3 ]
\delta(\omega+\omega_1+\omega_2-\omega_3)\nonumber\\
\!\!\!&+& \, \!\!\![ \, (1+n_1)(1+n_2)n_3-n_1n_2(1+n_3) ]
\delta(\omega-\omega_1-\omega_2+\omega_3)\nonumber\\
\!\!\!&+& \, \!\!\![\,  (1+n_1)n_2n_3-n_1(1+n_2)(1+n_3) ]
\delta(\omega-\omega_1+\omega_2+\omega_3)\nonumber\\
\!\!\!&+& \, \!\!\![\,  n_1(1+n_2)n_3-(1+n_1)n_2(1+n_3) ]
\delta(\omega+\omega_1-\omega_2+\omega_3)\nonumber\\
\!\!\!&+& \, \!\!\![ \, n_1n_2n_3-(1+n_1)(1+n_2)(1+n_3) ]
\delta(\omega+\omega_1+\omega_2+\omega_3)\ \},\nonumber
\end{eqnarray}
and $\quad{\bf q}_3:={\bf k}-{\bf q}_1-{\bf q}_2$,
$\quad\omega_i=\omega_{{\bf q}_i}$,
$\quad n_i=n(\omega_i)=\frac{1} {e^{\omega_i/T}-1},\quad i=1,2,3$.
The dispersion relation for the hard modes is taken as
$\omega_q=\sqrt{ q^2+m_p^2}$, where
$m_p$ denotes the dynamical mass. $i{\cal M}({\bf k},\omega)$
represents the net absorption rate for soft modes due to the interaction vertex
of a soft mode with three hard particles. The first and the last
term of (\ref{memory}) correspond to the decay of one soft mode
into three hard modes and the inverse
process. The other six terms correspond to the scattering process
$s+h \leftrightarrow h+h$.

For our study of stochastic DCC formation
we constructed an effective model for the chiral zero mode fields,
so that for the present purpose we take $k_c=0$ and
thus need only to calculate $i{\cal M}({\bf k}=0,\omega)$.
Evaluating the $\delta$-function
we will reduce the six dimensional integral of (\ref{memory}) to a
1-dimensional integral which we then treat further numerically.
In principle this task had
already been performed by Wang and Heinz \cite{EH95} investigating the 2-loop
resummed propagator for hot $\Phi^4$-theory. However, repeating the steps
in their tedious derivation we found out that some
particular kinematic boundaries of the integration
variables were not extracted correctly.
In the following we sketch the main strategy
and then state the final result for the considered dissipation kernel.

It is easy to see that $i{\cal M}({\bf k},\omega)$ is
antisymmetric in $\omega$ and therefore $\Gamma({\bf k},\omega)$ is
symmetric. We thus only have to consider the case for $\omega \ge 0$.
In this case the contributions from the 4th, 6th, 7th and the last line
of eq. (\ref{memory}) are identical to zero. In addition
one sees that the contributions from the 2nd, third and the 5th
line are the same.
Moreover one can
convince oneself that the yield of the respective absorption
processes just gives a common factor $e^{\frac{\omega}{T}}$ compared
to the respective emission processes due to the standard detailed balance
relation for systems at thermal equilibrium. With these observations we have
\begin{eqnarray}\label{a2}
& &i{\cal M}({\bf 0},\omega)=\frac{\pi}{24}\frac{g^4}{(2\pi)^6}\left (
e^{\frac{\omega}{T}}-1 \right )
\int\!d^3q_1d^3q_2\,\,\frac{1}{\omega_1\omega_2\omega_3}\,
\,\times \nonumber\\
& &\qquad\qquad\qquad\qquad\qquad\quad \{3(1+n_1)n_2n_3\,
\delta(\omega+\omega_1-\omega_2-\omega_3)\nonumber\\[2mm]
& &\qquad\qquad\qquad\qquad\qquad\quad + n_1n_2n_3\,
\delta(\omega-\omega_1-\omega_2-\omega_3)\}\,.
\end{eqnarray}

\begin{figure}[h]
\centerline{\epsfysize=6cm \epsfbox{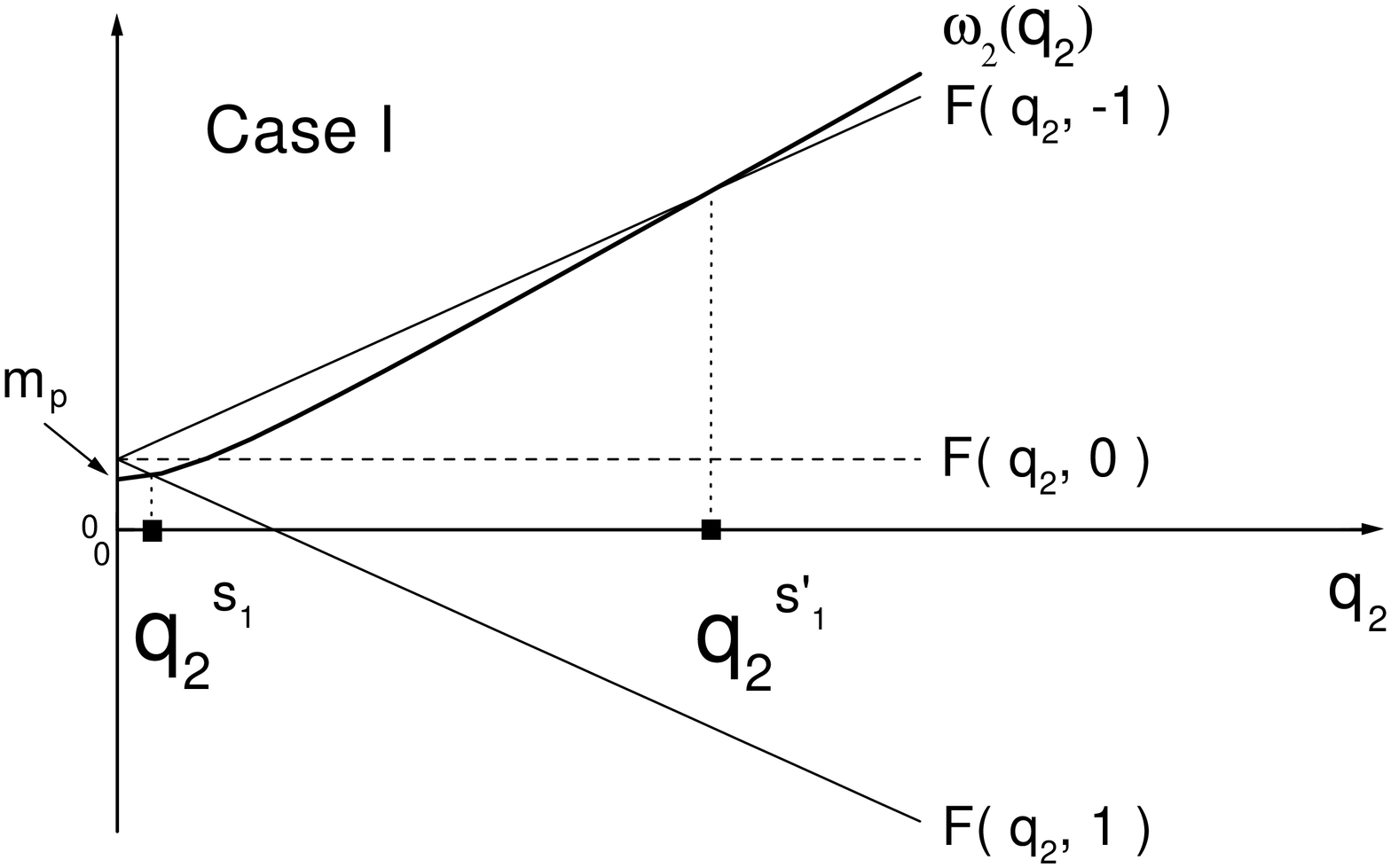}}
\centerline{\epsfysize=6cm \epsfbox{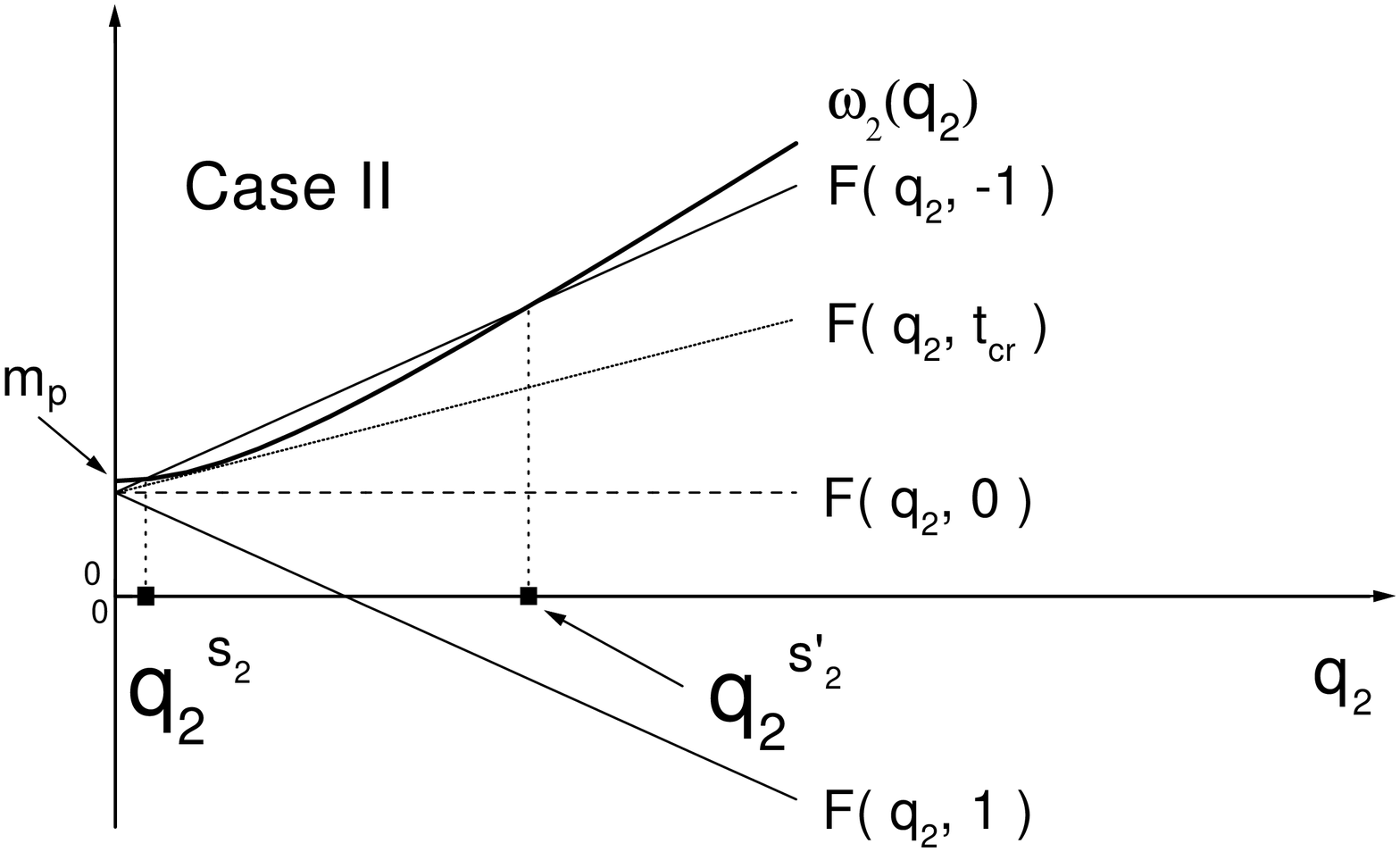}}
\centerline{\epsfysize=6cm \epsfbox{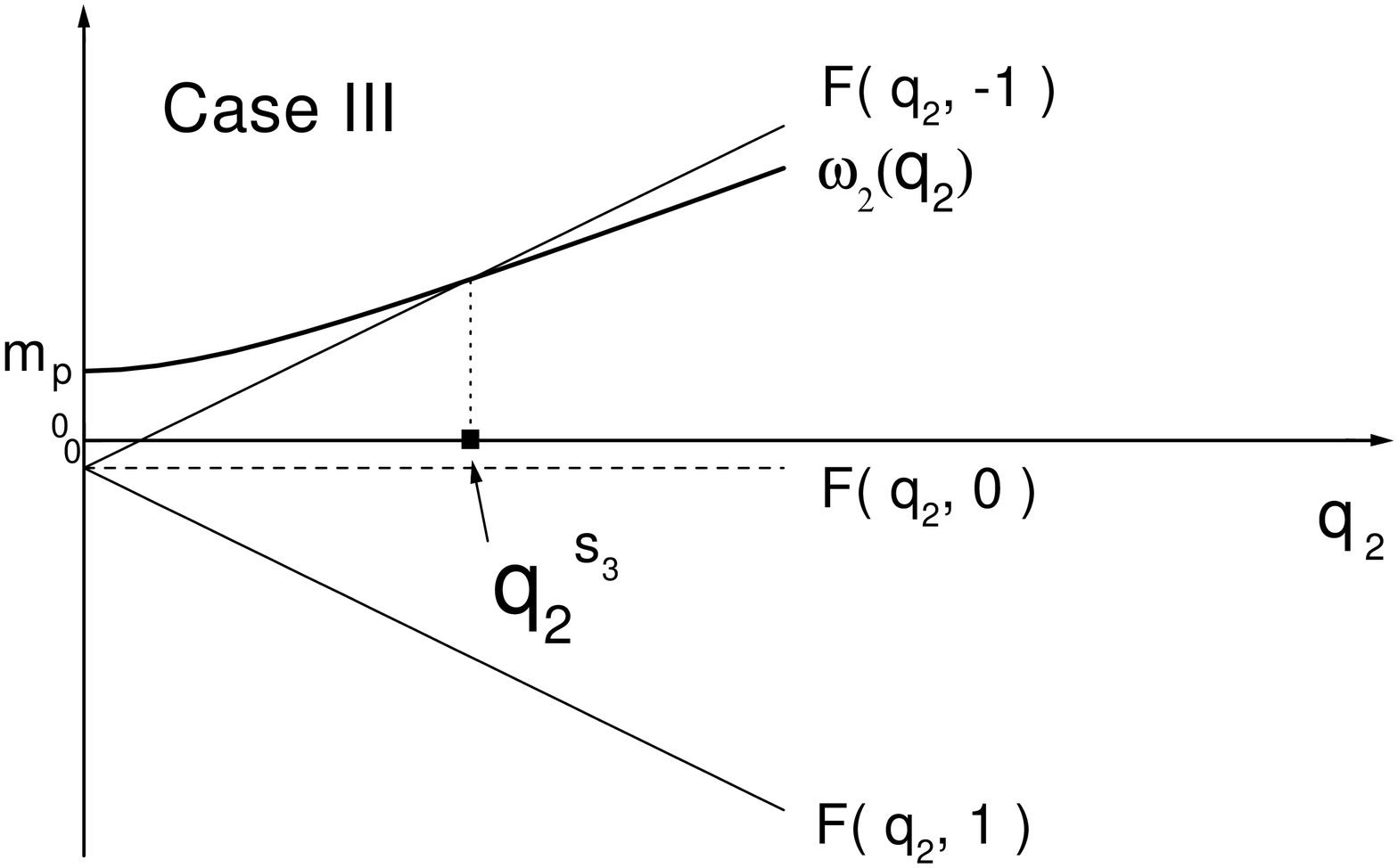}}
\caption{Illustration of the solutions and kinematical constraints for
$q_2$ and $t$
for the three cases (see text). For case I there exists one solution
for $q_2$ for every $t$. The
kinematical constraints $q_2^{s_1}$ and $q_2^{s'_1}$ are the points where the
lines $F(q_2,1)$ and $F(q_2,-1)$ cut $\omega_2(q_2)$. For case II there exist two
solutions for every $t\in [ t_{kr},-1]$. The kinematical constraints
$q_2^{s_2}$ and
$q_2^{s'_2}$ are the cut points between $F(q_2,-1)$ and $\omega_2(q_2)$. For
case III all solutions are larger than $q_2^{s_3}$. It is easy to show that
these solutions do not fulfill the equation (\ref{ec}).}
\label{DCCfig18}
\end{figure}

We now outline our strategy by manipulating the first integral
in (\ref{a2}) which corresponds to the emission rate of the scattering process
$h+h \rightarrow s+h$.
This integral can be reduced to a
3-dimensional integral
\begin{eqnarray}\label{gain}
g_1(\omega)&\!\!:=\!\!&3\int\!\!d^3q_1d^3q_2\,\,\frac{(1+n_1)n_2n_3}
{\omega_1\omega_2\omega_3}\,\delta(\omega+\omega_1-\omega_2-\omega_3)
\\
&\!\!=\!\!&24\pi^2\int_0^{\infty}\!dq_1dq_2\int_{-1}^1dt\,\,
q_1^2\,q_2^2\,\frac{(1+n_1)n_2n_3}{\omega_1\omega_2\omega_3}\,
\delta(\omega+\omega_1-\omega_2-\omega_3)\,, \nonumber
\end{eqnarray}
where $t=cos\theta$ and $\theta$ denotes the angle between ${\bf q}_1$ and
${\bf q}_2$. For the energy conservation stated by the $\delta$-function
one has
\begin{equation}\label{ec}
\omega+\omega_1-\omega_2\, = \, \omega_3\, = \,
\sqrt{q_1^2+q_2^2+2q_1q_2t +m_p^2} \,
\end{equation}
due to the momentum conservation ${\bf q}_1+{\bf q}_2={\bf q}_3$.
(\ref{ec}) represents the kinematical constraint
among the variables $q_1$,
$q_2$ and $t$. In order to determine from this equation
for a given frequency $\omega$
the kinematic boundaries for $q_1$,
$q_2$ and $t$ we take its square yielding
\begin{equation}\label{quadrat}
\left( F(q_2,t) \, := \right) \,
\frac{(\omega+\omega_1)^2-q_1^2}{2(\omega+\omega_1)}
-\frac{q_1}{\omega+\omega_1} q_2\,t=\sqrt{q_2^2+m_p^2}\,=\, \omega_2(q_2) \, .
\end{equation}
Taking $q_1$ as the most outer integration variable
we now consider it as a fixed constant and concentrate first on
the variables $q_2$ and $t$. The left side of
(\ref{quadrat}), which we define as a function $F(q_2,t)$, represents a
straight line in $q_2$ with different inclination for different values of $t$.
All straight lines for different $t\in [-1,1]$ cut at $q_2=0$. Then the solutions of
(\ref{quadrat}) for a fixed $q_1$ (and given $\omega $)
are the points where the bundle (in $t$) of
straight lines $F(q_2,t)$ cuts $\omega_2(q_2)$. There are three cases
to distinguish and which are classified by the position of
$F(q_2,t)$ at $q_2=0$: (case I) $F(q_2=0,t)\ge m_p$; (case II)
$0<F(q_2=0,t)<m_p$; and
(case III) $F(q_2=0,t)\le 0$. Fig. \ref{DCCfig18} illustrates the different
situations for the three cases and shows the kinematic boundaries of
$q_2$ and $t$.

We have to remark that the solutions of (\ref{quadrat}) do not necessarily
fulfill the original constraint (\ref{ec}). Therefore one has to
insert back the solutions into
(\ref{ec}) and check whether they indeed fulfill (\ref{ec}).
One finds out that the
solutions of case III do not obey (\ref{ec}) (as $\omega +\omega_1 - \omega_2
<0$).

The kinematic boundaries for the integration variable $q_2$ are
\begin{eqnarray*}
q_2^{s_1}\!\!\! \, &=& \, \!\!\!\frac{1}{2}\left ( \sqrt{B(\omega,q_1)}-q_1 \right )
\qquad,\qquad q_2^{s'_1}=\frac{1}{2}\left ( \sqrt{B(\omega,q_1)}+q_1 
\right )\,,\\
q_2^{s_2}\!\!\! \, &=&\, \!\!\!\frac{1}{2}\left (-\sqrt{B(\omega,q_1)}+q_1 \right )
\quad,\qquad q_2^{s'_2}=\frac{1}{2}\left ( \sqrt{B(\omega,q_1)}+q_1 
\right )\,,\\
\mbox{where\cite{EH95}}\!\!\!& &\!\!\!\qquad B(\omega,q_1)=\frac{(\omega+\omega_1)^2
\left [ (\omega+\omega_1)^2-q_1^2-4m_p^2 \right ]}
{(\omega+\omega_1)^2-q_1^2}\,.
\end{eqnarray*}
In order to fulfill the classification for case I (II) one finds
from the definition of $F(q_2=0,t)$ of (\ref{quadrat}) that the
energy $\omega$ has to be greater (less) than $m_p$.

One can now get the
kinematic boundaries of $q_1$ using the fact that the function $B(\omega,q_1)$
should not be negative. For case I one finds that $q_1$ has no further
constraints. For case II $q_1$
posesses a lower boundary $q_1^{cr}$:
\[
q_1^{cr}=\frac{1}{2\omega}\sqrt{(\omega^2-m_p^2)(\omega^2-9m_p^2)}\,.
\]

Eq. (\ref{gain}) can now be stated as
\begin{eqnarray}\label{gain1}
g_1(\omega)\!\!\! \, &=& \, \!\!\!24\pi^2\left \{ \theta(m_p-\omega)
\int_{q_1^{cr}}^{\infty}\!\!dq_1\int_{q_2^{s_2}}^{q_2^{s'_2}}\!\!dq_2
\int_{-1}^{t_{cr}}\!\!dt+\theta(\omega-m_p)\int_0^{\infty}\!\!dq_1
\int_{q_2^{s_1}}^{q_2^{s'_1}}\!\!dq_2\int_{-1}^1\!\!dt\right.\nonumber\\
& &\qquad\qquad\left.q_1^2\,q_2^2\,\frac{(1+n_1)n_2n_3}
{\omega_1\omega_2\omega_3}\,\delta(\omega+\omega_1-\omega_2-\omega_3)
\right \}\,.
\end{eqnarray}
By suitable substitutions for the integral variables,
\[
dt \to d\omega_3 = \frac{q_1q_2}{\omega_3}dt \ ,\ \mbox{and}\  dq_1, dq_2 \to dU_1, dU_2 \ \mbox{with}
\ U_i:=e^{-\frac{\omega_i}{T}}\ \  i=1,2\, ,
\]
one can carry out the integrations over $t$ and $q_2$. The result is
\begin{eqnarray}\label{result}
g_1(\omega)&=&24\pi^2T^2 \left \{ \theta(m_p-\omega)\int_0^{U(q_1^{cr})}\!\!
dU_1\,\,G_1(U_1;q_2^{s'_2},q_2^{s_2}) \right.\nonumber\\
& &\qquad\qquad\left.+\theta(\omega-m_p)\int_0^{U(0)}\!\!
dU_1\,\,G_1(U_1;q_2^{s'_1},q_2^{s_1}) \right \}
\end{eqnarray}
with
\begin{eqnarray*}
& &G_1(U_1;s_1,s_2):=\frac{1}{1-U_1}\,\frac{1}{1-U_1U_{\omega}}\,
ln\left [ \frac{(1-U(s_1))(U(s_2)-U_1 U_{\omega})}{(1-U(s_2))(U(s_1)
-U_1 U_{\omega})}\right ]\\
\mbox{and}\qquad \qquad & &U(s):=exp\left ( -\frac{\sqrt{s^2+m_p^2}}{T} \right )\,,\quad
U_{\omega}:=e^{-\omega/T}\,.
\end{eqnarray*}
The further evaluation of the second term in (\ref{a2}) corresponding to the
emission rate of the (inverse) off-shell decay process $h+h+h \rightarrow s $
follows in analogous but slightly more
complicated way to the strategy for the first term considered above.

We state the final result for $i{\cal M}({\bf 0},\omega)$:
\begin{eqnarray}\label{endresult}
\!\!\!& &\!\!\!i{\cal M}({\bf 0},\omega)=\frac{g^4T^2}{192\pi^3}
(1-U_{\omega})\times \nonumber\\
\!\!\!& &\!\!\!\left \{ 3\,\theta(m_p-\omega)\int_0^{U(q_1^{cr})}\!\!dU_1\,\,
G_1(U_1;q_2^{s'_2},q_2^{s_2})+3\,\theta(\omega-m_p)\int_0^{U(0)}\!\!dU_1\,\,
G_1(U_1;q_2^{s'_1},q_2^{s_1})\right.\nonumber\\
\!\!\!& &\!\!\!\left.+\theta(\omega-3m_p)\left [ 
\int_{U(q_1^*)}^{U(0)}\!\!dU_1\,\,G_2(U_1;q_2^{d'_1},q_2^{d_1})
+\int_{U(q_1^{cr})}^{U(q_1^*)}\!\!dU_1\,\,G_1(U_1;q_2^{d'_2},q_2^{d_2})
\right ] \right \}
\end{eqnarray}
with
\[
G_2(U_1;s_1,s_2):=\frac{1}{1-U_1}\,\frac{1}{U_1-U_{\omega}}\,
ln\left [ \frac{(1-U(s_1))(U_1 U(s_2)-U_{\omega})}{(1-U(s_2))(U_1 U(s_1)
-U_{\omega})}\right ]\,.
\]
Here the kinematic boundaries $q_2^{d_1}$, $q_2^{d'_1}$, $q_2^{d_2}$, $q_2^{d'_2}$
and $q_1^*$ are
\begin{eqnarray*}
q_2^{d_1}\!\!\! \, &=& \, \!\!\!\frac{1}{2}\left ( \sqrt{A(\omega,q_1)}-q_1 \right )
\qquad,\qquad q_2^{d'_1}=\frac{1}{2}\left ( \sqrt{A(\omega,q_1)}+q_1 
\right )\,,\\
q_2^{d_2}\!\!\! \, &=& \, \!\!\!\frac{1}{2}\left (-\sqrt{A(\omega,q_1)}+q_1 \right )
\qquad,\qquad q_2^{d'_2}=\frac{1}{2}\left ( \sqrt{A(\omega,q_1)}+q_1 
\right )\,,\\
q_1^*\!\!\! \, &=& \, \!\!\!\frac{1}{2}\sqrt{(\omega-m_p)^2-4m_p^2}\,,\\
\mbox{where\cite{EH95}}\!\!\!& &\!\!\!\qquad A(\omega,q_1)=\frac{(\omega-\omega_1)^2
\left ( (\omega-\omega_1)^2-q_1^2-4m_p^2 \right )}
{(\omega-\omega_1)^2-q_1^2}\,.
\end{eqnarray*}

(We note at this stage that we have obtained different results
for the lower kinematic boundaries $q_2^{s_2}$ and $q_2^{d_2}$
as compared to the ones given by Wang
and Heinz \cite{EH95}, which were there simply set as zero.)

\begin{figure}[h]
\centerline{\epsfysize=8cm \epsfbox{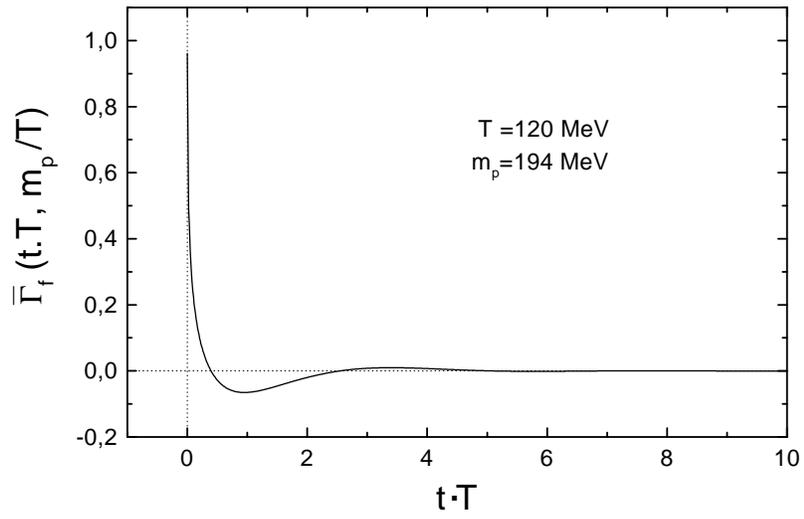}}
\caption{The reduced dissipation kernel ($\gamma_1$) in time
for $T=120$ MeV and a plasmon mass $m_p=194$ MeV.}
\label{DCCfig20}
\end{figure}

The dissipation kernel has then the form
\begin{equation}\label{a3}
\Gamma({\bf 0},\omega)=\frac{i{\cal M}({\bf 0},\omega)}
{2\omega}=:\frac{g^4T}{192\pi^3}\,{\bar \Gamma}\left ( \frac{\omega}{T},
\frac{m_p}{T} \right )
\end{equation}
where we have defined a reduced dissipation kernel
${\bar \Gamma}\left ( \frac{\omega}{T},\frac{m_p}{T} \right )$ which depends
only on $\omega/T$ and $m_p/T$. In fig. \ref{DCCfig19} we show the
reduced dissipation
kernel for $m_p/T=0.1$ and $m_p/T=1$.
Here $\gamma_1$ and $\gamma_2$ denote the two
different contributions to ${\bar \Gamma}$ from the
scattering and decay process. One recognizes that
$\gamma_2$ in fact diverges for $\omega \rightarrow \infty $.
One can furthermore show that $\gamma_1$ has an assymptotic behaviour
$\sim 1/\omega $ for sufficiently large $\omega$.
In our present study concerning stochastic formation of DCC
we have neglected the $\gamma_2$ contribution
and have thus only described the physical dominant scattering contribution
of $s+h \leftrightarrow h+h$. The dissipation kernel
$\gamma_1 (\omega )$ has its maximum very close to the on-shell frequency
$\omega = m_p $. Its shape with frequency is
thus effectively governed by two
characteristic and independent scales,
the temperature $T$ and the plasmon mass $m_p$.
The Fourier transform of
the reduced dissipation kernel,
i.e. ${\bar \Gamma}({\bf k}=0,t)\equiv \gamma_1(t)$,
for a temperature $T=120$ MeV near
the critical temperature $T_c$ is plotted in Fig. \ref{DCCfig20}. The mass
$m_p$ is
extracted as the transversal mass $\mu_{\bot }$ at that given temperature
from the right upper picture of Fig. \ref{DCCfig2}. By means of Fig.
\ref{DCCfig20} one can estimate that the
correlation in time of the kernel extents to about 5 fm/c.
 
On the plasmon mass shell $\omega=m_p$ the damping
coefficient $\Gamma({\bf 0},m_p)$ is obviously given
solely by the scattering contribution:
\[
\Gamma({\bf 0},m_p)=\frac{g^4\,T^2}{128\pi^3 m_p}\,(1-U(0))\int_0^{U(0)}
\!\!dU_1\,\,G_1(U_1;q_2^{s'_1}=q_1,q_2^{s_1}=0)\,.
\]
The above integration can be further simpified to
\[
(1-U(0))\int_0^{U(0)}\!\!dU_1\,\,G_1(U_1;q_1,0)=
f_{sp}\left ( 1-e^{-\frac{m_p}{T}} \right )
\]
where $f_{sp}(x)$ is the Spence function defined as
\[
f_{sp}(x):=-\int_1^x\!\!dy\,\,\frac{ln\,y}{y-1}\,.
\]
In the high temperature limit $m_p\ll T$ the dynamical mass $m_p$ is evaluated
by the tadpole diagram as $m_p^2=g^2T^2/24$. One thus recovers
\begin{equation}\label{damp}
\Gamma({\bf 0},m_p)=\frac{g^3T}{32\sqrt{24}\pi} \, ,
\end{equation}
which is twice the plasmon damping rate \cite{Par92,EH95}.

Generalizing from $O(1)$ to $O(N)$ one has an additional (pre-)factor
$\frac{N+2}{3}$ for the dissipation kernel and the dynamical mass.
Then for the case of the linear $O(4)$ $\sigma$-model one also has
to substitute $g^2 \rightarrow 6\lambda $.

\section{Numerical realisation of colored noise}
\label{app:Noise}

Here we outline a new numerical method for simulating nonwhite, i.e.
colored gaussian noise for an arbitrary (nonnegative and symmetric)
noise kernel $I(\omega )$.

For this consider the situation of a 1-dimensional Brownian particle
interacting with its thermal surrounding.
The force acting on the particle can be separated into a mean (dissipative)
part and a random, stochastic part.
This random force shall not depend on the state
of the particle and represents a fluctuating source
given as a particular noise sequence.

Suppose the noise in a time interval $[0,T]$ is composed of a series of pulses
which occur randomly \cite{Heer}.
Each puls can be written as $a\cdot b(\tau)$, where $b(\tau)$
has a certain uniform shape and $a$ denotes a random height
which is allowed to be either
positive or negative. Then a particular sequence of noise reads
\begin{equation}\label{noise}
\xi(t)=\sum_{i=1}^n a_i\,b(t-t_i)\,,\qquad t\in [0,T]\,.
\end{equation}
In (\ref{noise}) there are one random number $n$ and two random variables, the
height $a_i$ and the timing (center time) $t_i$. The random
number $n$ is now assumed to obey a Poisson distribution with the mean
value ${\bar n}=\mu T$, where $\mu$ denotes the mean counting rate. The
random distribution of $t_i$ shall occur uniformly.
Now one has to specify also the statistical distribution $p(a)$ of the height a.
In the limit of a large number of sufficiently weak pulses, i.e. large
$\mu$ and small mean-square value $\sigma^2$ of $p(a)$, the noise will then
approximately be given as a gaussian process due to the central limit theorem.
Gaussian noise
is solely characterized by the first two moments
\begin{eqnarray}\label{clt}
& & \expl \xi(t) \expr \,=0\nonumber\\
\mbox{and}\qquad \qquad & & \expl \xi(t)\xi(t') \expr \,=I(t,t')=\mu\sigma^2\int_0^T ds\,
b(t-s)b(t'-s)\,.
\end{eqnarray}
For this limiting case one can freely choose the distribution $p(a)$. Most commonly
one employs a gaussian distribution
\[
p(a)=\frac{1}{\sqrt{2\pi}\sigma}\,e^{-\frac{a^2}{2 \sigma^2}}\,.
\]
For a large $\mu$ the distribution of the puls number $n$ has a sharp maximum
centered at ${\bar n}$. Thus we set a fine time scale $\Delta t$ and assume
that on each time point one given pulse should occur. The number $n$ is then fixed
by $T/\Delta t$, and the timings $t_i$ of occurence of the pulses are also 
fixed. It remains to find out the form $b(\tau)$ which is related to the 
correlation function $I(t,t')$ of (\ref{clt}). If the correlation function
is a $\delta$-function then it is easy to show that $b(\tau)$
will be as well proportional to a $\delta$-function. For this case one denotes
the so constructed fluctuation $\xi(t)$ as white noise. In more general case
the noise is called colored noise.

For the simulation of gaussian colored noise we now assume
that $b(\tau)$ has a symmetric shape
within some time interval $[-\Delta,\Delta]$. Outside this interval $b(\tau)$
is taken 
to be zero. The noise $\xi(t)$ is a stationary process for 
$t\in[\Delta,T-\Delta]$. It means $I(t,t')=I(|t-t'|)=I(\tau)$ where
$\tau=t-t'$.(For $t\in[0,\Delta]$ and $t\in[T-\Delta,T]$ there are 
switching on/off artifacts.) Fourier transformation of (\ref{clt}) yields
\begin{equation}\label{sw}
I(\omega)=\mu\sigma^2\,|b(\omega)|^2\,.
\end{equation}
From the above equation it follows that one has to demand the Fourier transform
$I(\omega)$ of the correlation function of the noise should be nonnegative.
$b(\omega)$ is a real function due to the symmetry of $b(\tau)$. Further we
assume $b(\omega)$ to be positive. One thus ends with
\begin{eqnarray}\label{bt}
& &b\,(\tau)=\frac{1}{\sigma\sqrt{\mu}}\,G(\tau)\qquad
\mbox{for}\,\,\tau\in[-\Delta,\Delta]\\[0.12cm]
\mbox{with} \qquad \qquad & &
G(\tau):=\int_{-\infty}^{\infty}\frac{d\omega}{2\pi}\,
\sqrt{I(\omega)}\,e^{-i\omega \tau}\,.\nonumber
\end{eqnarray}

For the case of white noise with unit strength one has
$I(t-t')=\delta(t-t')$ and thus
\[
b(\tau)=\frac{1}{\sigma\sqrt{\mu}}\delta(\tau)\,.
\]
Then the sequence of white noise can be written as
\begin{eqnarray}\label{white}
\xi_w(t)&=&\sum_{i=1}^n a_i \frac{1}{\sigma\sqrt{\mu}}\delta(t-t_i)
\nonumber\\
&=&\sum_{i=1}^n \frac{1}{\sqrt{\mu}} {\bar a_i}\delta(t-t_i)\,,
\end{eqnarray}
where ${\bar a_i}$ can be sampled according to a gaussian distribution
with a unit mean-square value. Having fixed the pulse number $n$ by
$T/\Delta t=\mu T$, the mean counting rate is $\mu=1/\Delta t$. Furthermore
we approximate the $\delta$-functon as
\[
\delta(t_i)=\left \{
\begin{array}{r@{\quad:\quad}l}
\frac{1}{\Delta t} & t=t_i\\
0\, & t\neq t_i\,.
\end{array}
\right.
\]
Then the white noise at each time step can be simply generated as
\[
\xi_w(t_i)=\frac{{\bar a_i}}{\sqrt{\Delta t}}\,.
\]

Coming now back to the construction of a colored noise sequence, we
find that $\xi (t)$ can be obtained by an integral of
the history of a particular white noise sequence folded with the uniform pulse
$b(\tau)$ of (\ref{bt}), i.e.
\begin{eqnarray*}
\xi(t)\!\!\, &=& \, \!\!\sum_{i=1}^n a_i\,b(t-t_i)\\
\!\!&=& \, \!\!\sum_{i=1}^n a_i\int_0^T dt'\,\,b(t-t')\,\delta(t'-t_i)
=\int_0^T dt'\,\,b(t-t')\sum_{i=1}^n a_i\,\delta(t'-t_i)\\
\!\!&= \, &\!\!\int_0^T dt'\,{\sigma\sqrt{\mu}}\,b(t-t')\,\xi_w(t')\\
\!\!&=& \, \!\!\int_0^T dt'\,\,G(t-t')\,\xi_w(t')\,.
\end{eqnarray*}

\begin{figure}[h]
\centerline{\epsfysize=8cm \epsfbox{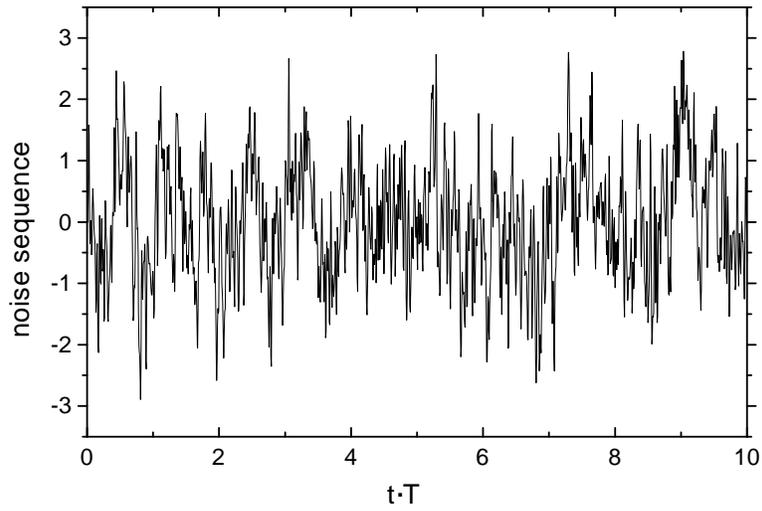}}
\centerline{\epsfysize=8cm \epsfbox{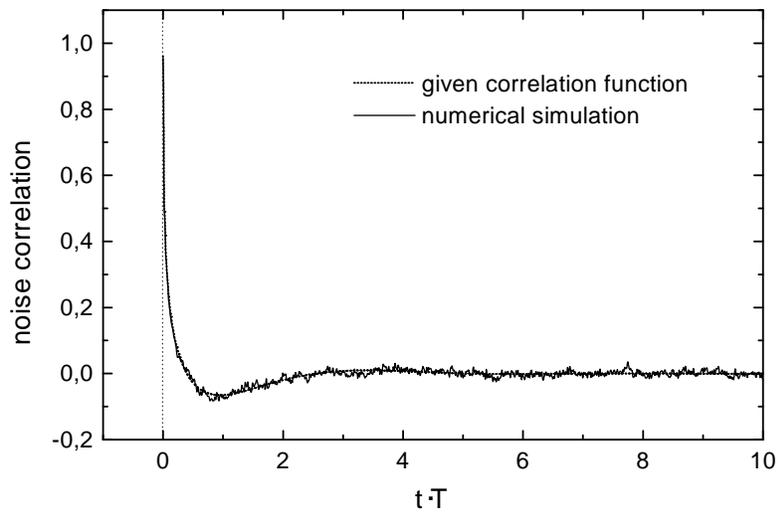}}
\caption{Comparsion of the correlation function
$I=\expl \xi(t) \xi(0) \expr$ of the numerically generated noise
with the given correlation function which is taken as
the reduced dissipation kernel
of Fig. \ref{DCCfig20} (lower part). The averaging is performed over an ensemble of
$10^4$ noise sequences.
In addition one exemplaric numerically generated noise sequence
dictated by the correlation function is also depicted (upper part).}
\label{DCCfig21}
\end{figure}

In order to check the reliability of our simulation we calculate numerically
the ensemble average of the noise correlation and compare it with the given 
correlation function for which we choose the reduced dissipation kernel 
plotted in Fig. \ref{DCCfig20}. The result is shown in Fig. \ref{DCCfig21}.
The depicted average was obtained by $10^4$
independently realized noise sequences.

\section{Cumulant expansion}
\label{app:Cumulant}

We give here a brief reminder of the factorial cumulant expansion
for discrete statistical distributions \cite{vKampen}. A stochastic number
$n$ is fully characterized by its
probability distribution $P(n), \, n=0,1,2,\ldots $.
An equivalent and convenient representation
of $P(n)$ is given by its probability generating function
\begin{equation}\label{generat}
F(1-x)\, := \, \sum_{n=0}^{\infty}\,(1-x)^n\,P(n)\,.
\end{equation}
If one defines the factorial moments $\phi_m$ by $\phi_0=1$ and
\begin{equation}\label{fmo}
\phi_m=<n(n-1)\cdots (n-m+1)>\qquad (m\ge 1)\,,
\end{equation}
then the
probability generating function becomes
\begin{equation}\label{generat1}
F(1-x)\, = \, \sum_{m=0}^{\infty}\,\frac{(-x)^m}{m!}\,\phi_m\,.
\end{equation}

The probability generating function also serves to generate the factorial
cumulants $\theta_m$, which are defined by
\begin{equation}\label{fcumu}
\ln \left( F(1-x) \right) \, := \,\sum_{m=1}^{\infty}\,\frac{(-x)^m}{m!}\,\theta_m\,.
\end{equation}
The factorial cumulants $\theta_m$ represent the nontrivial (`nonirreducible')
correlations at order $m$.

For a Poisson distribution,
\[
P(n)=\frac{\bar n^n}{n!}\,e^{-\bar n}\,,
\]
it follows immediately that its factorial moments read $\phi_m=\bar n^m$ and
consequently we have $F(1-x)=\exp (-x\bar n)$. Therefore the Poisson distribution
is characterized by the vanishing of all factorial cumulants except for
$\theta_1=\bar n$.

According to the definition (\ref{fcumu}) the factorial cumulants $\theta_m$ 
are generally combinations of
the factorial moments $\phi_i$ with ($i\le m$). For our use we list here the 
expressions up to order six, i.e.
\begin{eqnarray}\label{fcumu6}
\theta_1&=&\phi_1\nonumber\\
\theta_2&=&\phi_2-\phi_1^2\nonumber\\
\theta_3&=&\phi_3-3\phi_2\phi_1+2\phi_1^3\nonumber\\
\theta_4&=&\phi_4-4\phi_3\phi_1-3\phi_2^2+12\phi_2\phi_1^2-6\phi_1^4\nonumber\\
\theta_5&=&\phi_5-5\phi_4\phi_1-10\phi_3\phi_2+20\phi_3\phi_1^2
         +30\phi_2^2\phi_1-60\phi_2\phi_1^3+24\phi_1^5\nonumber\\
\theta_6&=&\phi_6-6\phi_5\phi_1-15\phi_4\phi_2+30\phi_4\phi_1^2-10\phi_3^2
           +120\phi_3\phi_2\phi_1\nonumber\\
& &-120\phi_3\phi_1^3+30\phi_2^3-270\phi_2^2\phi_1^2+360\phi_2\phi_1^4
   -120\phi_1^6\,.
\end{eqnarray}

Now we turn to the question of how - or whether it is
in general possible - to receive the probability
distribution $P(n)$, if all the factorial cumulants are given. Using
(\ref{generat}) one obtains
\begin{equation}
P(n)=\frac{(-1)^n}{n!}\left.\frac{d^n}{dx^n}F(1-x)\right |_{x=1}\,.
\end{equation}
Although one finds out from (\ref{generat}) and (\ref{fcumu})
\[
\sum_{n=0}^{\infty}\,P(n)=\left.F(1-x)\right |_{x=0}=1\,,
\]
the so inverted distribution $P(n)$ does not necessarily be positive
for all integers $n$. Therefore the general answer to the
question is `no'. For example assume the simple situation, where
all $\theta_m$ except $\theta_1$ and
$\theta_2$ are zero \cite{ES92,Mue71}. In this case $P(n)$
can serve as a probability
distribution only for $\theta_1\ge\theta_2$, i.e.
\begin{equation}\label{genGauss}
P(n)=\frac{\exp (\theta_2/2-\theta_1)}{n!}\,\left ( \frac{\theta_2}{2}
\right )^{n/2}\,(-i)^n\,H_n(i[(\theta_2-\theta_1)^2/2\theta_2]^{1/2})\,,
\end{equation}
where $H_n(z)$ is the Hermite polynomial of $n$th order. For 
$\theta_1<\theta_2$ it is easy to show that $P(1)$ is negative.
(This caveat has not been noticed in \cite{ES92,Mue71}.)
This means, that for situations, where
one finds that $\theta_2 > \theta_1$, there
exists no underlying statistical distribution
which can be expressed solely in terms of the first two factorial cumulants.

Another interesting example we want to discuss for the purpose
of analyzing the findings in subsection \ref{sec:critfluc} is
the case of a {\em shifted} Poisson distribution.
Here, to some upper limit, the resulting factorial cumulants
are situated in logarithmic representation on a curve,
which can be described to a good approximation by a straight line.
The shifted
poissonian distribution we introduce as
\begin{equation}\label{Xudistr}
P(n)=\frac{a^{n'}}{(n')!}\,e^{-a}\,,\quad n=B\,n'\,,
\quad n'=0,1\ldots\,,
\end{equation}
with $a$ and $B$ some positive constant.
According to (\ref{generat}) we have
\[
F(1-x)=\exp \left [
\sum _{m=1}^{B} \frac{(-x)^m}{m!}\,\frac{a\,B!}{(B-m)!} \right ]\,.
\]
For small $a\ll 1$ and large $B\gg 1$ the factorial
cumulants $\theta_m$ (with $m\le B$) are approximately given by
\begin{equation}
\label{appcumchar}
\theta_m \, \approx \,
a\cdot B^m \, .
\end{equation}
The higher factorial cumulants for $m>B$ vanish.

As a last reminder we consider a combined stochastic process
resulting in the discrete variable $n=n_1+n_2$ by two completely
independent stochastic processes $P_A(n_1)$ and $P_B(n_2)$.
The probability distribution $P_{A\cup B}(n)$ is then given by
\begin{equation}\label{pab}
P_{A\cup B}(n)=\sum_{\begin{array}{c} n_1,n_2;\\ n_1+n_2=n
\end{array}} \, P_A(n_1)\, P_B(n_2)\,.
\end{equation}
With this one finds from the definition (\ref{fcumu}) and (\ref{generat})
the factorial cumulants of the joined probability distribution $P_{A\cup B}(n)$
as
\begin{equation}\label{fcumuab}
\theta_m^{A\cup B}=\theta_m^A+\theta_m^B\,.
\end{equation}
The factorial cumulants of the independently combined variable are additive.



\newpage
\begin{table}[h]
\begin{tabular}{|c||c|c|}\hline
\multicolumn{3}{|l|}{(a) $D=1$} \\ \hline\hline
$\tau_0$ (fm/c) & 
\multicolumn{2}{c|}{$n_{\pi}$: the most prominent event/average}\\
\cline{2-3} & with dissipation & no dissipation \\
\hline
1 & 7/1.4 & 18/3.1 \\ \hline
0.5 & 20/3 & 34/5 \\ \hline
0.3 & 48/6 & 60/8 \\ \hline
\end{tabular}
\begin{tabular}{|c||c|c|}\hline
\multicolumn{3}{|l|}{(b) $D=2$} \\ \hline\hline
$\tau_0$ (fm/c) & 
\multicolumn{2}{c|}{$n_{\pi}$: the most prominent event/average}\\
\cline{2-3} & with dissipation & no dissipation \\
\hline
4 & 6/1.6 & 13/2.5 \\ \hline
2 & 22/3 & 24/3.7 \\ \hline
1 & 75/8.3 & 64/8.3 \\ \hline
\end{tabular}
\begin{tabular}{|c||c|c|}\hline
\multicolumn{3}{|l|}{(c) $D=3$} \\ \hline\hline
$\tau_0$ (fm/c) & 
\multicolumn{2}{c|}{$n_{\pi}$: the most prominent event/average}\\
\cline{2-3} & with dissipation & no dissipation \\
\hline
7 & 6.5/1.5 & 13/2.4 \\ \hline
3 & 32/3.9 & 28/4 \\ \hline
1.8 & 85/9 & 65/8.5 \\ \hline
\end{tabular}
\caption{The resulting pion yield for the most prominent
event and the average, respectively, obtained within different expansion
scenarios
simulated by the special choice of $D$ and $\tau_0$.
The calculations have been
performed by using the markovian Langevin equation (\ref{EOMDCC}).
The initial volume is chosen as $V(\tau_0)=100 \, fm^3$ for all cases.
For comparison we neglect
the dissipation (`no dissipation'), i.e. taking the damping
coefficient $\eta$ and the
noise as zero during the dynamical evolution of the order parameter.
The results are obtained within an ensemble of $10^3$ events.}
\label{dcctable1}
\end{table}

\newpage
\begin{table}[h]
\begin{tabular}{|c||c||c||c||c|}\hline
\multicolumn{5}{|l|}{$D=3$ and $T(\tau_0)=T_c$}\\ \hline\hline
$V(\tau_0)\setminus \tau_0$&3\,fm/c&5\,fm/c&7\,fm/c&10\,fm/c\\ \hline
10 $fm^3$& 33.2/2.9 & 13.6/1.8&6.5/1.5&7.2/1.4\\ \hline
25 $fm^3$& 62.8/3.7 & 13.4/2.0&7.0/1.6&5.3/1.4\\ \hline
100 $fm^3$&30.0/3.8 & 13.6/2.1&6.7/1.6&6.4/1.4\\ \hline
200 $fm^3$&25.0/3.7 & 11.1/2.0&7.4/1.7&5.5/1.4\\ \hline
\end{tabular}
\caption{Pion number of the most prominent event and the average
obtained within the
markovian Langevin scenario for different initial volumes $V(\tau_0)$. The
averages are taken over $2\cdot10^3$ events. The initial
proper time $\tau_0$ is also varied to simulate different expansion
scenarios.}
\label{dcctable2}
\end{table}

\newpage

\newpage

\newpage

\newpage

\newpage

\newpage

\newpage

\newpage

\newpage

\newpage

\newpage

\newpage

\newpage

\newpage

\newpage

\newpage

\newpage

\newpage

\newpage

\newpage

\clearpage

\end{document}